\pdfoutput=1
\documentclass{article}

\usepackage{arxiv}

\usepackage[utf8]{inputenc} 
\usepackage[T1]{fontenc}    
\usepackage{hyperref}       
\usepackage{url}            
\usepackage{booktabs}       
\usepackage{amsfonts}       
\usepackage{nicefrac}       
\usepackage{microtype}      
\usepackage{amsmath}
\usepackage{amssymb}
\usepackage{cleveref}       
\usepackage{graphicx}
\usepackage{subcaption}
\usepackage[numbers,sort&compress]{natbib}
\usepackage{doi}

\title{Data-Efficient Indentation Size Effect Mitigation in Steels Using Machine Learning and a Physics-Constrained Neural Network}

\newif\ifuniqueAffiliation
\uniqueAffiliationtrue

\ifuniqueAffiliation 
\author{ \href{https://orcid.org/0000-0002-0394-5574}{\includegraphics[scale=0.06]{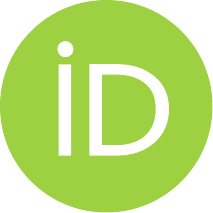}\hspace{1mm}Radmir Karamov}\thanks{corresponding author} \\
	Center for Material Technologies\\
	Skolkovo Institute of Science and Technology\\
	\texttt{r.karamov@skoltech.ru} \\
	\And
	\href{https://orcid.org/0000-0002-0413-9101}{\includegraphics[scale=0.06]{orcid.pdf}\hspace{1mm}Tagir Karamov} \\
	Center for Petroleum Science and Engineering\\
	Skolkovo Institute of Science and Technology\\
	\texttt{t.karamov@skoltech.ru} \\
}
\else
\usepackage{authblk}
\fi

\renewcommand{\shorttitle}{Data-Efficient ISE Mitigation in Steels Using ML and a PCNN}

\hypersetup{
pdftitle={Data-Efficient Indentation Size Effect Mitigation in Steels Using Machine Learning and a Physics-Constrained Neural Network},
pdfsubject={cond-mat.mtrl-sci},
pdfauthor={Radmir Karamov},
pdfkeywords={nanoindentation, indentation size effect, steel, machine
learning, physics-constrained neural network, small data, material-level validation},
}

\begin{document}
\maketitle

\begin{abstract}
Shallow nanoindentation, needed for thin layers, individual phases and small
volumes, gives hardness inflated by the indentation size effect (ISE). The
Nix--Gao correction reduces this inflation but requires deep indentations of
the same material, often impossible in such samples. We present a
data-efficient workflow, trained on a few reference materials, that predicts
a high-load reference hardness from a single shallow indentation. Seven
models were trained on 703 indentations (5--350 mN) on three steels
(2.0--6.5 GPa): Ridge regression, Random Forest, XGBoost, two feed-forward
networks, and two physics-constrained neural networks (PCNNs) that apply a
bounded learned correction \(q\) as \(H_{\mathrm{app}}/\sqrt{1+q}\). Over
five seeds, models were tested on an external steel not used for training
and indented at different loads, in leave-one-steel-out folds, and on fused
silica, an out-of-domain material where all models failed. Random Forest and
XGBoost, the best on the development set, transferred worst over all loads
(MAPE 10.2 and 8.5\%), whereas Ridge regression, the PCNNs and one
feed-forward network reached 3.7--4.8\%; with three training steels,
development-set accuracy thus did not indicate accuracy on an unseen steel.
From only the external steel's 3--15 mN indentations, the smooth (non-tree)
models estimated the reference within 0.03--0.19 GPa on average. The PCNNs
matched the linear baseline within the training hardness range and were the
most accurate for a held-out steel softer than the training steels, but
failed for a harder one. The workflow uses no steel-specific inputs and can
be retrained for other material classes.
\end{abstract}


\section{Introduction}
Instrumented indentation testing (IIT) is the established standard for
evaluating the mechanical properties of materials at the micro- and
nanoscale \cite{mencikDeterminationMechanicalProperties2007}. By
continuously recording the applied load (\(P\)) and the displacement
(\(h\)), IIT extracts hardness (\(H\)) and reduced elastic modulus
(\(E_{r}\)) from a single load--displacement curve, using the
Oliver--Pharr method to estimate the projected contact area (\(A_{c}\))
from the elastic unloading stiffness
\cite{oliverImprovedTechniqueDetermining1992}. Because the plastic zone
can be confined to sub-micron volumes, IIT can characterize thin films,
graded layers, individual phases of multiphase alloys and ion-irradiated
surface layers that are difficult to isolate in macroscopic tests
\cite{pelletierCharacterizationMechanicalProperties2006,
besharatlooInfluenceIndentationSize2021,
meiCharacterizationMechanicalProperty2022}.

However, this spatial confinement introduces fundamental physical
tension. Testing at low loads and shallow
contact depths, the condition that makes IIT useful for small volumes,
activates a depth-dependent hardening known as the indentation size
effect (ISE) \cite{wangComparativeStudyIndentation2024,
nixIndentationSizeEffects1998, voyiadjisReviewNanoindentationSize2017}.
The apparent hardness of shallow indentations is therefore higher than
that measured at large depths and cannot be compared directly with
hardness values obtained at higher loads or by macroscopic tests.

Although the ISE is observed in most crystalline materials, its
mechanism is not universal. In conventional metallic alloys and
multiphase steels, it is usually attributed to strain-gradient
plasticity: the self-similar Berkovich geometry imposes a plastic strain
gradient that scales inversely with contact depth and requires
geometrically necessary dislocations (GNDs) in addition to the
statistically stored dislocations of the undeformed material
\cite{nixIndentationSizeEffects1998,
elmustafaNanoindentationIndentationSize2003,
aifantisModelingDislocationGrain2007}. In nanocrystalline metals and
high-entropy alloys, grain-boundary plasticity and lattice distortion
add depth-dependent strengthening that does not reduce to a single GND
density \cite{wuElasticPlasticDeformations2016}, and in ceramics and
covalently bonded materials cracking, densification and
pressure-induced phase transformations contribute
\cite{sahinAnalysisLoadpenetrationDepth2008}. No single analytical
correction is therefore universally applicable.

Steels were chosen for this study because their ISE is well described
by the Nix--Gao model \cite{nixIndentationSizeEffects1998}, in which
GND-mediated hardening gives
\[H^{2} = H_{0}^{2}\left( 1+\frac{h^{*}}{h_{c}} \right),\]
where \(H_{0}\) is the depth-independent hardness, \(h_{c}\) the contact
depth and \(h^{*}\) a characteristic length that depends on the indenter
geometry and the dislocation density. Plotting \(H^{2}\) against
\(1/h_{c}\) and extrapolating the linear regime to
\(1/h_{c}\rightarrow 0\) gives \(H_{0}\). This correction has two
requirements: a series of indentations of the same material over a
range of depths, and a linear regime that is reached only at
sufficiently large depths. At shallow depths, tip rounding and pile-up
add deviations from the linear relation
\cite{pharrIndentationSizeEffect2010, quIndenterTipRadius2004};
modified models account for individual contributions but introduce
additional fitting parameters \cite{wangComparativeStudyIndentation2024,
quIndenterTipRadius2004, abualrubAnalyticalExperimentalDetermination2004}.
In practice, shallow measurements are therefore excluded from the fit.
For thin layers, individual phases and other small volumes, the
permissible depth is often below the linear regime, and a reliable
Nix--Gao extrapolation is not possible.
This work addresses the question of how a high-load reference hardness,
the hardness at the highest accessible load, can be estimated from
shallow indentations alone, without a load series
of the material being tested.

Inverse finite element analysis (FEA) offers an alternative: the
elastoplastic constitutive parameters are updated until the simulated
\(P\)--\(h\) response matches the experimental curve
\cite{daoComputationalModelingForward2001}. The inverse problem,
however, requires many non-linear simulations per material
\cite{kangDeterminingElasticPlastic2012}, and its solution is not unique:
different combinations of yield strength and strain hardening can
produce nearly identical \(P\)--\(h\) curves
\cite{thoUniquenessReverseAnalysis2004,
chenUniquenessMeasuringElastoplastic2007}. Size effects can be included
through strain-gradient plasticity, for example by training a neural
surrogate on several hundred strain-gradient finite element simulations
and inverting it with a genetic algorithm
\cite{zhuPredictingTensileBehavior2025}, but this route requires a
constitutive model of the material class and a simulation campaign
before any material can be characterized.

Machine learning offers a faster mapping from indentation data to
material properties \cite{bockReviewApplicationMachine2019,
huberDeterminationConstitutiveProperties1999,
jinRecentAdvancesApplications2023, puchi-cabreraMachineLearningAided2023}. Beyond indentation, neural networks
and tree-based models have predicted fatigue life, fracture toughness
and alloy hardness \cite{genelApplicationArtificialNeural2004,
karamovPredictionFractureToughness2022,
heInterpretableMachineLearning2023}. In indentation, neural networks
extract elastoplastic parameters from Berkovich and spherical
indentation curves, mostly trained on finite element data
\cite{thoArtificialNeuralNetwork2004, jiaoMachineLearningPerspective2024,
marimuthuPhysicsinformedNeuralNetworks2023}, or on simulations combined
with a few experiments in multifidelity schemes
\cite{luExtractionMechanicalProperties2020}; regression models have
also extracted elastic moduli from atomic force microscopy
force--indentation curves without fitting a contact model
\cite{nguyenMachineLearningFramework2022}, and neural networks convert
nanoindentation hardness into Vickers hardness for case-hardened steel
\cite{niuLeveragingDeepNeural2024}. Machine-learning treatments of the
ISE itself have modeled hardness as a function of load and
composition, either directly \cite{mukherjeeLoadDependentHardness2026}
or through an empirical size-effect law whose parameters are learned
\cite{mannanGlassHardnessPredicting2023}, or have learned the residuals
of a Nix--Gao fit within the tested alloys
\cite{kumarMechanisticDatadrivenInterpretation2026}. None of these
estimates a high-load reference hardness from a single shallow
instrumented indentation of a material that was not used for training.
Moreover, most were evaluated on random splits within the same
materials, which measure interpolation rather than transfer to a new
material \cite{meredigCanMachineLearning2018}; transfer has been examined only qualitatively, for three
glasses \cite{mannanGlassHardnessPredicting2023}. Whether a size-effect
correction learned from a few reference materials transfers to a new
material, and which type of model does so, therefore remains open.

The approaches described above rely on large datasets, and addressing
the ISE through data-driven means therefore faces a second constraint:
a ``small data'' problem that limits the routine experimental
application of machine learning in solid mechanics
\cite{xuSmallDataMachine2023}. Simulation-trained networks typically
require thousands to tens of thousands of computed curves
\cite{jiaoMachineLearningPerspective2024, parkDeepLearningBased2023} and
transfer poorly to real indentations without experimental data
\cite{luExtractionMechanicalProperties2020}. The machine-learning
studies of the ISE were likewise based on large datasets compiled
mostly from the literature, about 3,000 glasses
\cite{mannanGlassHardnessPredicting2023} and about 2,500 Vickers
hardness records \cite{mukherjeeLoadDependentHardness2026}, or on tens
of thousands of points generated from fitted hardness profiles
\cite{niuLeveragingDeepNeural2024}. A laboratory characterizing its own
materials can usually test only a few reference specimens: hundreds of
indentations describe the scatter and load dependence within each
specimen, but provide only as many material-level reference values as
there are specimens. Learning from such small datasets has been made
possible by restricting the model with dimensional analysis, scaling
laws or other prior physical knowledge \cite{xuSmallDataMachine2023,
kumarMachineLearningConstrained2019, wangPhysicsinformedFewshotDeep2023}.
In this work, a workflow is called data-efficient when it needs only
standard experimental indentations of a few reference materials for
training, without finite element simulations, and a single shallow
load--displacement curve of the new material for prediction, without a
load series of that material. The efficiency refers to the material being
tested: building the reference set still required load series of
238--303 indentations on each reference material. With so few material-level values,
flexible models can fit the training materials without learning a
correction that transfers, which makes simple and structured models
natural candidates \cite{jinRecentAdvancesApplications2023}.

A further choice concerns the output of the model. Direct regressors
predict hardness as an unconstrained scalar, which can fit the training
materials closely but need not respect known relations when applied
outside them \cite{beuclerEnforcingAnalyticConstraints2021}. An
alternative is to fix the form of the output in the architecture and
learn only part of it \cite{linkaConstitutiveArtificialNeural2021,
sukumarExactImpositionBoundary2022}. A related approach for the ISE was
reported for glasses, where a network learned the parameters of an
empirical size-effect law from composition and load and transferred to
new compositions better than a direct regressor
\cite{mannanGlassHardnessPredicting2023}. The physics-constrained neural
network (PCNN) used here differs in that it learns a bounded correction
of the hardness of a single indentation from its load--displacement
curve, with an output form motivated by the Nix--Gao relation
(Section~\ref{pcnn-methods}).

The present study tests whether the high-load reference hardness of a
steel can be estimated from its shallow indentations alone, without a
load series of that steel, using corrections learned from other
steels. The ISE is mitigated rather than fully removed: the target is
the mean hardness at the highest load of each material (350 or
375 mN), a finite-load value, because in many practical cases the bulk
hardness is not known and only the hardness at the highest accessible
load is available. Seven models, Ridge regression, Random
Forest, XGBoost, two feed-forward neural networks and two PCNNs, are
trained on 703 indentations of three steels and compared under the
same protocol: five training seeds; an external steel not used for
training and indented at different loads; a shallow-only test against
a Nix--Gao fit; leave-one-steel-out folds; fused silica as an
out-of-domain test; and an ablation, feature-attribution and
latent-space analysis of the models. The contribution is threefold: the task of estimating a
high-load reference hardness from single shallow indentations of an
unseen material; a material-level evaluation that tests transfer
rather than interpolation; and the resulting evidence on which models
transfer with only three training materials, and where the bounded
correction stops transferring. Steels serve as a controlled case in
which the ISE is well described; retraining the workflow for other
material classes is not tested here.

\section{Materials and Methods}
\label{materials-and-methods}

\subsection{Materials Preparation and Microstructural Characterization}
\label{materials-preparation-and-microstructural-characterization}
Four steel specimens covering a range of hardness and elastoplastic
response were used. The training dataset comprises three certified steel
hardness reference blocks, traceable to national metrology standards,
with certified Vickers hardness values of 212 ± 11 HV0.025,
426 ± 22 HV0.025, and 841 ± 9 HV0.025 (25 gf load). These specimens are
referred to hereafter as E212, E426, and E841, respectively; their
Oliver--Pharr hardness at the highest indentation load (350 mN) is
approximately 2.0, 3.6, and 6.5 GPa. The certified values identify the blocks and
were not used as the supervised learning target, because they
were also measured at low load and are affected by the ISE. Based on
their hardness, composition, and expected processing, the blocks
nominally range from a soft ferritic/pearlitic microstructure to a hard
martensitic one, and are therefore expected to differ in strain
hardening and pile-up tendency. These phase assignments are nominal;
phases and phase fractions were not quantified by phase-sensitive
characterization.

The three training blocks provide low, intermediate, and high hardness
levels within the investigated interval. Repeated indentations at
several loads characterize the within-specimen scatter and the load
dependence of each block, but the supervised dataset contains only three
independent material states and three specimen-level reference hardness
values. This compact design allows the proposed reconstruction to be
evaluated on contrasting indentation responses; it does not establish
the minimum number of materials required for reliable prediction.

A fourth steel reference block from a different manufacturer (certified
value \(H_{\mathrm{cert}} \approx 6\) GPa at 5 N; no HV0.025 value is
available) was used as an external test material: a steel not used for training
and indented at loads different from the training schedule
(Section~\ref{nanoindentation-protocol}). It was not used for model
training, for fitting any preprocessing constant, or for choosing any
model setting; it was used only to evaluate the final models, and all
settings reported here were fixed before it was scored.

A fused silica calibration specimen was used for nanoindentation system
calibration and was included as instrumental boundary control. As an
amorphous material, fused silica does not exhibit the classical
GND-mediated indentation size effect characteristic of crystalline
metals \cite{pharrIndentationSizeEffect2010}; any measured depth dependence is expected to be
dominated by densification, tip geometry, cracking if present, and
calibration artifacts \cite{oliverMeasurementHardnessElastic2004}, allowing these instrumental
contributions to be isolated from genuine strain-gradient hardening in
the steel specimens.

All metallic specimens were mechanically polished to a mirror finish,
culminating in colloidal silica suspension to remove the near-surface
Beilby layer \cite{wuInvestigationMorphologyChemistry2023}. Elemental compositions of all steel grades were
measured by X-ray fluorescence (done on XRF Olympus Vanta C) and energy
dispersive X-ray spectroscopy (EDS). Nanoindentation marks and the
surrounding microstructure were imaged by scanning electron microscopy
(SEM); together with the compositions, these observations are consistent
with the nominal phase assignments but do not quantify phase fractions.
SEM and EDS were done on ThermoFisher
Quattro S in Skoltech's Advanced Imaging Core Facility. These
characterization results are presented in Section~\ref{microstructural-and-compositional-analysis}.

\hypertarget{nanoindentation-protocol}{%
\subsection{Nanoindentation
Protocol}\label{nanoindentation-protocol}}

All instrumented indentation experiments were performed on a Nanovea
PB1000 (Nanovea, USA) mechanical tester equipped with a calibrated
Berkovich diamond indenter. Prior to each indentation cycle, a thermal
equilibration hold was maintained until the displacement drift rate fell
below 0.05 nm/s, at which point the drift rate was recorded and
subtracted as a linear correction from the subsequent displacement
signal \cite{oliverImprovedTechniqueDetermining1992}.

Tests were conducted in load-controlled mode at a constant loading rate
(\({\text{dP}}/{\text{dt}} = \ 2 \cdot P_{\max}\) mN per minute) to
ensure quasi-static deformation conditions. Upon reaching each target
peak load (\(P_{\max}\)), a 15-second dwell period was applied before
unloading to allow time-dependent plastic displacement to stabilize,
thereby preventing the upward curvature (``nose'' artifact) in the
initial unloading segment that would otherwise distort the contact
stiffness calculation (S) \cite{oliverMeasurementHardnessElastic2004, fengEffectsCreepThermal2002}.

For the three reference steel specimens constituting the primary
training dataset, a loading schedule of seven discrete peak loads was
applied: 5, 10, 20, 50, 100, 200, and 350 mN. This range was selected to
span the size-affected range, from shallow contacts (mean \(h_c\) of about
76--172 nm at 5 mN) to deep contacts at 350 mN (\(h_c > 1\ \mu\)m),
where the hardness still decreases by about 10\% over the last load
interval. At each load level, indentations were placed in regular grids
with sufficient spacing to avoid interaction between neighboring
indents. Indentations were excluded if the unloading segment
deviated from a power-law fit
(\(P\  = \ \alpha\left( h\  - \ h_{f} \right)^{m}\)) with a coefficient
of determination below R² = 0.98, or if a discontinuous load drop
exceeding 5\% of \(P_{\max}\) was present in the loading curve,
indicative of cracking or surface debris contact. Indentations for which
the Oliver--Pharr analysis gave no finite contact depth or hardness were
also excluded. After these criteria, 782 valid indentations of the three
training steels remained (238, 241, and 303 for E212, E426, and E841),
with at least 27 indentations per load level and specimen.

The fourth steel specimen was tested at a different set of peak loads
(3, 7, 15, 35, 75, 150, 275, and 375 mN) deliberately offset from the
training load levels, yielding 346 valid indentations after the same
exclusion criteria. The loads from 7 to 275 mN lie within the training
load range and require interpolation between training load levels; 3 mN
lies below the shallowest training load (5 mN) and requires
extrapolation; 375 mN is the reference load of this specimen. The model
evaluation therefore does not rely on recall of training load
conditions.

The fused silica calibration standard was tested at four to six
indentations per load level (35 indentations) using the same protocol as for the
training dataset. As established in 
Section~\ref{materials-preparation-and-microstructural-characterization},
the purpose of this evaluation is to characterize the depth dependence
of purely instrumental geometric artifacts, tip-apex rounding and
compliance calibration residuals, in the absence of any
dislocation-mediated plasticity, and to test the models on a material
outside the steel training domain.

\hypertarget{data-processing-and-feature-engineering}{%
\subsection{Data processing and feature
engineering}\label{data-processing-and-feature-engineering}}

Raw load--displacement (\(P\)--\(h\)) data were processed using custom Python
routines following the Oliver--Pharr methodology \cite{oliverImprovedTechniqueDetermining1992}.
For each indentation, the contact stiffness \(S = dP/dh\) was determined by fitting
the upper 50--95\% segment of the unloading curve to the power-law
relation \(P\  = \ \alpha\left( h\  - \ h_{f} \right)^{m}\). The contact
depth was computed as
\(h_{c} = \ h_{\max} - \ \varepsilon\left( {P_{\max}}/{S} \right)\)
here \(\varepsilon\) was evaluated individually for each indentation
\cite{oliverMeasurementHardnessElastic2004},
 and the projected contact area \(A_{c}\) was obtained from the
indenter area function \(A_{c}\left( h_{c} \right)\) calibrated on the
fused silica standard over contact depths of 86~nm to 1.23~\(\mu\)m.
The reference-load indentations of the steels (\(h_c\) =
1.47--2.67~\(\mu\)m) lie above this range, where the calibrated area
already equals the ideal Berkovich area within 2\%
(\(A_c/24.5h_c^2\) = 0.98--1.00); at 5 mN the calibrated area is
1.9--3.2 times the ideal area, reflecting rounding of the tip apex. The apparent hardness
\(H\  = \ {P_{\max}}/{A_{c}}\) and reduced elastic modulus
\(E_{r}\  = \ \left( {\sqrt{\pi}}/{2} \right) \cdot {S}/{\sqrt{A_{c}}}\)
were extracted for each curve.

Because \(A_{c}\) is derived exclusively from the elastic unloading
stiffness, the Oliver--Pharr procedure cannot account for elastoplastic
pile-up at the contact periphery. In low-strain-hardening
microstructures, pile-up increases the true contact area relative to
the Oliver--Pharr estimate and therefore inflates the apparent \(H\).
This bias depends on strain hardening and microstructure and may also
vary with depth; it was not quantified here and cannot be separated
from the depth-dependent ISE with the present data. The feature engineering described below
therefore includes descriptors that do not use the per-indent contact
area \(A_{c}\), so that this bias does not enter them directly; the
predictor as a whole is not free of the area function.

As an analytical baseline, Oliver--Pharr \(H\) and \(h_{c}\) values were
used to construct the classical Nix--Gao linearization \cite{nixIndentationSizeEffects1998}:

\[H^{2} = H_{0}^{2}\left( 1+\frac{h^{*}}{h_{c}} \right)\]

where \(h^{*}\) encodes indenter geometry and dislocation density.
Extrapolating the linear \(H^{2}\) vs. \({1}/{h_{c}}\) regime to
\({1}/{h_{c}} \rightarrow 0\) recovers~\(H_0^2\). Below a contact depth of
about 200 nm, the data deviate upward from this linear relation; tip
rounding and pile-up are likely contributions. The linear fit was therefore restricted
to \(h_{c} \geq 215\) nm (\(1/h_{c} \leq 4.65\ \mu\text{m}^{-1}\)),
applied identically to every material without fallback to the full data
set (deep-only fit). The need for such a truncation, and for deep
indentations to perform it, is the limitation the present framework
addresses.

Physics-guided feature engineering was implemented to create the machine
learning input vector, which was populated with three categories of
descriptors extracted from each \(P\)--\(h\) curve. Primary load-geometric
parameters: \(P_{\max}\), \(h_{\max}\), \(h_{c}\), and \(S\). Energetic
parameters obtained by numerical integration: \(W_{\text{tot}}\),
\(W_{\text{el}}\), and \(W_{p} = \ W_{\text{tot}} - \ W_{\text{el}}\).

The descriptors included the dimensionless plastic-work ratio
\({W_{p}}/{W_{\text{tot}}}\) (energy partitioning between plastic
flow and elastic recovery \cite{chengRelationshipsHardnessElastic1998}), 
the dimensionless hardness-to-modulus ratio \({H}/{E_{r}}\) 
(scales with pile-up propensity \cite{bolshakovInfluencesPileupMeasurement1998}), 
and the area-function-free compliance quantity \(P_{\max}/S^{2}\). 
The last quantity follows from the
Joslin-Oliver relation \cite{joslinNewMethodAnalyzing1990}: since
\(\text{H\ } \propto {P_{\max}}/{A_{c}}\) and
\(E_{r}\  \propto {S}/{\sqrt{A_{c}}}:\)

\[\frac{P_{\max}}{S^{2}} \propto \frac{H}{E_{r}^{2}}\]

The input feature vector was extended beyond the core physics-guided
descriptors to include additional Oliver--Pharr output quantities,
power-law unloading fit parameters, and geometric ratios, with the
expectation that less informative features would be down-weighted
during model training; SHAP analysis
(Section~\ref{feature-attribution-and-latent-space-analysis}) was used
to describe how the trained models use these inputs. The complete
input feature vector is:

\begin{multline*}
\mathbf{X} = \Bigl\{ P_{\max},\ h_{\max},\ h_{c},\ S,\ A_{c},\ H,\ E_{r},\ W_{\text{tot}},\ W_{\text{el}}, \\
\ W_{p},\; \frac{W_{p}}{W_{\text{tot}}},\; \frac{H}{E_{r}},\; \frac{P_{\max}}{S^{2}},
\frac{h_{c}}{h_{\max}},\; h_{f},\; \frac{h_{f}}{h_{\max}},\; A,\; m,\; \varepsilon,\; R^{2},\; n \Bigr\}
\end{multline*}

where~\(A\),~\(m\), and~\(h_{f}\)~are the power-law coefficient,
exponent, and residual depth from the unloading fit
respectively,~\(\varepsilon\)~is the geometry factor used in the contact
depth calculation, and~\(R^{2}\)~is the goodness of fit of the unloading
power-law. The loading exponent \(n\) is the energy-equivalent exponent
of the loading curve,
\[
n = \frac{P_{\mathrm{hold}}\,h_{\mathrm{hold}}}{W_{\mathrm{load}}} - 1,
\]
where \(W_{\mathrm{load}}\) is the loading work integrated up to the
start of the dwell (the first point with \(P \geq 0.98\,P_{\max}\)) and
\(P_{\mathrm{hold}}\), \(h_{\mathrm{hold}}\) are the load and depth at
that point. For an exact power law \(P = C h^{n}\), this is the
power-law exponent; for a measured curve it is a descriptive shape
parameter, interpreted as a departure from Kick's law (\(n = 2\)). The
integration stops before the dwell because creep during the 15 s hold
would bias \(n\) low. The complete vector contains 21 inputs.

For the PCNN, a second feature frame of eight descriptors was constructed
from the same indentation quantities. It contains depth, energy and
curve-shape descriptors, while the apparent hardness \(H_{\mathrm{app}}\)
is kept separate for the final reconstruction
(Section~\ref{pcnn-methods}). The bounded depth coordinate was defined as
\[
t=\frac{h_s}{h_c+h_s},
\]
where \(h_s\) is a characteristic depth scale, taken as the median
contact depth of the training split. Shallow contacts therefore map
toward \(t\rightarrow 1\), and deeper contacts toward \(t\rightarrow 0\).
The PCNN-DI feature frame is
\[
\mathbf{X}_{\mathrm{DI}} = \Bigl\{\,
t,\;
\frac{W_p}{W_{\mathrm{tot}}},\;
\frac{h_c}{h_{\max}},\;
\frac{h_f}{h_{\max}},\;
\log z_{\mathrm{JO}},\;
\log\frac{E_r}{E_{r,\mathrm{ref}}},\;
m,\;
n
\,\Bigr\},
\]
where
\[
z_{\mathrm{JO}}=E_{r,\mathrm{ref}}\frac{P_{\max}}{S^2}
\]
is a Joslin--Oliver-type dimensionless compliance descriptor, with unit
conversions applied so that \(z_{\mathrm{JO}}\) is dimensionless. It is
proportional to the area-free hardness estimate
\(H_{\mathrm{JO}}=(4/\pi)\,E_{r,\mathrm{ref}}^2 P_{\max}/S^2\). The
reference modulus \(E_{r,\mathrm{ref}}\) is the median reduced modulus of
the training split. Both constants, \(h_s\) and \(E_{r,\mathrm{ref}}\),
were computed once from the training split pooled over the three
training steels and kept fixed for all predictions. No PCNN-DI input
depends on other indentations of the same material, so each indentation
is predicted from its own curve alone.

The descriptor \(\log(E_r/E_{r,\mathrm{ref}})\) is the only PCNN-DI input
evaluated from the per-indent contact area. Because
\((E_r/E_{r,\mathrm{ref}})^2 = H_{\mathrm{app}}/H_{\mathrm{JO}}\), it
measures the disagreement between the Oliver--Pharr and the area-free
hardness. For a material of approximately constant modulus, such as
steel, this disagreement is consistent with contact-area error such as
pile-up; true modulus differences, compliance errors and unloading creep
also contribute. The descriptor set is not fully independent:
\(h_c/h_{\max}\) is determined by \(t\), \(\log z_{\mathrm{JO}}\),
\(\log(E_r/E_{r,\mathrm{ref}})\) and \(m\) through the area function, and
\(h_f/h_{\max}\) is strongly correlated with it (\(r = 0.999\)); both are
retained as familiar contact-mechanics descriptors. \(H_{\mathrm{app}}\)
itself is not an input of the PCNN-DI network; it enters only the final
reconstruction. The material label was not supplied as a predictor; it
was used only to group indentations in the load-independence loss
(Section~\ref{pcnn-methods}).

The supervised learning target was the high-load reference hardness
\(H_{\mathrm{ref}}\), defined as the mean Oliver--Pharr hardness of the
valid indentations at the highest load of each material: 350 mN for the
training steels and 375 mN for the fourth steel. The resulting values
(mean ± standard deviation over indentations) are
2.003 ± 0.032 GPa for E212 (\(N = 27\)), 3.605 ± 0.070 GPa for E426
(\(N = 27\)), 6.491 ± 0.274 GPa for E841 (\(N = 53\)), and
6.080 ± 0.112 GPa for the fourth steel (\(N = 27\)). All indentations of a
material share its \(H_{\mathrm{ref}}\) as the target. \(H_{\mathrm{ref}}\)
is a finite-load value within the accessible experimental load range,
not an ISE-free or bulk hardness: the mean hardness of each steel still
decreases by 7--12\% over its last load interval. A high-load reference
was chosen because in many practical cases the bulk hardness is not known
and only the hardness at the highest accessible load is available, for
example when the indentation depth is limited by film thickness or grain
size. The learning objective
is therefore
\[
\mathbf{X}_{\mathrm{shallow}} \rightarrow H_{\mathrm{ref}},
\]
recovering the high-load reference hardness from indentation features
measured in the shallow, size-affected regime.

\hypertarget{physics-informed-tabular-data-augmentation}{%
\subsection{Tabular Data
Augmentation}\label{physics-informed-tabular-data-augmentation}}

The experimental dataset comprised 782 valid indentations of three
steels, sufficient for analytical fitting or ensemble architectures but
limited for constraining neural networks against overfitting
\cite{morganOpportunitiesChallengesMachine2020}. To regularize training
without synthetic finite element simulation, a tabular augmentation
coupled to the measured indentation quantities was applied. These
records were split 90/10 (split seed 42) into 703 training and 79
development indentations before augmentation; all synthetic rows
were generated exclusively from training indentations, and every
quantity used by the augmentation was estimated from them only. The
effect of the augmentation is tested in
Section~\ref{augmentation-comparison}.

The augmentation operates on the measured features in physical units
(mN, nm, GPa, pJ, and dimensionless ratios and exponents). For each
original training indentation, three synthetic rows were added: a noise
copy, a drift copy, and a MixUp copy, expanding the training set from
703 to 2812 rows. The original rows were always kept. For the PCNN, the
input descriptors of every synthetic row were then computed from its
synthetic measured features with the same code and the same training
constants (\(h_s\), \(E_{r,\mathrm{ref}}\)) as for a measured
indentation. Finally, the input scaler (mean and standard deviation) was
fitted on the original training rows only and applied unchanged to
synthetic, development, and external rows. The same procedure was used
for all models.

(1) Noise copy: Gaussian scatter with the measured covariance. Each
training indentation belongs to a group \(g\) of indentations of one
material at one nominal load (21 groups). Because all measured features
are positive, noise was applied to their logarithms as a Gaussian-kernel
smoothed bootstrap with variance correction
\cite{silvermanDensityEstimationStatistics1986}:
\[
\log \mathbf{x}_{\mathrm{syn},i} = \boldsymbol{\mu}_g
+ \frac{\mathbf{r}_i + h\,\boldsymbol{\epsilon}_i}{\sqrt{1+h^2}},
\qquad \boldsymbol{\epsilon}_i \sim \mathcal{N}(\mathbf{0}, \boldsymbol{\Sigma}_g),
\]
where \(\boldsymbol{\mu}_g\) is the group mean of \(\log\mathbf{x}\),
\(\mathbf{r}_i\) the residual of the parent indentation,
\(\boldsymbol{\Sigma}_g\) the measured within-group covariance of
\(\log\mathbf{x}\), and \(h = 1\). The noise is one correlated vector per
row rather than independent noise per feature, so the synthetic rows
reproduce the size and the correlations of the measured
indentation-to-indentation scatter of each material at each load,
including its larger magnitude at shallow loads. Products and ratios
such as \(H = P_{\max}/A_c\), \(H/E_r\), \(h_c/h_{\max}\), and
\(P_{\max}/S^2\) remain exact in every synthetic row. The target is
unchanged. The noise represents the measured scatter (instrumental
precision, surface and microstructural variation, pop-ins, as far as
they occur in the repeated indentations), not a model of a specific
error source; being Gaussian, it does not reproduce rare large
deviations.

(2) Drift copy: area-function error. One systematic error of the
contact-area function, \(\delta_A \sim \mathcal{N}(0, 0.02^2)\), was drawn
per drift copy and applied to all its rows, representing a
session-level error such as tip blunting
\cite{lashgariDataAugmentationDeeplearningbased2020}. The 2\% amplitude
is consistent with reported tip-wear tolerances for diamond Berkovich
indenters \cite{herrmannProgressDeterminationArea2000}. The error was
propagated through the area-derived quantities:
\(A_c \rightarrow A_c(1+\delta_A)\), \(H \rightarrow H(1+\delta_A)^{-1}\),
and \(E_r\) and \(H/E_r\) scaled by \((1+\delta_A)^{-1/2}\). Load,
stiffness, depths, energies, exponents, and \(P_{\max}/S^2\) are
unchanged, and so are the area-free PCNN descriptors; only
\(\log(E_r/E_{r,\mathrm{ref}})\) and \(H_{\mathrm{app}}\) shift. The
target is unchanged.

(3) MixUp copy. Each training row \(i\) was blended with a randomly
chosen training partner \(j\) \cite{zhangMixupEmpiricalRisk2018}:
\[
\widetilde{\mathbf{x}} = (1-\lambda)\,\mathbf{x}_i + \lambda\,\mathbf{x}_j,
\qquad
\widetilde{H}_{\mathrm{ref}} = (1-\lambda)\,H_{\mathrm{ref},i} + \lambda\,H_{\mathrm{ref},j},
\qquad \lambda \sim \mathcal{U}(0.1,\,0.2).
\]
The blending is linear in the measured features, and the PCNN
descriptors were computed afterwards from the blended features. Partners
were drawn without restriction, so pairs of indentations from different
steels and different loads are included. Linear blending does not keep
products and ratios exact for such pairs, and MixUp rows are therefore
treated primarily as a heuristic interpolation between measured
indentations that regularizes the model. The restriction to low
\(\lambda\) keeps each synthetic row close to its parent indentation. A
possible physical reading, that linear mixing of area-invariant
descriptors (\(W_p/W_{\mathrm{tot}}\), \(P_{\max}/S^2\)) approximates
to first order the composite elastoplastic response of an indentation
sampling a phase boundary \cite{constantinidesGridIndentationAnalysis2006},
is a hypothesis that was not tested here and requires further
investigation with directly characterized phase-boundary indentations.

Synthetic rows were screened before training. For the PCNN, the
correction target \(q = (H_{\mathrm{app}}/H_{\mathrm{ref}})^2 - 1\) was
computed from the final synthetic pair, and a synthetic row was removed
if any value was non-finite or non-positive, or if \(q\) fell outside
the bounds defined in Section~\ref{pcnn-methods}; original rows were
never removed. The load-independence loss of the PCNN uses the original
training rows only. The synthetic rows are interpreted as a training
regularizer, not as additional material states or independently
simulated indentation curves. No transformation was applied to the
development set, the fourth steel specimen, or fused silica.

\hypertarget{machine-learning-architectures-and-training-strategy}{%
\subsection{Conventional Machine Learning Architectures and 
Training Strategy}\label{machine-learning-architectures-and-training-strategy}}

Five conventional regression architectures were implemented in Python
using the scikit-learn \cite{pedregosaScikitlearnMachineLearning2011}
and XGBoost libraries and benchmarked with
increasing representational capacity: Ridge Regression, Random Forest,
XGBoost, a standard feed-forward neural network, and a bottlenecked
neural network. All five use the 21 inputs listed in
Section~\ref{data-processing-and-feature-engineering}, standardized with
the scaler fitted on the original training rows.

(1) Ridge Regression~(\(\alpha\  = \ 1.0\)) serves as a linear baseline
\cite{hoerlRidgeRegressionBiased1970}. L2 regularization stabilizes
coefficient estimates under the severe feature collinearity inherent to
indentation data, where \(P_{\max}\), \(h_{\max}\), \(W_{\text{tot}}\),
and \(h_{c}\) scale concurrently with load.

(2) Random Forest~(\(N_{\text{estimators}}\  = \ 100\), unconstrained
depth, min samples leaf  =  2) reduces variance through
bootstrap aggregation \cite{breimanRandomForests2001}. The minimum leaf
constraint limits memorization of individual augmented points.

(3) XGBoost~(\(N_{\text{estimators}}\  = \ 100\), \(\eta\  = \ 0.05\),
\({\text{max depth}}{} = \ 6\), row subsample = 0.8) employs sequential
residual minimization \cite{chenXGBoostScalableTree2016}. It was trained
for a fixed 100 boosting rounds without early stopping.

(4 and 5) Two feed-forward neural network architectures were evaluated
with ReLU activations and the Adam optimizer (up to 10,000 iterations
or until the training loss converged, scikit-learn default tolerance;
without early stopping) \cite{kingmaAdamMethodStochastic2017}. The
unconstrained baseline, a multilayer perceptron with two hidden
layers of 64 units (MLP 64×64), provides a standard non-linear mapping
reference. The bottlenecked architecture (MLP 64-8-64) introduces a narrow
8-neuron intermediate layer \cite{hintonReducingDimensionalityData2006}.
Because \(P_{\max}\), \(W_{\text{tot}}\), and \(h_{\max}\) reflect the
same loading history at different integration levels, the 8-dimensional
layer is intended to compress this collinear load-scaling information.
How the latent activations of this layer are analyzed by Principal
Component Analysis (PCA) is described in
Section~\ref{feature-attribution-and-latent-space-analysis}.

The model settings are fixed values chosen during development, partly by
manual comparison on the development set; no automated hyperparameter
search was performed, and no setting was chosen using the fourth steel.

As described in Section~\ref{physics-informed-tabular-data-augmentation},
the original experimental records were split 90/10 before augmentation;
the 703 training indentations yield 2812 augmented training rows, and
the remaining 79 indentations formed the development set, which contains
only experimental rows. Because the development set contains
indentations of the same specimens and loads as the training set, it
measures interpolation within the training steels, not transfer to a new
material, and its metrics are interpreted accordingly. Transfer is
assessed on the fourth steel (Section~\ref{transfer-to-the-training-steels-and-the-fourth-steel}),
which was not used in any part of model development. Predictive
performance was quantified by R², root mean square error (RMSE, GPa),
and mean absolute percentage error (MAPE, \%).

These conventional models directly regress \(H_{\mathrm{ref}}\) from the
input feature vector. They therefore provide a baseline for evaluating
whether constraining the network output to a correction of the measured
hardness changes transfer to an unseen steel and the load dependence of
the prediction.

\hypertarget{pcnn-methods}{%
\subsection{Physics-Constrained Neural Network}
\label{pcnn-methods}}

The physics-constrained neural network (PCNN) constrains its output to a
bounded reconstruction motivated by the Nix--Gao relation
\cite{nixIndentationSizeEffects1998}. Instead of predicting
\(H_{\mathrm{ref}}\) directly, the network predicts a signed correction
\(q\) and reconstructs the high-load reference hardness as
\[
\widehat{H}_{\mathrm{ref}} =
\frac{H_{\mathrm{app}}}{\sqrt{1+q}} ,
\]
which corresponds to the hardness-ratio relation
\[
\frac{H_{\mathrm{app}}^2}{H_{\mathrm{ref}}^2}=1+q.
\]
In the Nix--Gao relation, \(q = h^{*}/h_c > 0\); here \(q\) is learned from
the indentation descriptors and may take either sign. The constraint
therefore concerns only the algebraic form of the output: the network
does not enforce the Nix--Gao inverse-depth scaling or its dislocation
mechanism. Throughout this work, ``physics-constrained'' refers to this
output form.

For a classical indentation size effect, \(q>0\) and
\(\widehat{H}_{\mathrm{ref}}<H_{\mathrm{app}}\). Negative values of \(q\)
permit upward corrections when the shallow hardness lies below the
high-load reference, as for an inverse indentation size effect, or when
contact-area error, indentation scatter, or variability of the reference
require it. The ability to represent negative corrections does not by
itself show that the model predicts inverse-ISE materials correctly;
such materials were not tested here. The correction was bounded
according to
\[
q_{\min} \le q \le q_{\max}, \qquad q_{\min}>-1,
\]
which keeps the square-root reconstruction finite. The bounds were
derived from the training data: with \(q_P\) denoting the \(P\)-th
percentile of \((H_{\mathrm{app}}/H_{\mathrm{ref}})^2 - 1\) over the
training indentations,
\[
q_{\max} = \max(5\,q_{99},\ 20), \qquad
q_{\min} = \mathrm{clip}\bigl(\min(5\,q_{1},\ -0.5),\ -0.95,\ -0.05\bigr).
\]
For the training data (\(q_{1} = -0.086\), \(q_{99} = 3.57\); observed
range \(-0.155\) to \(19.4\)), this gives \(q_{\min} = -0.5\) and
\(q_{\max} = 20\) for all fits. No predicted \(q\) came within 1\% of
either bound for any material or seed. With these values, the bound
fixes the direction and the admissible range of the correction; it is
not intended to limit the size of the correction within that range.

Two variants of this architecture were evaluated. PCNN-DI (descriptor
input) applies the correction network to the eight descriptors
\(\mathbf{X}_{\mathrm{DI}}\) defined in
Section~\ref{data-processing-and-feature-engineering} (\(N = 8\)).
PCNN-FI (full input) applies it to the same 21 inputs as the
conventional models (\(N = 21\)), without the PCNN-DI descriptors. In
both variants, \(H_{\mathrm{app}}\) enters the final algebraic
reconstruction; in PCNN-DI it is not a network input. The network inputs
were standardized with means and standard deviations fitted on the
original training rows.

The correction network had the architecture
\[
N \rightarrow 16 \rightarrow 8 \rightarrow 4 \rightarrow 1,
\]
with SiLU activations between linear layers. If \(g_\theta\) denotes the
raw network output, the signed correction signal was
\[
s=\tanh(g_\theta),
\]
and the bounded correction was
\[
q =
\begin{cases}
s\,q_{\max}, & s \ge 0,\\
s\,|q_{\min}|, & s < 0.
\end{cases}
\]
This mapping is continuous and bounded; because \(|q_{\min}| \ne q_{\max}\),
it is differentiable everywhere except at \(s = 0\).

Training minimized a composite loss
\[
\mathcal{L}
=
\mathcal{L}_{H}
+
\lambda_q \mathcal{L}_{q}
+
\lambda_{\mathrm{load}}\mathcal{L}_{\mathrm{load}} .
\]
The hardness loss \(\mathcal{L}_{H}\) was a Smooth L1 loss between
\(\widehat{H}_{\mathrm{ref}}\) and the high-load reference target
\(H_{\mathrm{ref}}\). The correction loss \(\mathcal{L}_{q}\) is an
auxiliary supervision derived from the same target: it penalized
deviations of the predicted \(q\) from
\[
q_{\mathrm{target}} =
\left(\frac{H_{\mathrm{app}}}{H_{\mathrm{ref}}}\right)^2 - 1 .
\]
The load-independence loss penalized the within-material variance of
\(\widehat{H}_{\mathrm{ref}}\),
\[
\mathcal{L}_{\mathrm{load}}
=
\frac{1}{M}
\sum_{m=1}^{M}
\mathrm{Var}
\left(
\widehat{H}_{\mathrm{ref},i}
\,\middle|\,
i \in m
\right),
\]
where \(m\) indexes the training materials and only original training
rows are included. This term encodes the expectation that the predicted
high-load reference hardness should be approximately independent of the
indentation load within a given specimen.

The PCNN was implemented in PyTorch \cite{paszkePyTorchImperativeStyle2019}
and trained with the AdamW optimizer \cite{loshchilovDecoupledWeightDecay2019}
(learning rate \(10^{-3}\), weight decay \(10^{-5}\)), Smooth L1
parameter \(\beta=0.5\), \(\lambda_q=0.1\), and
\(\lambda_{\mathrm{load}}=0.5\), using full-batch gradient descent. Training
ran for a fixed 600 epochs and the parameters after the last epoch were
used, so no data were used to decide when to stop (training curves in
Figure~\ref{fig:S_history}). A fixed number of
epochs was preferred to early stopping: in the leave-one-steel-out check
(Section~\ref{evaluation-protocol}), early stopping on held-out records
of the training steels improved the fit to those records but increased
the error on the left-out steel in every fold (PCNN-DI MAPE 7.3, 10.6,
and 36.2\% with 600 epochs versus 12.9, 39.5, and 53.0\% with early
stopping in folds A, B, and C), and also on the fourth steel (PCNN-DI
4.84 versus 5.69\%, PCNN-FI 3.66 versus 6.50\%). The fourth steel was
excluded from loss minimization, augmentation, and all settings.

\subsection{Evaluation Protocol}
\label{evaluation-protocol}

Each model was trained with five run seeds (42--46) on the same split.
The seeds change the initialization, the augmentation draws, and the
tree and row sampling. Metrics were computed per run and are reported as
mean ± standard deviation over the five runs; this spread describes the
variability of the training procedure for fixed data, not a confidence
interval. All models were scored on the same valid indentations, against
the high-load reference hardness of each material, using RMSE, MAPE,
mean bias, and the coefficient of variation of the predictions over
all indentations of a material (CV). To assess how much of the
indentation-to-indentation variation of the measured hardness a model
retains, the CV of its predictions within each material and load was
divided by the CV of the measured apparent hardness and averaged over
steels, loads, and seeds (within-load scatter ratio).

Three additional evaluations were performed with the same settings.
(i) Shallow-only test: the fourth steel was predicted from its 3, 7, and
15 mN indentations alone (190 indentations, \(h_{c} = 42\)--203 nm),
from an input frame that contained no deeper record of this steel. The
estimate of each model is its mean prediction over these indentations;
the per-indentation RMSE against the 375 mN reference is also reported.
As an analytical baseline, the Nix--Gao relation was fitted by ordinary
least squares to the same indentations, with a fit to the three
per-load means as a sensitivity check. (ii) Augmentation comparison: the PCNN-DI and the 64-8-64
network were trained without augmentation (C0), with noise and drift
copies (C1), with an additional within-material MixUp copy (C2), and
with the full recipe of Section~\ref{physics-informed-tabular-data-augmentation}
(C3). C2 and C3 differ only in how MixUp partners are chosen. The
recipe C3 was fixed before the fourth steel was scored; the comparison
of the conditions on the fourth steel is reported for information and
was not used to select the recipe. Because all rows carry equal weight
in the data loss, the original records carry all of it in C0, one third
in C1, and one quarter in C2 and C3.
(iii) Leave-one-steel-out check: each training steel was left out in
turn, the models were trained on the other two, and the left-out steel
was scored. Because each fold uses only two training steels, this check
is a robustness test and its errors are not compared with the main
results.

\subsection{Ablation Study}
\label{pcnn-ablation-methods}
An ablation study was performed to test how the PCNN-DI depends on each
of its components. Every variant was trained with the same settings,
split, augmentation, and five seeds as the full PCNN-DI and differs from
it only in the named item:
\begin{enumerate}
    \item a direct neural network that predicts \(H_{\mathrm{ref}}\)
    without the bounded \(q\)-based reconstruction and without the
    auxiliary losses;
    \item a PCNN-DI without the \(q\)-target loss (\(\lambda_q = 0\));
    \item a PCNN-DI without the load-independence loss
    (\(\lambda_{\mathrm{load}} = 0\));
    \item a PCNN-DI without the curve-shape exponents \(m\) and \(n\);
    \item a PCNN-DI without the area-free descriptors
    \(\log z_{\mathrm{JO}}\), \(h_c/h_{\max}\), \(h_f/h_{\max}\), and
    \(W_p/W_{\mathrm{tot}}\);
    \item a PCNN-DI without the modulus/reference discrepancy
    \(\log(E_r/E_{r,\mathrm{ref}})\);
    \item a PCNN-DI without the depth coordinate \(t\) as a network
    input.
\end{enumerate}
Because some descriptors are partly redundant
(Section~\ref{data-processing-and-feature-engineering}), removing one of
them leaves part of its information available through the others. The
ablation therefore shows how the model depends on an input or component,
not a unique physical mechanism.

\hypertarget{feature-attribution-and-latent-space-analysis}{%
\subsection{Feature Attribution and Latent Space
Analysis}\label{feature-attribution-and-latent-space-analysis}}

Model behavior was examined with SHAP feature attribution and a
principal component analysis of hidden-layer activations. Both describe
how the trained models use their inputs; they are not evidence of a
physical mechanism.

SHapley Additive exPlanations
\cite{lundbergUnifiedApproachInterpreting2017} were computed for the
Random Forest, XGBoost, the 64-8-64 network, PCNN-FI, and PCNN-DI,
explaining the development-set indentations. The background
distribution was a k-means summary (50 centers) of the original
training rows, without augmented rows. For the conventional models, SHAP
values were computed for the predicted \(H_{\mathrm{ref}}\). For the
PCNNs, they were computed both for the predicted
\(\widehat{H}_{\mathrm{ref}}\), where \(H_{\mathrm{app}}\) enters through
the reconstruction and is therefore expected to carry high weight, and
for the correction \(q\), which shows how the network itself uses its
inputs. Because several inputs are correlated or partly redundant, SHAP
credit is shared between dependent inputs and is interpreted at the
level of descriptor groups. Single-run plots use seed 42, the first
prespecified seed; the stability of the input ranking was checked over
all five seeds.

For the 64-8-64 network and PCNN-DI, the activations of the 8-unit
hidden layer were extracted for the training indentations, and the
fourth-steel indentations were projected into the same space. PCA
\cite{abdiPrincipalComponentAnalysis2010} with two components was
applied, and the projections were colored by material, apparent
hardness, and contact depth \(h_c\) to examine whether the
hidden representation is ordered by hardness and how depth is
represented.

\hypertarget{results-and-discussion}{%
\section{Results and Discussion}\label{results-and-discussion}}

\hypertarget{microstructural-and-compositional-analysis}{%
\subsection{Microstructural and Compositional
Analysis}\label{microstructural-and-compositional-analysis}}

The elemental compositions of all steel specimens were analyzed by XRF
and by EDS for near-surface stoichiometry. The
consolidated results are summarized in Table~\ref{tab:bulkXRF} and Table~\ref{tab:surfaceEDS}.

\begin{table*}[t]
  \caption{XRF compositions (wt.\%), not renormalized. Elements below the detection limit are indicated by ``---''. Residual Mg, K, Al, and Ba from polishing media are excluded.}
  \label{tab:bulkXRF}
  \centering
  \begin{tabular}{lrrrr}
    \toprule
    \textbf{Element} & \textbf{E212} & \textbf{E426} & \textbf{E841} & \textbf{Fourth steel} \\
    \midrule
    Fe & 93.08 & 91.59 & 91.83 & 90.88 \\
    Si & 0.34 & 0.35 & 0.30 & 0.51 \\
    Cr & 0.25 & 0.25 & 0.25 & 0.41 \\
    Mn & 0.21 & 0.17 & 0.19 & 0.96 \\
    Ni & 0.25 & 0.24 & 0.18 & 0.14 \\
    Cu & 0.19 & 0.19 & 0.23 & 0.10 \\
    Mo & 0.03 & 0.03 & 0.02 & 0.08 \\
    V  & ---   & ---   & ---   & 0.09 \\
    W  & ---   & ---   & ---   & 0.31 \\
    \bottomrule
  \end{tabular}
\end{table*}

\begin{table*}[t]
  \caption{Near-surface EDS compositions (mass-normalized, wt.\%). Carbon values reflect adventitious surface contamination and are not quantitative (see text)~\cite{newburyScanningElectronMicroscopy2013}.}
  \label{tab:surfaceEDS}
  \centering
  \begin{tabular}{lrrrr}
    \toprule
    \textbf{Element} & \textbf{E212} & \textbf{E426} & \textbf{E841} & \textbf{Fourth steel} \\
    \midrule
    Fe & 96.44 & 95.82 & 97.02 & 96.1 \\
    C  & 2.95  & 3.06  & 1.89  & 2.3  \\
    Si & 0.24  & 0.26  & 0.31  & 0.45 \\
    Mn & 0.27  & 0.32  & ---   & 0.72 \\
    Cr & 0.10  & 0.12  & 0.30  & 0.28 \\
    Cu & ---   & 0.25  & 0.45  & ---  \\
    Ni & ---   & 0.18  & ---   & ---  \\
    Al & ---   & ---   & 0.03  & 0.05 \\
    \bottomrule
  \end{tabular}
\end{table*}

The two techniques exhibit consistent elemental hierarchies for all
major transition-metal constituents. Minor quantitative offsets between
XRF and EDS are expected from the difference in excitation volume and
the EDS mass-normalization procedure, which redistributes weight
fractions when light elements absent from XRF are included. Elevated Mg
(2.9-4.6 wt.\%), K (\textasciitilde0.8 wt.\%), and Al
(\textasciitilde1.6 wt.\%) in the XRF spectra are attributed to residual
particulate embedment from colloidal silica and alumina polishing
suspensions and are excluded from all subsequent analysis. The handheld
XRF analyser samples a near-surface layer of the order of tens of
micrometres, and light-element lines originate from the uppermost
micrometres, so embedded polishing particles contribute to the spectra;
the values are therefore not renormalized and are used to compare the
specimens, not as certified bulk compositions. Carbon is not detectable
by XRF. The EDS
carbon values do not increase monotonically with hardness across the
specimen set, indicating that the measured carbon signal is dominated by
adventitious surface contamination and does not reflect the bulk carbon
content \cite{newburyScanningElectronMicroscopy2013}. The carbon content,
the main determinant of steel hardness, was therefore not measured.

The compositional data show that E212, E426, and E841 share
closely matched substitutional alloying with Si (0.30-0.35 wt.\%), Cr
(\textasciitilde0.25 wt.\%), and Mn (0.17-0.21 wt.\%), yet their
high-load reference hardness ranges from approximately 2.0 to 6.5 GPa
(350 mN). This difference is therefore attributed to differences in carbon
content and heat treatment and the resulting microstructures, nominally ferritic/pearlitic to martensitic
(Section~\ref{materials-preparation-and-microstructural-characterization}),
as inferred from the hardness levels and established CCT behavior for
steels of comparable chemistry \cite{kraussSteelsProcessingStructure2015}.
Because the substitutional alloying of the training steels is nearly
constant and
composition is not a model input, the models learn from the
indentation response alone.

The fourth steel, used only as the external test material, differs in
composition from the training steels: its Mn content (0.96 wt.\%) is
about 4.6 times the training maximum, Si and Cr are higher, and V
(0.09 wt.\%) and W (0.31 wt.\%) are present, whereas both are below the
XRF detection limit in the training steels. Its prediction
(Section~\ref{external-test-on-the-fourth-steel})
therefore shows that the prediction did not require a composition
matched to the training steels; with a single external steel and unknown
carbon content, it does not establish transfer across compositions or
steel grades.

\hypertarget{baseline-nanoindentation-and-limitations-of-the-nix-gao-model}{%
\subsection{Baseline Nanoindentation and the Nix--Gao
Reference}\label{baseline-nanoindentation-and-limitations-of-the-nix-gao-model}}

Baseline nanoindentation of the fused silica standard, which was also
used to calibrate the area function, gave a hardness of
9.21 ± 0.15 GPa at 350 mN, which changed by only 0.2\% over the last
load interval and is consistent with the accepted reference value
\cite{oliverImprovedTechniqueDetermining1992}. Because calibration and
check use the same specimen, this confirms the internal consistency of
the calibration rather than verifying it independently. An independent
check is the reduced modulus of the steels at the reference load: 186,
202, and 205 GPa for E841, E426, and the fourth steel, close to the
184--192 GPa expected for steel against diamond, and 223 GPa for E212,
consistent with pile-up. At the shallowest loads, \(E_r\) rises to
217--250 GPa, consistent with area-function error at the rounded tip
apex; the descriptor \(\log(E_r/E_{r,\mathrm{ref}})\) captures this
deviation.

\begin{figure*}[htbp]
  \centering
  \begin{subfigure}[b]{0.49\textwidth}
    \centering
    \includegraphics[width=\textwidth]{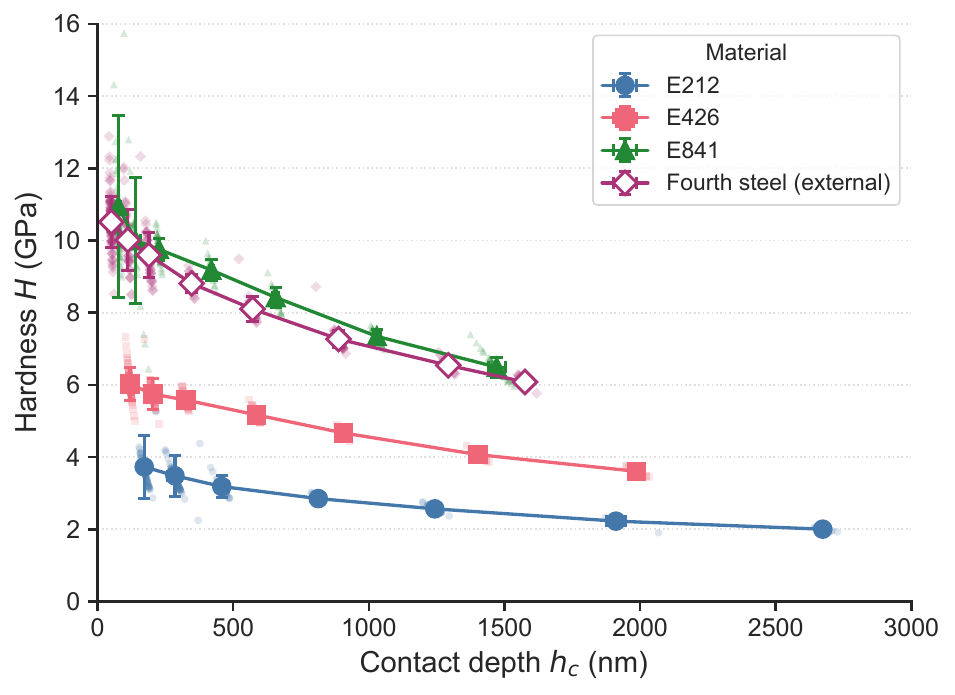}
    \caption{}
    \label{fig:ise_trend}
  \end{subfigure}
  \hfill
  \begin{subfigure}[b]{0.49\textwidth}
    \centering
    \includegraphics[width=\textwidth]{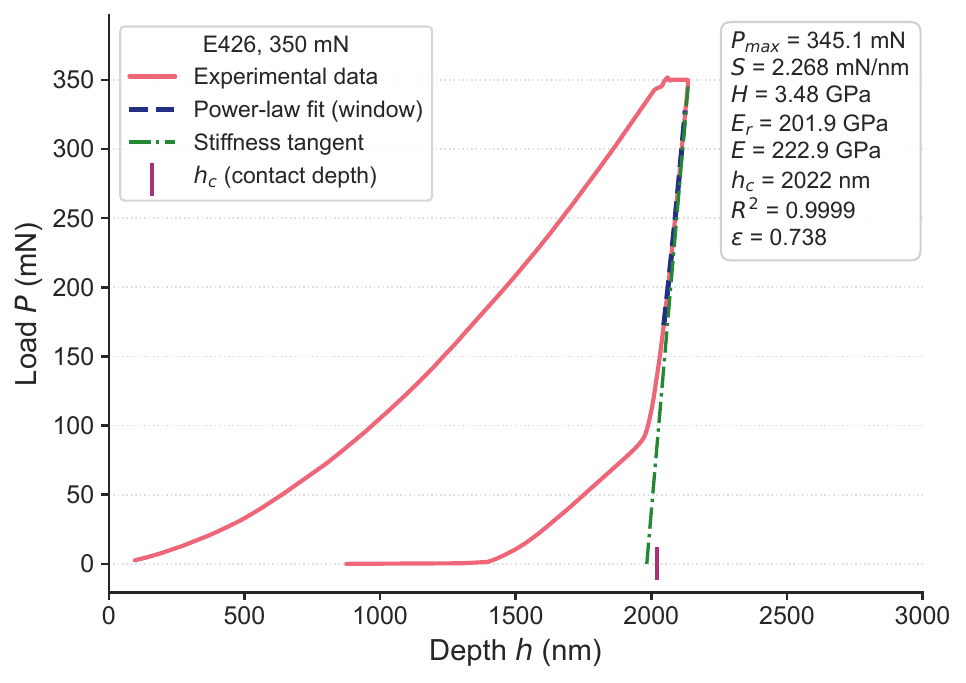}
	\caption{}
    \label{fig:solo}
  \end{subfigure}
  
  \vspace{\floatsep} 
  
  \begin{subfigure}[b]{0.99\textwidth}
    \centering
    \includegraphics[width=\textwidth]{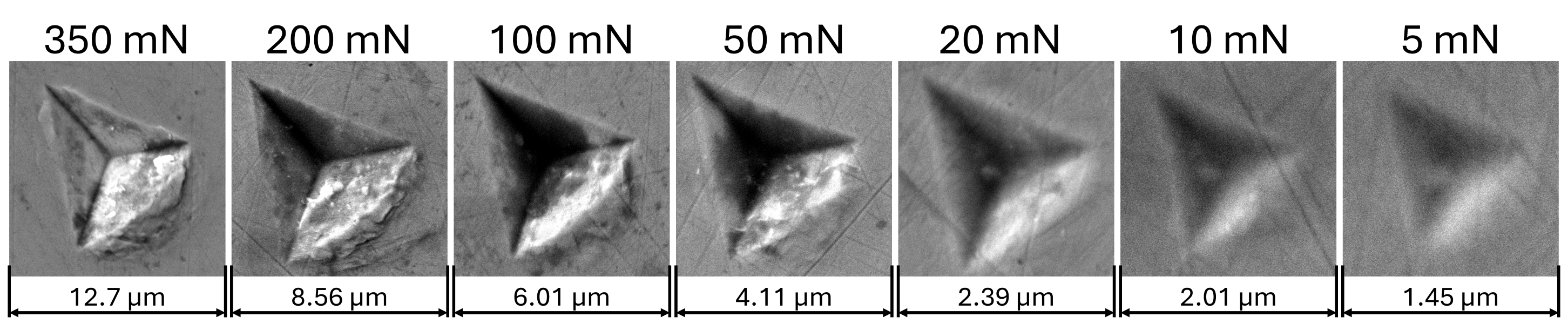}
    \caption{}
    \label{fig:SEM}
  \end{subfigure}
  
  \caption{Depth-dependent apparent hardness and representative reference-load
	indentation curve. (a) Apparent Oliver--Pharr hardness versus contact
	depth for the three training steels and the fourth steel (open
	diamonds); markers and error bars show the mean ± SD per load, faint
	symbols individual indentations. (b) Representative 350 mN
	\(P\)--\(h\) curve of E426 showing the unloading fit, stiffness
	tangent, and contact-depth estimate. (c) SEM images of indents on
	E426; image blur at 10 and 5 mN reflects imaging conditions at this
	magnification.}
  \label{fig:combined}
\end{figure*}

Figure~\ref{fig:ise_trend} presents the depth-dependent apparent
hardness of the three training steels (5--350 mN) and of the fourth
steel (3--375 mN). The ISE is pronounced in all four materials. At the
shallowest training load (\(P_{\max} = 5\) mN, \(h_{c} = 76\)--172 nm),
the mean apparent hardness exceeds the high-load reference hardness by
factors of 1.67--1.86: from 2.00 to 3.73 GPa for E212 and from 6.49 to
10.95 GPa for E841. The fourth steel shows a similar inflation at 3 mN
(\(h_{c} \approx 51\) nm, 10.51 GPa, 1.73 times its 375 mN reference)
and follows a trend close to that of E841 at shallow depths. The
representative \(P\)--\(h\) curve in Figure~\ref{fig:solo}, acquired at
the reference load on E426 (\(P_{\max} = 345\) mN,
\(h_{c} \approx 2.02\ \mu\)m, \(H = 3.48\) GPa, \(E_r = 202\) GPa),
illustrates the Oliver--Pharr analysis from which the reference is
obtained. The SEM micrographs of
residual indents on E426 (Figure~\ref{fig:SEM}), representative of the
other steels, show material piled up at the contact periphery. At
350 mN, the mean Oliver--Pharr contact area of E426
(95.8~\(\mu\mathrm{m}^2\)) corresponds to a Berkovich impression about
12.9~\(\mu\)m high, consistent with the 12.7~\(\mu\)m measured in
Figure~\ref{fig:SEM}. Pile-up
was not quantified; because the Oliver--Pharr area estimate does not
include piled-up material, it may bias the apparent hardness at all
depths.

The high-load reference hardness \(H_{\mathrm{ref}}\) (Section~\ref{data-processing-and-feature-engineering})
is a finite-load value rather than a depth-independent hardness: the
mean hardness still decreases by 10.0--11.6\% between 200 and 350 mN for
the training steels and by 7.0\% between 275 and 375 mN for the fourth
steel. \(H_{\mathrm{ref}}\) and the Nix--Gao intercept \(H_{0}\) are
therefore compared below but are not treated as the same quantity. Both
also retain any pile-up bias of the Oliver--Pharr contact area; a bulk
hardness additionally requires a pile-up correction and comparison with
macroscale hardness \cite{zhuAccurateEvaluationBulk2022}, which was not
attempted here.

\begin{figure*}[htbp]
  \centering
  \includegraphics[width=\textwidth]{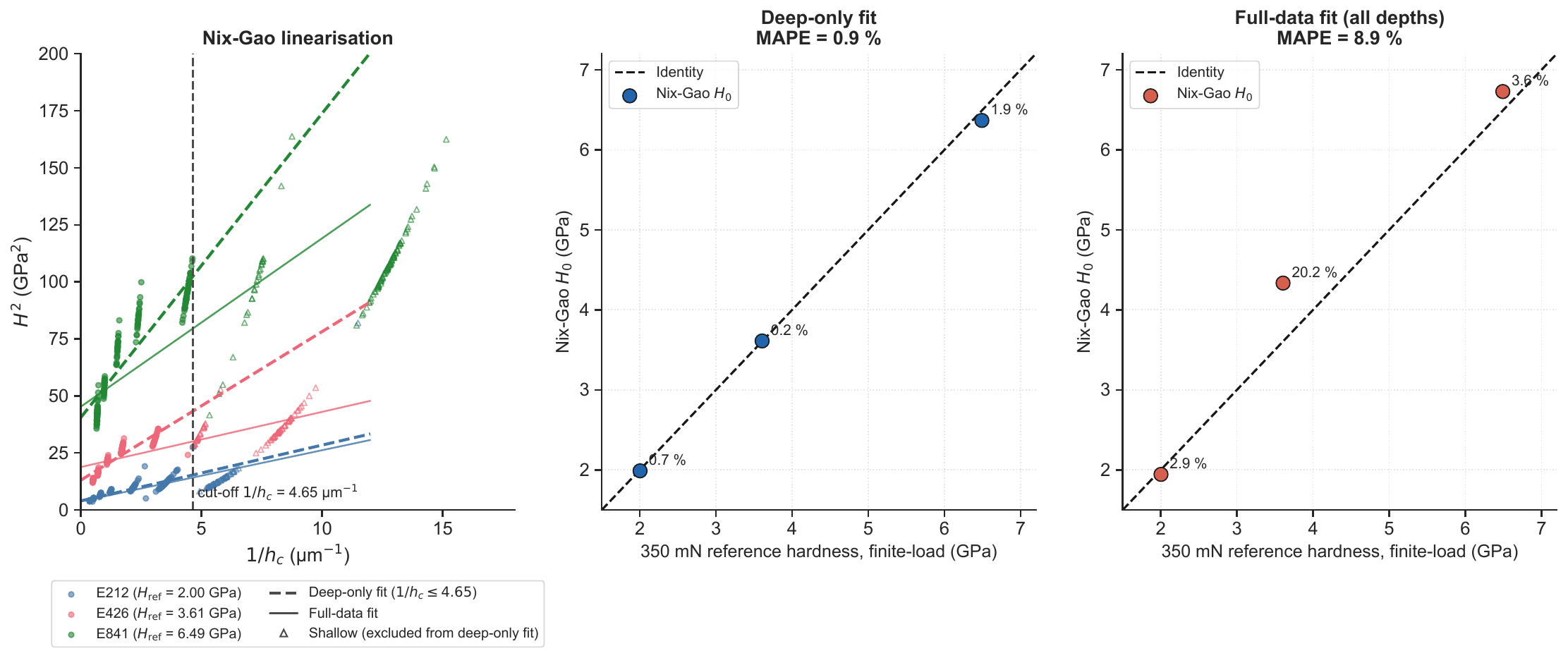}

  \caption{Nix--Gao baseline for the training steels. (a) \(H^{2}\)
	versus \(1/h_{c}\) with deep-only fits (\(1/h_{c} \leq 4.65\
	\mu\mathrm{m}^{-1}\), dashed) and full-data fits (solid); triangles
	mark shallow indentations excluded from the deep-only fit. (b)
	Deep-only \(H_{0}\) versus the 350 mN high-load reference hardness
	(MAPE 0.9\%). (c) Full-data \(H_{0}\) versus the same reference
	(MAPE 8.9\%).}
  \label{fig:nix_gao_full}
\end{figure*}

The classical Nix--Gao linearization (\(H^{2}\) versus \(1/h_{c}\)) was
applied as the analytical baseline (Figure~\ref{fig:nix_gao_full}).
The data follow an approximately linear trend in the deep contact regime
(\(h_{c} \geq 215\) nm, \(1/h_{c} \leq 4.65\ \mu\mathrm{m}^{-1}\)) and
deviate upward at shallower depths; tip rounding and pile-up are likely
contributions, but they are not separated here. The cut-off was chosen
once, before any model was trained, by inspection of the
\(H^{2}\)--\(1/h_{c}\) plots of the training steels, and was applied
unchanged to every material without fallback to the full data set. The
deep-only fit (\(R^{2} = 0.76\)--0.85) gives \(H_{0}\) within 0.7, 0.2,
and 1.9\% of \(H_{\mathrm{ref}}\) for E212, E426, and E841 (MAPE
0.9\%). A linear fit through all depths is distorted by the shallow
upward deviation and gives a MAPE of 8.9\%, with E426 overestimated by
20.2\%. Because the deep-only fit includes the reference-load
indentations, its agreement with \(H_{\mathrm{ref}}\) is partly built
in. For the fourth steel, the same deep-only fit gives
\(H_{0} = 5.49\) GPa, 10\% below its 375 mN reference (6.08 GPa);
excluding the reference-load indentations raises \(H_{0}\) to 5.79 GPa.
The agreement between \(H_{0}\) and \(H_{\mathrm{ref}}\) is thus
sensitive to the depth range and reference load. The deep-only fit is
the analytical reference used in this study.

The deep-only fit requires a series of indentations at several loads,
including deep ones, on the material being tested. The task considered
here is different: to estimate the high-load reference hardness from
single indentations, including shallow ones, using a size-effect
correction learned from other steels, without a load series of the
material being tested and without discarding the shallow depth range.
A Nix--Gao fit restricted to the shallow indentations of the fourth
steel is compared with the learned corrections in the shallow-only test
(Section~\ref{shallow-only-test-on-the-fourth-steel}).

\hypertarget{machine-learning-predictive-performance}{%
\subsection{Machine Learning Predictive
Performance}\label{machine-learning-predictive-performance}}

\begin{figure*}[htbp]
  \centering
  \includegraphics[width=\textwidth]{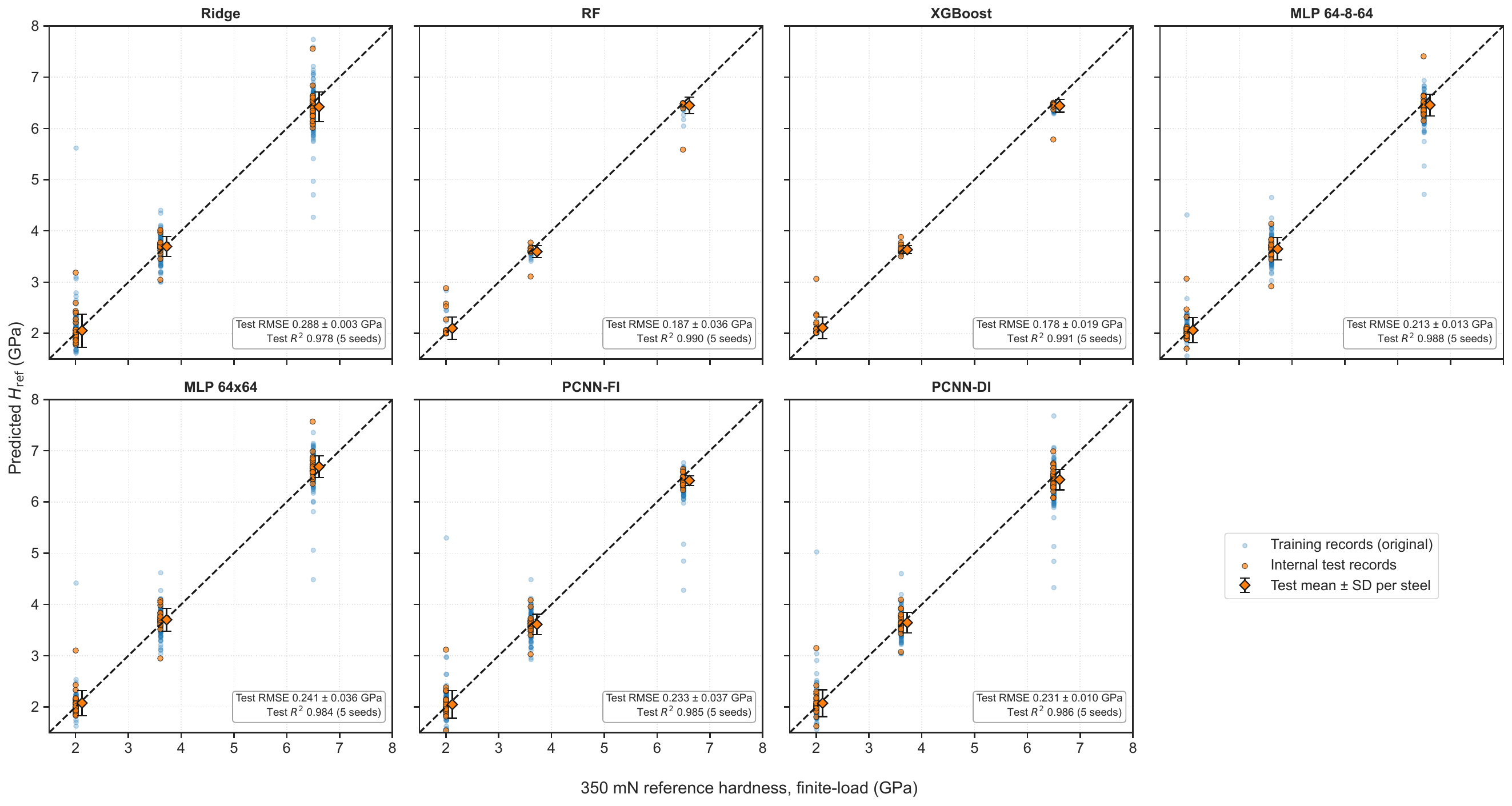}
  \caption{Parity plots of the predicted versus the 350 mN high-load
  reference hardness for the seven models (seed 42, the first
  prespecified seed). Faint symbols: original training records; orange
  symbols: development-set records; diamonds: development-set mean ± SD
  per steel. The panel insets give the five-seed test RMSE and \(R^{2}\).}
  \label{fig:parity_plots}
\end{figure*}

Seven regression models were evaluated on the development set: the
five conventional models (Ridge regression, Random Forest, XGBoost, and
the two multilayer perceptrons) and the two PCNN variants. PCNN-FI uses
the same 21-input feature vector as the conventional models, whereas
PCNN-DI uses the eight dimensionless descriptors described in
Section~\ref{data-processing-and-feature-engineering}. Table~\ref{tab:ml_metrics}
reports the mean ± SD over the five run seeds, and
Figure~\ref{fig:parity_plots} shows the parity plots for the first
prespecified seed (42). The development set contains indentations of the
same three steels and loads as the training set, and every indentation
of a steel shares the same target value; its metrics therefore measure
interpolation within the training steels rather than transfer to a new
material (Section~\ref{evaluation-protocol}).

\begin{table*}[t]
  \caption{Development-set performance. Training metrics are computed on
  the original (non-augmented) training records. Values are mean ± SD
  over five run seeds (42--46) on the same split; the SD describes
  training-pipeline variability, not a confidence interval. RMSE in GPa.}
  \label{tab:ml_metrics}
  \centering
  \begin{tabular}{lcccc}
    \toprule
    Model       & Train $R^2$       & Test $R^2$        & Train RMSE        & Test RMSE         \\
    \midrule
    Ridge       & 0.974 ± 0.000 & 0.978 ± 0.000 & 0.308 ± 0.001 & 0.288 ± 0.003 \\
    Random Forest & 0.999 ± 0.000 & 0.990 ± 0.003 & 0.054 ± 0.007 & 0.187 ± 0.036 \\
    XGBoost     & 1.000 ± 0.000 & 0.991 ± 0.002 & 0.040 ± 0.001 & 0.178 ± 0.019 \\
    MLP 64-8-64 & 0.993 ± 0.004 & 0.988 ± 0.001 & 0.150 ± 0.044 & 0.213 ± 0.013 \\
    MLP 64×64   & 0.989 ± 0.005 & 0.984 ± 0.004 & 0.196 ± 0.045 & 0.241 ± 0.036 \\
    PCNN-FI     & 0.983 ± 0.002 & 0.985 ± 0.005 & 0.246 ± 0.016 & 0.233 ± 0.037 \\
    PCNN-DI     & 0.983 ± 0.002 & 0.986 ± 0.001 & 0.249 ± 0.013 & 0.231 ± 0.010 \\
    \bottomrule
  \end{tabular}
\end{table*}

All seven models reach a test \(R^{2}\) of 0.978--0.991 and a test RMSE
of 0.18--0.29 GPa. Because each steel has a single reference value, the
parity plots consist of three columns, and these metrics mainly reflect
the per-indentation scatter of the predictions around each steel's
reference. Ridge regression, a linear model on the same inputs, already
reaches a test RMSE of 0.288 GPa, so the development-set task does not
require a strongly non-linear model.

Random Forest and XGBoost give the lowest development-set errors (test
RMSE 0.187 and 0.178 GPa) and fit the original training records almost
exactly (train RMSE 0.054 and 0.040 GPa). Their test RMSE is about 3.5 and
4.5 times their train RMSE, the largest gap among the models. The two
multilayer perceptrons follow (0.213 and 0.241 GPa); their difference is
comparable to the seed SD. Ridge regression and both PCNN variants have
similar train and test RMSE (0.23--0.31 GPa), and the two PCNN variants
are almost identical on this set (0.233 ± 0.037 and 0.231 ± 0.010 GPa).

\hypertarget{transfer-to-the-training-steels-and-the-fourth-steel}{%
\subsection{Performance on the Training Steels (In-Sample) and on the
Fourth Steel}\label{transfer-to-the-training-steels-and-the-fourth-steel}}

All models were scored on all eligible indentations of each material
(238, 241, 303, and 346 for E212, E426, E841, and the fourth steel)
against the high-load reference hardness of that material. For the
training steels these include the training records; the fourth steel
was not used in model training. Table~\ref{tab:mape_all} reports the
MAPE for every material, and Table~\ref{tab:fourth_steel} gives further
metrics for the fourth steel. The fused silica row is discussed in
Section~\ref{boundary-testing-on-fused-silica}.

\begin{table*}[t]
  \caption{MAPE (\%) against the high-load reference hardness (350 mN;
  375 mN for the fourth steel) on all eligible indentations of each
  material. Mean ± SD over five run seeds.}
  \label{tab:mape_all}
  \centering
  \footnotesize
  \setlength{\tabcolsep}{2pt}
  \begin{tabular}{@{}lccccccccc@{}}
    \toprule
    Material & Train. & \(N\) & Ridge & RF & XGBoost & MLP 64-8-64 & MLP 64×64 & PCNN-FI & PCNN-DI \\
    \midrule
    E212 & Yes & 238 & 8.37 ± 0.08 & 1.26 ± 0.19 & 1.83 ± 0.12 & 3.52 ± 0.78 & 5.10 ± 1.07 & 6.46 ± 0.40 & 6.17 ± 0.13 \\
    E426 & Yes & 241 & 4.49 ± 0.22 & 0.38 ± 0.05 & 0.59 ± 0.04 & 2.99 ± 0.62 & 3.44 ± 0.48 & 3.51 ± 0.16 & 3.68 ± 0.08 \\
    E841 & Yes & 303 & 4.13 ± 0.05 & 0.35 ± 0.04 & 0.61 ± 0.01 & 1.91 ± 0.78 & 2.62 ± 1.02 & 2.60 ± 0.64 & 3.05 ± 0.44 \\
    Fourth steel & No & 346 & 3.65 ± 0.09 & 10.19 ± 2.21 & 8.51 ± 1.53 & 4.28 ± 0.99 & 5.31 ± 0.61 & 3.66 ± 0.27 & 4.84 ± 0.28 \\
    Fused silica & No & 35 & 15.88 ± 4.37 & 57.67 ± 3.17 & 53.10 ± 11.43 & 59.30 ± 47.91 & 67.18 ± 50.56 & 46.53 ± 26.98 & 31.29 ± 24.70 \\
    \bottomrule
  \end{tabular}
\end{table*}

\hypertarget{training-steels}{%
\subsubsection{Training Steels}\label{training-steels}}

\begin{figure*}[htbp]
  \centering
  \includegraphics[width=\textwidth]{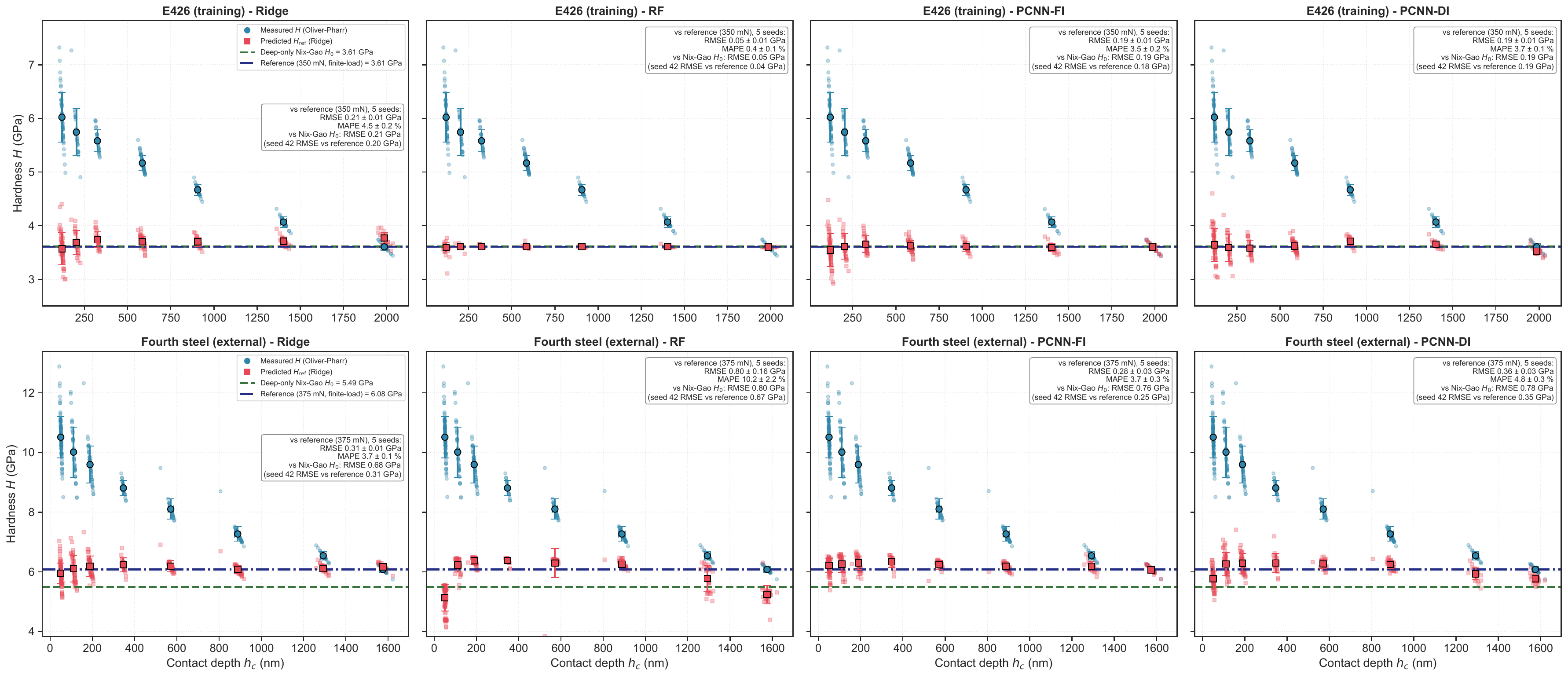}
  \caption{Measured apparent hardness and predicted high-load reference
  hardness versus contact depth for Ridge regression, Random Forest,
  PCNN-FI, and PCNN-DI. Top row: E426 (training steel, 350 mN
  reference); bottom row: fourth steel (external, 375 mN reference).
  Horizontal lines mark the high-load reference hardness and the
  deep-only Nix--Gao \(H_{0}\). Points: seed 42 (the first prespecified
  seed); insets: five-seed RMSE and MAPE.}
  \label{fig:error_e426}
\end{figure*}

On the three training steels, Random Forest and XGBoost give the lowest
errors (MAPE 0.35--1.83\%), consistent with their development-set
results; the other models reach 1.9--8.4\%. All models remove most of
the size effect at every load, although the apparent hardness at 5 mN is
up to 1.86 times the reference (Figure~\ref{fig:error_e426}, top row).
The highest MAPE values occur for E212 with Ridge regression and the two
PCNN variants (6.2--8.4\%), mainly because its reference hardness is
only 2.0 GPa, so that absolute errors of about 0.3 GPa correspond to
large relative errors.

\hypertarget{external-test-on-the-fourth-steel}{%
\subsubsection{External Test on the Fourth
Steel}\label{external-test-on-the-fourth-steel}}

The fourth steel was indented at loads offset from the training schedule
(3, 7, 15, 35, 75, 150, 275, and 375 mN). Its apparent hardness falls
from 10.5 GPa at 3 mN to the reference of 6.08 GPa at 375 mN; the
deep-only Nix--Gao fit gives \(H_{0} = 5.49\) GPa
(Section~\ref{baseline-nanoindentation-and-limitations-of-the-nix-gao-model}).
Predictions at 7--275 mN interpolate between training loads, those at
3 mN extrapolate below the shallowest training load (5 mN), and 375 mN
is the reference load of this steel.

The development-set ranking reverses on this steel
(Tables~\ref{tab:ml_metrics} and~\ref{tab:mape_all};
Figure~\ref{fig:S_reversal}). Random Forest and
XGBoost, the most accurate models on the development set, give the
largest errors (MAPE 10.19 and 8.51\%). Ridge regression, which had the
largest development-set error, gives the lowest MAPE (3.65\%), together
with PCNN-FI (3.66\%), followed by the MLP 64-8-64 (4.28\%) and PCNN-DI
(4.84\%). These four models differ by about one percentage point, which
is smaller than the ambiguity of the reference itself: the mean hardness
of this steel still decreases by 7\% between 275 and 375 mN, and its
Nix--Gao \(H_{0}\) lies 10\% below the 375 mN reference. Against
\(H_{0}\), the four models give similar RMSE values of 0.68--0.78 GPa
(Table~\ref{tab:fourth_steel}). They are therefore not ranked.

\begin{table}[t]
  \caption{Fourth steel: RMSE against the 375 mN high-load reference
  hardness (mean ± SD over five run seeds), mean bias (prediction minus
  reference), RMSE against the deep-only Nix--Gao \(H_{0}\), and
  coefficient of variation of the predictions over all indentations
  (CV). Values without SD are five-seed means.}
  \label{tab:fourth_steel}
  \centering
  \small
  \begin{tabular}{@{}lcccc@{}}
    \toprule
    Model & RMSE (GPa) & Bias (GPa) & RMSE vs \(H_{0}\) (GPa) & CV (\%) \\
    \midrule
    Ridge       & 0.308 ± 0.006 & \(+\)0.021 & 0.682 & 5.0 \\
    RF          & 0.796 ± 0.163 & \(-\)0.305 & 0.799 & 12.7 \\
    XGBoost     & 0.649 ± 0.145 & \(-\)0.212 & 0.730 & 10.5 \\
    MLP 64-8-64 & 0.328 ± 0.079 & \(+\)0.105 & 0.762 & 5.0 \\
    MLP 64×64   & 0.408 ± 0.052 & \(+\)0.144 & 0.823 & 5.9 \\
    PCNN-FI     & 0.275 ± 0.026 & \(+\)0.128 & 0.756 & 3.8 \\
    PCNN-DI     & 0.359 ± 0.027 & \(+\)0.109 & 0.776 & 5.4 \\
    \bottomrule
  \end{tabular}
\end{table}

\begin{figure}[htbp]
  \centering
  \includegraphics[width=\columnwidth]{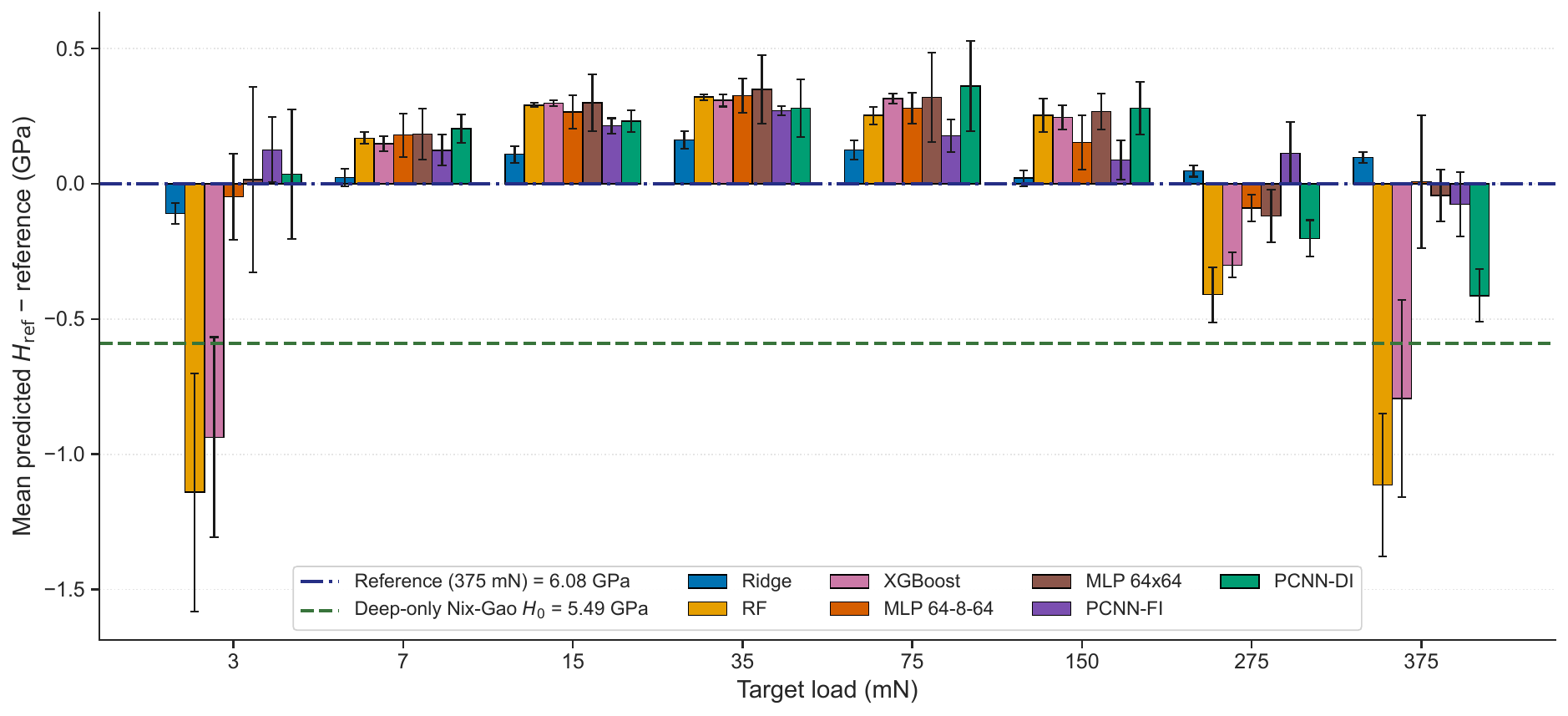}
  \caption{Fourth steel: per-load mean prediction minus the 375 mN
  high-load reference hardness for all models. Bars: mean over five run
  seeds; error bars: between-seed SD.}
  \label{fig:loads_valid}
\end{figure}

PCNN-FI and PCNN-DI agree at every load of the training steels and
differ on the fourth steel mainly at the ends of the load range
(Figure~\ref{fig:loads_valid}). PCNN-FI receives the applied load among
its inputs and can learn that the correction vanishes at the highest
load of each training steel, where the target is defined; at 375 mN its
mean prediction is 6.01 GPa. PCNN-DI sees depth only through the
normalized coordinate \(t\), and similar \(t\) values correspond to the
reference load of one training steel and an intermediate load of
another. It therefore retains a correction of about 7--10\% at the deepest
loads (Figure~\ref{fig:S_q}), and its mean prediction at 375 mN (5.67 GPa) lies between the
reference and the Nix--Gao \(H_{0}\). Part of the PCNN-FI advantage thus
reflects the high-load reference convention rather than a better
size-effect correction. Because each PCNN prediction is reconstructed
from the measured \(H_{\mathrm{app}}\) of the individual indentation,
the per-indentation scatter of the measurements carries over to the
predictions and is largest at the shallowest loads
(Figure~\ref{fig:error_e426}; the other three models are shown in
Figure~\ref{fig:S_fourth_other}); this dispersion is not a calibrated
uncertainty.

The per-load errors (Figure~\ref{fig:loads_valid}) show where the
models differ. At 7--150 mN, Ridge regression, the two MLPs, and the two
PCNN variants overpredict by 0.02--0.36 GPa. At 3 mN, the tree ensembles
underpredict by about 1 GPa (\(-0.94\) and \(-1.14\) GPa), because the
load- and size-dependent inputs lie below their training range and tree
ensembles cannot extrapolate; the other models stay within
\(\pm 0.13\) GPa. At 275 and 375 mN, the tree ensembles again
underpredict (\(-0.30\) to \(-1.11\) GPa), although all inputs at 275 mN and all but \(P_{\max}\) and
\(W_{\mathrm{el}}\) at 375 mN lie within the training range, so this is
not mainly an input-range effect. PCNN-DI underpredicts at these loads by 0.20 and 0.41 GPa, for
the reason given above, whereas Ridge regression stays within
\(\pm 0.16\) GPa of the reference at every load.

\hypertarget{shallow-only-test-on-the-fourth-steel}{%
\subsection{Shallow-Only Test on the Fourth
Steel}\label{shallow-only-test-on-the-fourth-steel}}

The fourth-steel errors in Section~\ref{external-test-on-the-fourth-steel}
include deep indentations up to the reference load. The shallow-only
test (Section~\ref{evaluation-protocol}) estimates the reference from
the 3, 7, and 15 mN indentations alone; Table~\ref{tab:shallow_only}
compares the learned models with Nix--Gao fits.

\begin{table}[t]
  \caption{Shallow-only test on the fourth steel (3, 7, and 15 mN; 190
  indentations). Estimate: mean prediction over the indentations; the
  difference is taken from the 375 mN high-load reference hardness
  (6.08 GPa). Machine-learning values: mean ± SD over five run seeds.}
  \label{tab:shallow_only}
  \centering
  \small
  \setlength{\tabcolsep}{3pt}
  \begin{tabular}{@{}lccc@{}}
    \toprule
    Method & Estimate (GPa) & \(|\)Difference\(|\) (GPa) & RMSE (GPa) \\
    \midrule
    Ridge       & 6.056 ± 0.034 & 0.032 ± 0.023 & 0.382 ± 0.005 \\
    RF          & 5.598 ± 0.232 & 0.482 ± 0.232 & 0.900 ± 0.255 \\
    XGBoost     & 5.701 ± 0.196 & 0.379 ± 0.196 & 0.733 ± 0.243 \\
    MLP 64-8-64 & 6.164 ± 0.086 & 0.105 ± 0.048 & 0.348 ± 0.117 \\
    MLP 64×64   & 6.206 ± 0.211 & 0.194 ± 0.133 & 0.462 ± 0.045 \\
    PCNN-FI     & 6.229 ± 0.078 & 0.148 ± 0.078 & 0.312 ± 0.043 \\
    PCNN-DI     & 6.204 ± 0.123 & 0.151 ± 0.078 & 0.341 ± 0.039 \\
    \midrule
    Nix--Gao, shallow only & 9.18 & 3.10 & -- \\
    Nix--Gao, deep only\textsuperscript{a} & 5.49 & 0.59 & -- \\
    \bottomrule
  \end{tabular}

  \raggedright\footnotesize \textsuperscript{a}\,Context only: requires
  35--375 mN indentations, including the reference indentations.
\end{table}

From the 3--15 mN indentations alone, the Nix--Gao fit is physically
admissible (positive slope and intercept, \(R^{2} = 0.32\)), but it
overestimates the reference by 3.1 GPa (Table~\ref{tab:shallow_only},
Figure~\ref{fig:S_nixgao_shallow}); the fit to the per-load means
gives the same result (9.37 GPa). A straight line through contacts of
42--203 nm does not extrapolate to the deep regime. The machine-learning
models estimate the reference within 0.03--0.48 GPa from the same
indentations. This is not a like-for-like comparison: the Nix--Gao fit
extrapolates within one material from its shallow points, whereas the
learned models apply a correction learned from the deep-load behavior of
three other steels.

The smooth models (Ridge regression, the two MLPs, and the two PCNN
variants) stay within 0.3 GPa of the reference at each of the three
loads (Figure~\ref{fig:loads_valid}). The tree ensembles underestimate
the 3 mN indentations by about 1 GPa and therefore give the largest
shallow-only differences. The differences among the smooth models
(0.03--0.19 GPa) are smaller than the ambiguity of the reference itself
and are not ranked. Two limitations apply. The 3 mN indentations lie
below the shallowest training load, so their predictions are
extrapolations in the depth coordinate (PCNN-DI) and in the load- and
size-dependent inputs (the other models). They also lie below the
shallowest calibration depth of the area function (86 nm), within the
rounded tip apex, so they are an extrapolation in contact geometry as
well. In addition, at shallow loads
the fourth steel resembles E841: its hardness at 3 mN (10.5 GPa) lies
between that of E841 at 5 and 10 mN (10.9 and 10.0 GPa), and its reference (6.08 GPa) lies just below that
of E841 (6.49 GPa). Part of this agreement may therefore reflect this resemblance. Fold A
of the leave-one-steel-out check (Section~\ref{leave-one-steel-out})
tests a steel without such a close training neighbour; the agreement
should not be generalized to steels outside the hardness range of the
training set.

\hypertarget{leave-one-steel-out}{%
\subsection{Leave-One-Steel-Out Check}\label{leave-one-steel-out}}

The fourth steel is a single external material. To test whether the
pattern seen there holds for other unseen steels, each training steel
was left out in turn (Section~\ref{evaluation-protocol}): fold A leaves
out E426 (interpolation in hardness), fold B leaves out E212 (a softer
steel than any training steel), and fold C leaves out E841 (a harder
one). Each fold trains on two steels only, so its errors are larger
than those of the main models and are not compared with
Tables~\ref{tab:mape_all}--\ref{tab:shallow_only}. Table~\ref{tab:loso}
gives the MAPE on the left-out steel.

\begin{table}[t]
  \caption{Leave-one-steel-out check: MAPE (\%) on the left-out steel
  against its 350 mN high-load reference hardness, all eligible
  indentations. Each fold trains on the other two steels. Mean ± SD over
  five run seeds.}
  \label{tab:loso}
  \centering
  \small
  \setlength{\tabcolsep}{3pt}
  \begin{tabular}{@{}lccc@{}}
    \toprule
    Model & A: E426 out & B: E212 out & C: E841 out \\
          & (3.61 GPa)  & (2.00 GPa)  & (6.49 GPa)  \\
    \midrule
    Ridge       & 6.40 ± 0.25  & 18.18 ± 0.69  & 12.34 ± 0.43 \\
    RF          & 24.15 ± 6.75 & 84.66 ± 4.20  & 53.43 ± 2.40 \\
    XGBoost     & 24.15 ± 5.99 & 91.89 ± 15.46 & 49.12 ± 1.46 \\
    MLP 64-8-64 & 12.19 ± 2.39 & 67.76 ± 16.10 & 28.08 ± 7.44 \\
    MLP 64×64   & 9.88 ± 2.76  & 59.42 ± 6.15  & 29.20 ± 11.09 \\
    PCNN-FI     & 6.48 ± 1.48  & 12.85 ± 7.72  & 38.91 ± 15.88 \\
    PCNN-DI     & 7.33 ± 1.38  & 10.57 ± 5.79  & 36.17 ± 10.96 \\
    \bottomrule
  \end{tabular}
\end{table}

In fold A, Ridge regression, PCNN-FI, and PCNN-DI agree within about one
percentage point (6.4--7.3\%), whereas the MLPs (9.9--12.2\%) and the
tree ensembles (24\%) are clearly worse. Because E426 lies between the two remaining training steels and
resembles neither of them, fold A is an interpolation test in hardness
without a close training neighbour; it repeats the fourth-steel pattern
with the same load schedule and reference definition.

\begin{figure*}[htbp]
  \centering
  \begin{subfigure}[b]{\textwidth}
    \centering
    \includegraphics[width=\textwidth]{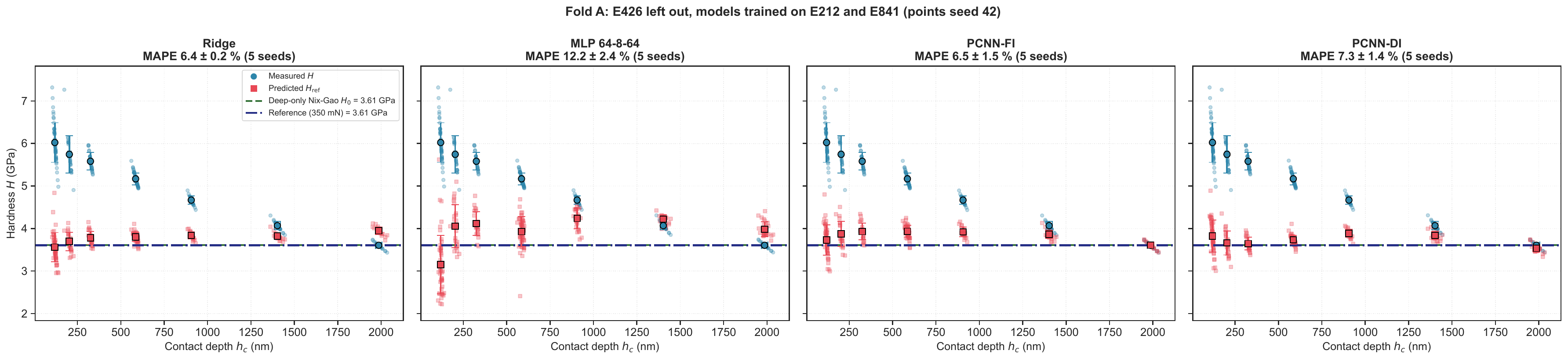}
    \caption{}
    \label{fig:loso_A}
  \end{subfigure}

  \vspace{\floatsep}

  \begin{subfigure}[b]{\textwidth}
    \centering
    \includegraphics[width=\textwidth]{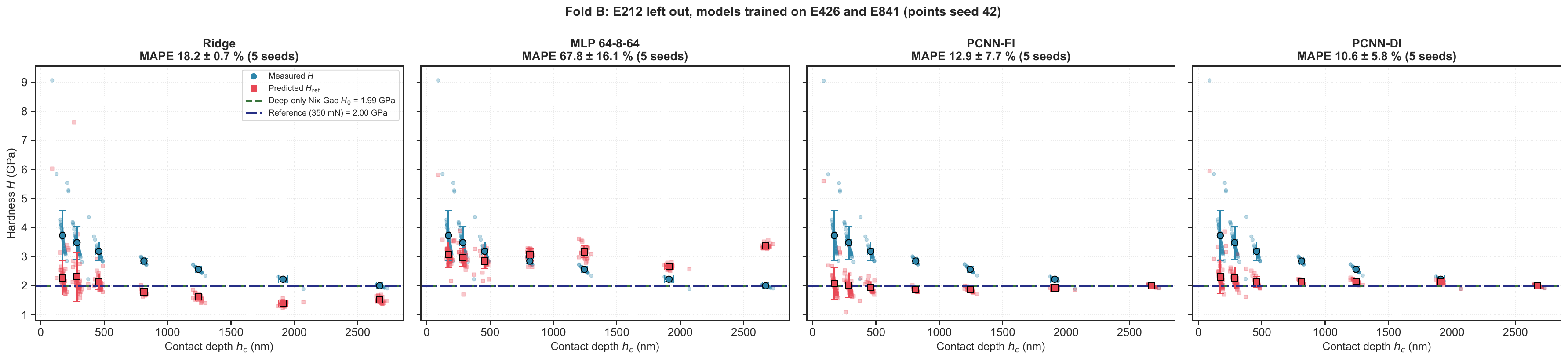}
    \caption{}
    \label{fig:loso_B}
  \end{subfigure}
  \caption{Leave-one-steel-out check: measured apparent hardness and
  predicted high-load reference hardness versus contact depth for Ridge
  regression, the MLP 64-8-64, PCNN-FI, and PCNN-DI on the left-out
  steel. (a) Fold A, E426 left out; (b) fold B, E212 left out. Points:
  seed 42 (the first prespecified seed); panel titles: five-seed MAPE.}
  \label{fig:loso}
\end{figure*}

In fold B, the two PCNN variants give the lowest errors
(Figure~\ref{fig:loso}b; per-load errors of all folds in
Figure~\ref{fig:S_loso_perload}). Their predictions stay close to the 2.00 GPa
reference at every load (PCNN-DI 2.01--2.26 GPa), although no training
steel was this soft, because the reconstruction
\(H_{\mathrm{app}}/\sqrt{1+q}\) scales with the measured hardness. The
tree ensembles stay near the lowest hardness in their training targets
(3.6--3.9 GPa at every load), and the MLPs overpredict by up to 1.7 GPa,
most strongly at shallow loads. Ridge regression reaches
the right level at shallow loads but continues to correct at deep loads,
where the measured hardness already equals the reference, and falls to
1.4--1.6 GPa at 200--350 mN; its prediction is therefore not
load-independent.

In fold C, only Ridge regression extrapolates upward to the harder steel
(12.3\%; Figure~\ref{fig:S_foldC}). All non-linear models underpredict strongly, including the
PCNN variants (36--39\%), which predict 3.6--4.6 GPa (4.4--4.6 GPa even at 350 mN),
where the measured hardness already equals the reference. For inputs
outside the training range, the network outputs a large correction: the
bounded form fixes the direction of the correction, not its size.
Upward extrapolation in hardness is therefore outside the domain of
validity of the PCNNs trained here. With only two training steels per
fold and large seed SDs in folds B and C, these results are robustness
evidence rather than precise effect sizes.

\hypertarget{augmentation-comparison}{%
\subsection{Augmentation Comparison}\label{augmentation-comparison}}

Table~\ref{tab:augmentation} compares the four augmentation conditions
(Section~\ref{evaluation-protocol}) for PCNN-DI and, as a conventional
comparator, for the MLP 64-8-64 under C0 and C3.

\begin{table}[t]
  \caption{Augmentation conditions: C0, no augmentation; C1, noise and
  drift copies; C2, plus within-material MixUp; C3, plus unrestricted
  MixUp (the recipe of all reported models). RMSE (GPa) on the
  development set and on the fourth steel against its 375 mN reference,
  for the shallow (3--15 mN, 190 indentations) and all (346)
  indentations. Mean ± SD over five run seeds.}
  \label{tab:augmentation}
  \centering
  \small
  \setlength{\tabcolsep}{3pt}
  \begin{tabular}{@{}llccc@{}}
    \toprule
    Model & Cond. & Development set & Fourth steel, shallow & Fourth steel, all \\
    \midrule
    PCNN-DI     & C0 & 0.237 ± 0.019 & 0.349 ± 0.059 & 0.337 ± 0.035 \\
                & C1 & 0.231 ± 0.014 & 0.334 ± 0.019 & 0.337 ± 0.036 \\
                & C2 & 0.230 ± 0.011 & 0.341 ± 0.024 & 0.342 ± 0.033 \\
                & C3 & 0.231 ± 0.010 & 0.341 ± 0.039 & 0.359 ± 0.027 \\
    \midrule
    MLP 64-8-64 & C0 & 0.190 ± 0.019 & 0.429 ± 0.081 & 0.392 ± 0.058 \\
                & C3 & 0.213 ± 0.013 & 0.348 ± 0.117 & 0.328 ± 0.079 \\
    \bottomrule
  \end{tabular}
\end{table}

On the development set, augmentation had no measurable effect on
PCNN-DI: the paired difference C3 \(-\) C0 was \(-0.007 \pm 0.017\) GPa,
and all differences between conditions were smaller than the spread
over seeds. For the MLP, C3 \(-\) C0 was \(+0.023 \pm 0.031\) GPa, also
within the seed spread.

On the shallow indentations of the fourth steel, the four conditions
gave similar errors for PCNN-DI (C3 \(-\) C0 \(= -0.008 \pm 0.055\) GPa),
and no condition was clearly better; the augmented conditions had a
smaller spread over seeds than C0. Over all loads, C3 was slightly less
accurate than the controls (C3 \(-\) C0 \(= +0.022 \pm 0.031\) GPa).
Allowing cross-material MixUp pairs (C3 versus C2) made no difference on
the shallow indentations (\(+0.000 \pm 0.019\) GPa) and increased the
all-load error slightly but in all five seeds
(\(+0.018 \pm 0.010\) GPa). For the MLP, augmentation lowered the
shallow-indentation error in four of five seeds
(C3 \(-\) C0 \(= -0.081 \pm 0.128\) GPa), a difference within the seed
spread.

Augmentation therefore acts at most as a mild training regularizer; it
does not add independent material states, and the PCNN-DI results do
not depend on it. These observations are specific to one steel of
intermediate hardness, and the leave-one-steel-out check
(Section~\ref{leave-one-steel-out}) used only the fixed recipe C3.

\hypertarget{boundary-testing-on-fused-silica}{%
\subsection{Out-of-Domain Test on Fused
Silica}\label{boundary-testing-on-fused-silica}}

Fused silica (reference 9.21 GPa, \(E_{r} \approx 72\) GPa) is an
amorphous material outside the training domain in deformation
mechanism, elastic modulus, and hardness. Its apparent hardness shows
almost no size effect and is already within 2.2\% of the reference, so
any correction makes the prediction worse. All models fail
(Table~\ref{tab:mape_all}, Figure~\ref{fig:S_glass}): Ridge regression gives a MAPE of 15.9\%,
PCNN-DI \(31 \pm 25\)\%, and the other models 47--67\%, with large
spreads over seeds. Because fused silica differs from the steels in
several respects at once, this result marks a limit of the domain of
validity but does not identify which difference causes the failure. The
models give no indication that an input lies outside their training
domain, so they should not be applied to other material classes without
an out-of-domain check or retraining.

\subsection{Ablation of the PCNN-DI}
\label{boundedpinn-ablation-results}

Table~\ref{tab:pcnn_ablation_results} compares the full PCNN-DI with the
seven variants defined in Section~\ref{pcnn-ablation-methods}
(shown graphically in Figure~\ref{fig:S_ablation}). The
fourth steel is the transfer test; the development set is reported for
reference.

\begin{table*}[t]
  \caption{Ablation of the PCNN-DI. Development-set RMSE and, for the
  fourth steel, RMSE against the 375 mN high-load reference hardness and
  against the deep-only Nix--Gao \(H_{0}\), CV, and MAPE.
  Mean ± SD over five run seeds.}
  \label{tab:pcnn_ablation_results}
  \centering
  \small
  \setlength{\tabcolsep}{4pt}
  \begin{tabular}{@{}lccccc@{}}
    \toprule
    & Development set & \multicolumn{4}{c}{Fourth steel} \\
    \cmidrule(lr){2-2}\cmidrule(lr){3-6}
    Variant & RMSE (GPa) & RMSE (GPa) & RMSE vs \(H_{0}\) (GPa) & CV (\%) & MAPE (\%) \\
    \midrule
    PCNN-DI (full)                          & 0.231 ± 0.010 & 0.359 ± 0.027 & 0.776 ± 0.061 & 5.42 ± 0.62 & 4.84 ± 0.28 \\
    Direct NN                               & 0.292 ± 0.028 & 0.506 ± 0.053 & 0.948 ± 0.047 & 6.93 ± 0.72 & 7.07 ± 0.68 \\
    No \(q\)-target loss                    & 0.232 ± 0.010 & 0.345 ± 0.023 & 0.776 ± 0.058 & 5.13 ± 0.53 & 4.63 ± 0.28 \\
    No load-independence loss               & 0.222 ± 0.010 & 0.348 ± 0.052 & 0.744 ± 0.051 & 5.46 ± 0.92 & 4.62 ± 0.71 \\
    No \(m\), \(n\)                         & 0.238 ± 0.019 & 0.373 ± 0.048 & 0.618 ± 0.027 & 6.04 ± 0.78 & 5.07 ± 0.69 \\
    No area-free descriptors                & 0.310 ± 0.023 & 0.463 ± 0.042 & 0.823 ± 0.042 & 7.31 ± 0.61 & 5.83 ± 0.40 \\
    No \(\log(E_r/E_{r,\mathrm{ref}})\)     & 0.269 ± 0.025 & 0.394 ± 0.044 & 0.868 ± 0.044 & 5.24 ± 0.54 & 4.76 ± 0.54 \\
    No depth coordinate \(t\)               & 0.324 ± 0.047 & 0.727 ± 0.063 & 1.028 ± 0.063 & 11.37 ± 1.06 & 9.34 ± 0.50 \\
    \bottomrule
  \end{tabular}
\end{table*}

Three components clearly matter on the fourth steel. Replacing the
bounded reconstruction \(H_{\mathrm{app}}/\sqrt{1+q}\) by a direct
network increases the fourth-steel RMSE from 0.359 to 0.506 GPa and the
MAPE from 4.8 to 7.1\%, and also worsens the development-set error. Because no prediction
reached the bounds of \(q\) (Section~\ref{pcnn-methods}), this gain is
attributed to the multiplicative reconstruction, which scales the
prediction with the measured hardness; the bounds act only through the
scaling of the tanh output.
Removing the depth coordinate \(t\) causes the largest loss (MAPE
9.3\%, CV 11.4\%), and removing the area-free descriptors also worsens
all metrics (MAPE 5.8\%, CV 7.3\%). Removing
\(\log(E_r/E_{r,\mathrm{ref}})\) or the exponents \(m\) and \(n\)
changes the fourth-steel error by less than or about one seed SD;
without \(m\) and \(n\), the RMSE against \(H_{0}\) is lower
(0.618 GPa), which again reflects the ambiguity between the two
references rather than a clear improvement.

The two auxiliary losses are not supported by this test. Without the
\(q\)-target loss or without the load-independence loss, the
fourth-steel errors are slightly lower than those of the full model
(MAPE 4.63 and 4.62\% versus 4.84\%) and the CV is unchanged, all within
the seed SD. The losses were kept in the reported model as specified
before the evaluation, but they are not shown to improve transfer.
Because several descriptors are partly redundant, these results show
how the model depends on its components, not a unique physical
mechanism.

\hypertarget{physics-guided-feature-importance-and-latent-space-analysis}{%
\subsection{Feature Attribution and Latent Space
Analysis}\label{physics-guided-feature-importance-and-latent-space-analysis}}

\hypertarget{shap-analysis}{%
\subsubsection{SHAP Analysis}\label{shap-analysis}}

\begin{figure*}[htbp]
  \centering
  \begin{subfigure}[b]{0.49\textwidth}
    \centering
    \includegraphics[width=\linewidth]{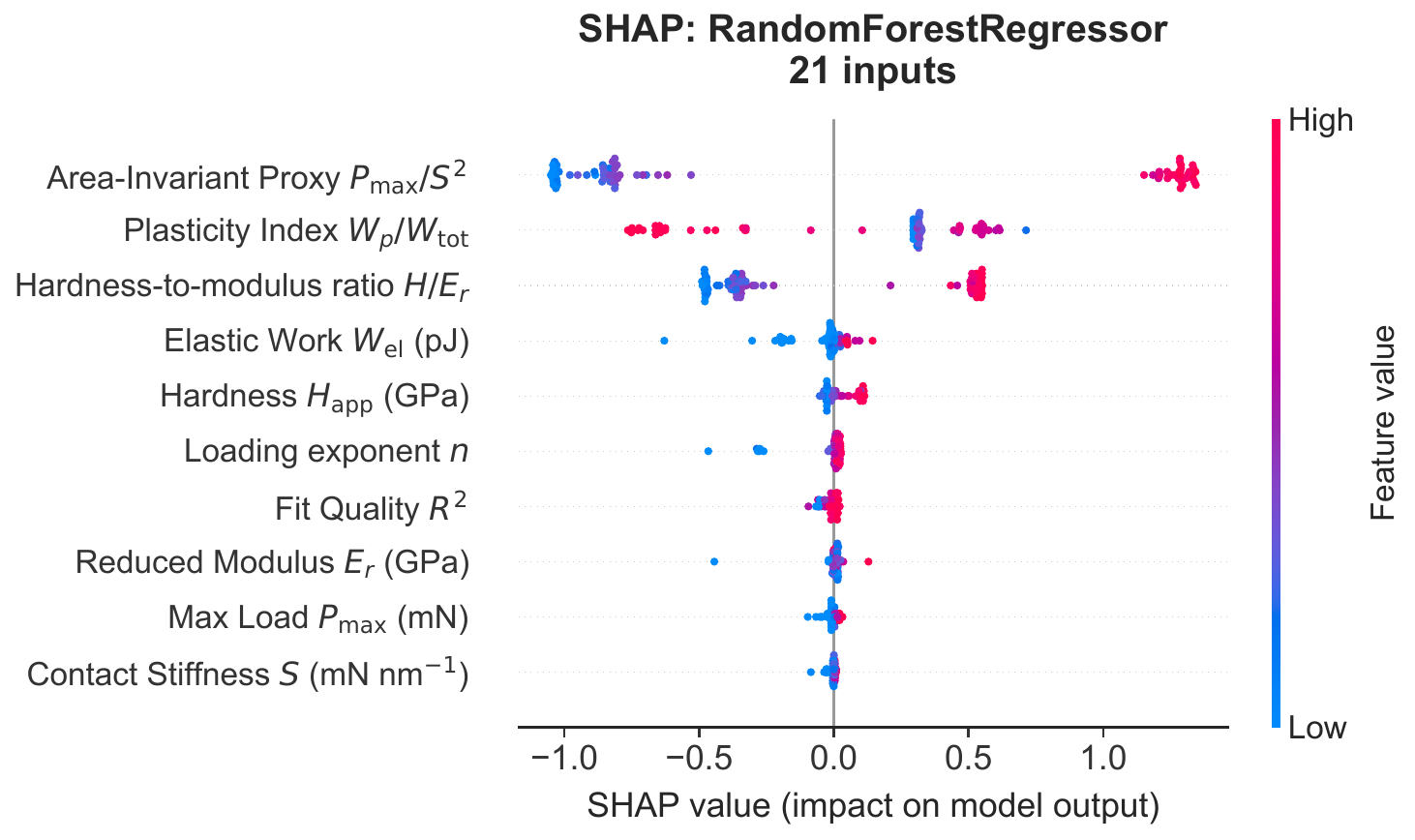}
    \caption{}
    \label{fig:RF_SHAP}
  \end{subfigure}
  \hfill
  \begin{subfigure}[b]{0.49\textwidth}
    \centering
    \includegraphics[width=\linewidth]{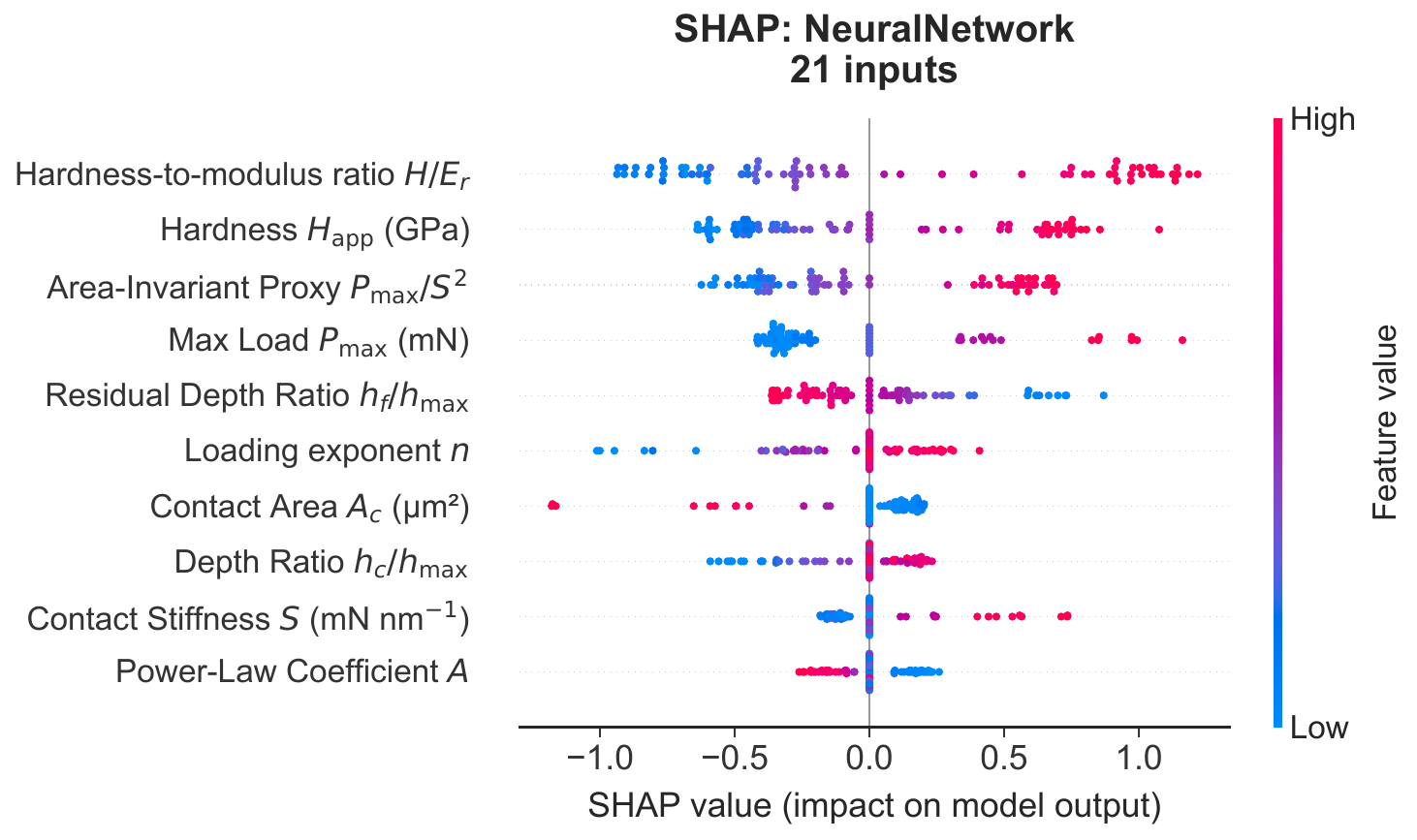}
    \caption{}
    \label{fig:NN_SHAP}
  \end{subfigure}

  \vspace{\floatsep}

  \begin{subfigure}[b]{0.6\textwidth}
    \centering
    \includegraphics[width=\linewidth]{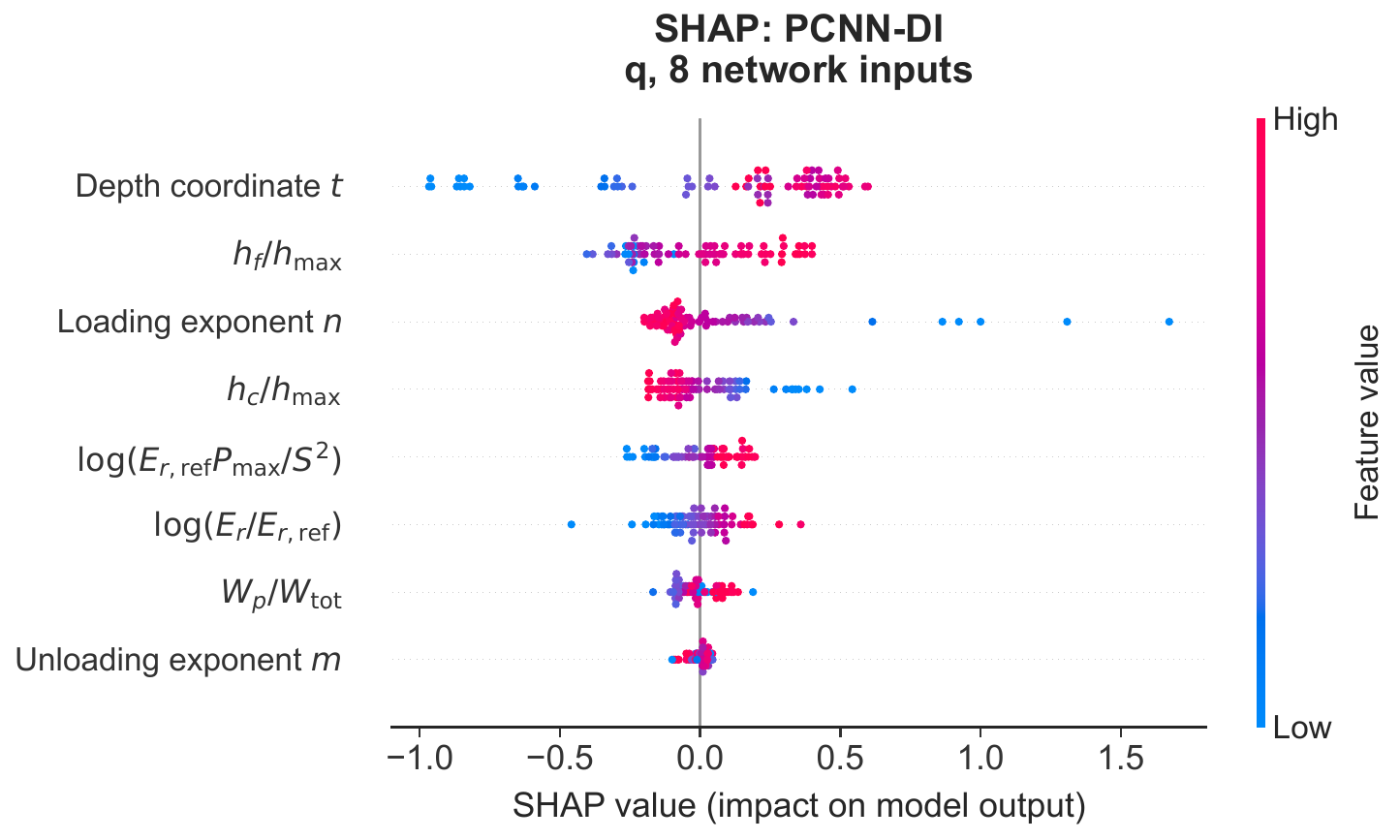}
    \caption{}
    \label{fig:PCNN_SHAP}
  \end{subfigure}

  \caption{SHAP summary plots (seed 42, development-set indentations):
  (a) Random Forest and (b) MLP 64-8-64, both for the predicted
  high-load reference hardness; (c) PCNN-DI, for the correction \(q\)
  (eight network inputs). The SHAP values describe model behavior, not
  physical mechanisms.}
  \label{fig:SHAP}
\end{figure*}

Figure~\ref{fig:SHAP} shows the SHAP summary plots of the seed-42 Random
Forest, MLP 64-8-64, and the PCNN-DI correction \(q\); input rankings
quoted below are mean absolute SHAP values averaged over the five seeds
(Figure~\ref{fig:S_shap_seeds}).
SHAP describes how each trained model uses its inputs, and credit is
shared between correlated inputs; it is not evidence of a physical
mechanism.

For the Random Forest, \(H/E_r\), \(P_{\max}/S^{2}\), and
\(W_p/W_{\mathrm{tot}}\) carry 87\% of the total mean absolute SHAP value
and are the three leading inputs in every seed; XGBoost uses the same
three inputs together with \(H\). The MLP 64-8-64 spreads its
attribution over more inputs, led by \(H/E_r\), \(H\), \(P_{\max}\),
\(P_{\max}/S^{2}\), and \(A_c\), and its ranking varies more between
seeds. Of these inputs, only \(P_{\max}/S^{2}\) is independent of the
area function; \(H/E_r\), \(H\), and \(A_c\) all depend on it. None of
the conventional models therefore avoids area-dependent inputs.

For the PCNN-DI, \(H_{\mathrm{app}}\) carries 57\% of the attribution
of the predicted \(\widehat{H}_{\mathrm{ref}}\), as expected from the
reconstruction \(H_{\mathrm{app}}/\sqrt{1+q}\). The attribution of the
correction \(q\) itself (Figure~\ref{fig:PCNN_SHAP}) is led by the depth
coordinate \(t\) (28\%), the loading exponent \(n\) (18\%), the
compliance descriptor \(\log z_{\mathrm{JO}}\) (16\%), and
\(h_f/h_{\max}\) (12\%); \(t\) and \(n\) are among the three leading
inputs in all five seeds. In the seed-42 model, shallow contacts (high
\(t\)) increase \(q\), that is, the downward correction, as expected
for a size effect; low \(n\) and low \(h_c/h_{\max}\) also increase
\(q\). The dominant role of \(t\) agrees with the ablation, in which
removing \(t\) caused the largest loss
(Section~\ref{boundedpinn-ablation-results}). For PCNN-FI, \(q\) is led
by \(n\) in every seed, followed by absolute quantities such as
\(P_{\max}\), \(E_r\), \(H\), and \(A_c\), consistent with its use of
the applied load noted in
Section~\ref{external-test-on-the-fourth-steel}.

\hypertarget{latent-space-analysis-of-the-neural-network-bottleneck}{%
\subsubsection{Latent Space Analysis}\label{latent-space-analysis-of-the-neural-network-bottleneck}}

\begin{figure*}[htbp]
  \centering
  \begin{subfigure}[b]{\textwidth}
    \centering
    \includegraphics[width=\linewidth]{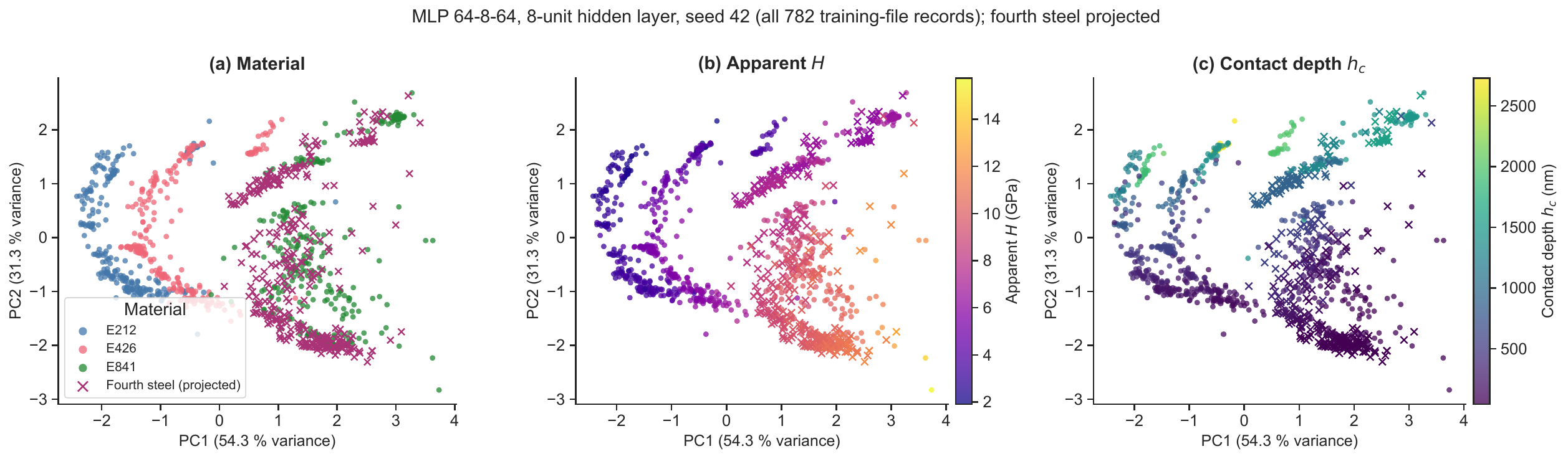}
    \caption{}
    \label{fig:NN_PCA}
  \end{subfigure}

  \begin{subfigure}[b]{\textwidth}
    \centering
    \includegraphics[width=\linewidth]{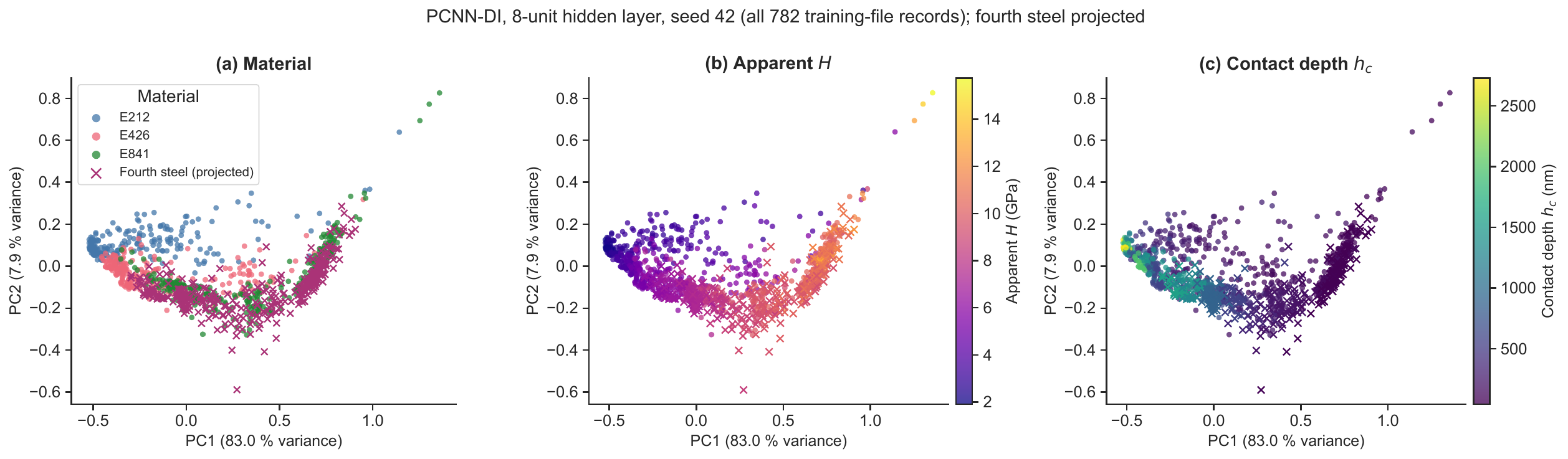}
    \caption{}
    \label{fig:PCNN_PCA}
  \end{subfigure}
  \caption{PCA of the 8-unit hidden-layer activations (seed 42) for (a)
  the MLP 64-8-64 and (b) the PCNN-DI, colored by material, apparent
  hardness, and contact depth; crosses mark the projected fourth-steel
  indentations.}
  \label{fig:PCA}
\end{figure*}

Figure~\ref{fig:PCA} shows the first two principal components of the
8-unit hidden layer of the MLP 64-8-64 and the PCNN-DI (seed 42), with
the fourth steel projected into the same space. In both networks, the
activations are ordered by apparent hardness and contact depth. The MLP
forms separate branches for the three training steels, whereas the
PCNN-DI activations lie on a single curved band on which the steels
overlap. In both, the shallow fourth-steel indentations project onto the
region of E841, consistent with the resemblance noted in
Section~\ref{shallow-only-test-on-the-fourth-steel}. The share of
variance on PC1 varies strongly between seeds (MLP 54--82\%, PCNN-DI
46--83\%), so only this qualitative structure is interpreted; the
projections do not show that the size effect, pile-up, and material
strength are separated inside the networks.

\hypertarget{broader-implications-for-materials-characterization-and-theory}{%
\subsection{Implications and
Limitations}\label{broader-implications-for-materials-characterization-and-theory}}

The results support a narrower claim than accuracy alone. With only
three training steels, the models that fit the development set best,
Random Forest and XGBoost, transferred worst to unseen steels, whereas
low-capacity or structured models transferred best: on the fourth steel
and in fold A, Ridge regression and the two PCNN variants agree within
about one percentage point, a difference smaller than the ambiguity of
the high-load reference itself. In these steels the size-effect
correction appears to be simple enough for a linear model to
approximate it within the training hardness range. The PCNN matches this
linear baseline while giving each indentation a correction of fixed
form, \(\widehat{H}_{\mathrm{ref}}=H_{\mathrm{app}}/\sqrt{1+q}\). Its
clearest advantage appeared for a steel softer than all training steels
(fold B), where it stayed close to the reference at every load, whereas
Ridge regression drifted with load and the other non-linear models
failed; this fold, however, used only two training steels and showed a
large spread over seeds. In the
opposite direction, for a steel harder than all training steels (fold
C), only Ridge regression extrapolated, because the bounded form fixes
the direction of the correction but not its size.

These findings agree with related work when the type of test is taken
into account. Machine-learning corrections of indentation hardness have
reported coefficients of determination of 0.98--1.00 on random or
synthetic splits within the same materials, for example a Random Forest
learning the residuals of a Nix--Gao fit in refractory high-entropy
alloys \cite{kumarMechanisticDatadrivenInterpretation2026} and networks
converting nanoindentation hardness to Vickers hardness in case-hardened
steel \cite{niuLeveragingDeepNeural2024}. Such splits measure
interpolation within the tested materials; in the present data, the same
model family was also the most accurate on the development set but the
least accurate on unseen steels. A subsequent preprint that cites an
earlier version of the present work found a similar pattern with
simulated indentation curves of superconductors: the Nix--Gao length
scale was recovered from single curves under random cross-validation,
a direct single-curve Nix--Gao fit failed, and family classification
degraded unevenly when whole material families were withheld
\cite{yesilkanatPhysicsinformedMachineLearning2026}. For glasses, a network that predicts
the parameters of an empirical size-effect law transferred to new
compositions, whereas a direct data-driven regressor did not
\cite{mannanGlassHardnessPredicting2023}; the present results show the
same pattern for steels, with inputs taken from single instrumented
indentation curves instead of composition and load. Finally, the
competitive Ridge baseline follows the recommendation to test whether a
simple model is sufficient before applying machine learning
\cite{jinRecentAdvancesApplications2023}: for these steels it is
sufficient within the training hardness range, and the bounded network
adds a correction of fixed form and the behavior seen in fold B.

In practice, the learned corrections estimate the high-load reference
hardness of a steel within the training hardness range from shallow
indentations alone (Section~\ref{shallow-only-test-on-the-fourth-steel}),
where a Nix--Gao fit to the same indentations, a within-material
extrapolation and not a like-for-like comparison, overestimates it by
3.1 GPa. Each prediction uses the load--displacement curve of a single
indentation, and no input depends on other indentations of the same
material (Section~\ref{data-processing-and-feature-engineering}). For the
smooth models, a single shallow indentation gives the reference within a
per-indentation RMSE of 0.31--0.46 GPa (5--8\%; Table~\ref{tab:shallow_only});
averaging over several indentations reduces this scatter. This requires training steels with known high-load reference
hardness; it removes the need for a load series on the material being
tested, not the need for reference data. The bounded correction form is
a weak assumption, that a depth-affected apparent hardness is mapped to
a high-load reference by a finite multiplicative correction; whether it also applies to
materials in which size effects
arise from mechanisms beyond a single geometrically necessary
dislocation population, such as irradiated or nanocrystalline
microstructures \cite{hardieUnderstandingEffectsIon2015,
yangDependenceNanohardnessIndentation2007,
trelewiczHallPetchBreakdown2007}, remains to be tested. The trained models, however, do not
transfer to such materials: the fused silica test shows that they fail
outside their training domain without any warning, and every new
material class, deformation mechanism, or hardness range requires new
reference data and validation. The workflow itself, however, is not
specific to steels: it requires only standard instrumented indentations
on a small number of reference materials with known high-load hardness,
uses inputs from single load--displacement curves, and involves no
material-specific tuning. Retraining it for another material class
would require reference materials of that class and validation on
unseen materials of that class; how many
reference materials are needed for reliable transfer remains to be
established.

The study has several limitations. (i) Only three independent training
materials and one external steel were available, complemented by
leave-one-steel-out folds with two training steels each; differences
between the best models are therefore not precise effect sizes. (ii)
The target is a finite-load reference hardness: the mean hardness still
decreases by 7--12\% over the last load interval, and, being defined at
a fixed load rather than a fixed depth, it corresponds to contact depths
of 1.5--2.7~\(\mu\)m across the steels; the load- and depth-dependent
inputs allow the models to learn partly that the correction vanishes at
the reference load. The fourth-steel reference at 375 mN lies about
1.7\% below the value interpolated at 350 mN (6.18 GPa), and for this
steel the Nix--Gao \(H_{0}\) lies 10\% below the reference, so rankings
by this reference carry that ambiguity. For a material that cannot be
indented at the reference load, such as a thin layer, the prediction is
the high-load reference that a bulk specimen with the same indentation
response would give. (iii) The domain of
validity is limited to steels within the training hardness range and to
loads within the training schedule; predictions at 3 mN are
extrapolations, upward extrapolation in hardness failed (fold C), and
other material classes are excluded. (iv) The predictions are anchored
to the Oliver--Pharr apparent hardness, so any pile-up or area-function
bias remains embedded in them; pile-up was not quantified. (v) The
ablation does not support the auxiliary \(q\)-target and
load-independence losses on the transfer test, and augmentation acted
at most as a mild regularizer. (vi) The target is a specimen-level
value, whereas shallow indentations in the multiphase steels, in
particular the nominally ferritic/pearlitic E212, can sample individual
phases. Part of the shallow-depth scatter therefore reflects real local
hardness differences rather than the ISE, and training each indentation
toward the specimen-level reference partly removes this variation. The
PCNNs, which reconstruct each prediction from the measured hardness of
the indentation, and Ridge regression retain most of this
indentation-to-indentation variation (within-load scatter ratio
0.83--1.00), whereas the tree ensembles reduce it to about half
(Figures~\ref{fig:error_e426} and \ref{fig:S_fourth_other}). The
variation was not separated into phase contrast and measurement
scatter. Applied to an individual phase, the prediction estimates the
high-load hardness of a specimen with the same indentation response,
not a phase-specific property.

Several directions follow. An out-of-domain indicator, for example a
distance measure in the descriptor space, would turn the silent failure
seen on fused silica into a flagged rejection, which is a prerequisite
for routine use. More training steels, in particular harder ones, are
needed to extend the hardness range and to test whether the model
ranking holds with more independent materials. Other indenter
geometries would test whether the dimensionless descriptor frame
transfers. Retraining and validating the workflow for other material
classes, and for thin layers, graded layers and individual phases, where
substrate effects, property gradients and other size-effect mechanisms
are present, is needed before it is applied there. Finally, because the PCNN correction \(q\) is an explicit
function of a few dimensionless descriptors, fitting a closed-form
expression to it could be compared with strain-gradient models of the
shallow-depth size effect.

\hypertarget{conclusion}{%
\section{\texorpdfstring{Conclusion
}{Conclusion}}\label{conclusion}}

This study tested whether the high-load reference hardness of a steel
(350 or 375 mN, a finite-load value) can be estimated from single
instrumented indentations, including shallow ones, of a steel that was
not used for training. Seven models were trained on only three
independent steels: their 703 indentations describe the scatter and
load dependence within each steel, but provide only three
material-level reference values. The models were evaluated over five
seeds on an external steel not used for training and indented at
different loads, in leave-one-steel-out
folds, and on fused silica.

The deep-only Nix--Gao fit reproduced the reference of the training
steels (MAPE 0.9\%), but it requires deep indentations, including the
reference indentations themselves; a fit through all depths gave a MAPE
of 8.9\%, and a fit to the 3--15 mN indentations of the external steel
overestimated its reference by 3.1 GPa.

The development-set ranking did not carry over to unseen steels. Random
Forest and XGBoost were the most accurate on the development set but the
least accurate on the external steel (MAPE 10.2 and 8.5\%), whereas Ridge
regression, PCNN-FI, the MLP 64-8-64, and PCNN-DI reached 3.7--4.8\%,
differences smaller than the ambiguity of the reference itself. From the
shallow indentations alone, the smooth models estimated the reference
within 0.03--0.19 GPa on average, with a single-indentation RMSE of
0.31--0.46 GPa; the 3 mN indentations were an extrapolation below the
training loads.

The physics-constrained neural network, which reconstructs the
reference as \(H_{\mathrm{app}}/\sqrt{1+q}\) with a bounded correction
\(q\), matched the linear baseline on steels within the training
hardness range and remained accurate for a steel softer than all
training steels, where the other models failed or drifted with load. It did not extrapolate to a harder steel. The
ablation supported the bounded reconstruction, the depth coordinate, and
the area-free descriptors, but not the auxiliary \(q\)-target and
load-independence losses.

These conclusions rest on three training materials and one external
steel, and they hold only for steels within the training hardness range
and load schedule; the fused silica test showed that the models fail
outside this domain without warning. The workflow uses no steel-specific inputs, but its retraining for
other material classes, and its use for thin layers and individual
phases, remain to be tested.
More training steels, in particular harder ones, and an out-of-domain
indicator are needed before routine use.

\section*{Data and Code Availability}
The data and code used in this study are available at
\url{https://doi.org/10.5281/zenodo.23472790}.

\hypertarget{acknowledgements}{%
\section{Acknowledgements}\label{acknowledgements}}

The research was supported by the Russian Science Foundation grant No.
25-79-00327 (\url{https://rscf.ru/project/25-79-00327/})

\bibliographystyle{unsrtnat}
\bibliography{ML_NI}

\clearpage

\ifdefined\suppstandalone\else
\section*{Supplementary Material}
\addcontentsline{toc}{section}{Supplementary Material}
\fi

\setcounter{figure}{0}
\setcounter{table}{0}
\renewcommand{\thefigure}{S\arabic{figure}}
\renewcommand{\thetable}{S\arabic{table}}
\renewcommand{\theHfigure}{S\arabic{figure}}
\renewcommand{\theHtable}{S\arabic{table}}

\begin{figure}[htbp]
  \centering
  \includegraphics[width=\textwidth]{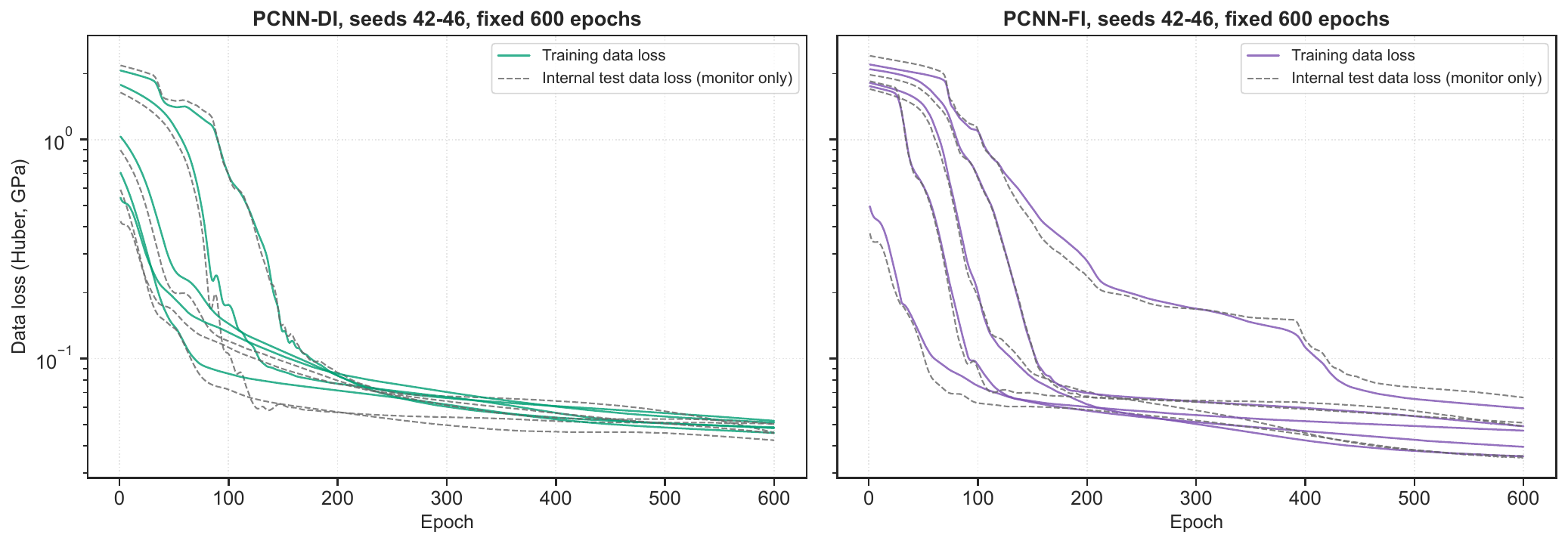}
  \caption{Training of the PCNN-DI (left) and PCNN-FI (right) for the
  five run seeds: data loss on the training rows (solid) and on the
  development set (dashed) per epoch. The development loss was only
  recorded; training ran for a fixed 600 epochs and the last epoch was
  used.}
  \label{fig:S_history}
\end{figure}

\begin{figure}[htbp]
  \centering
  \includegraphics[width=\textwidth]{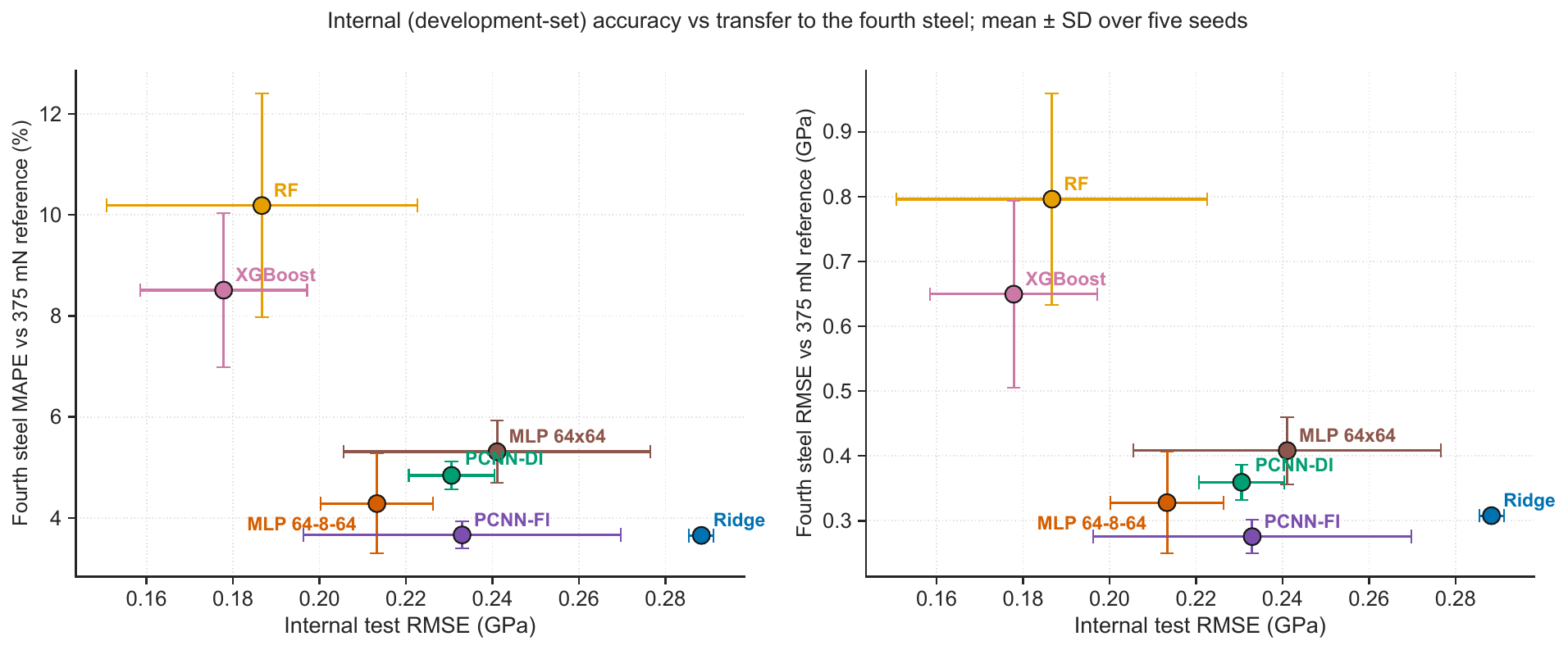}
  \caption{Development-set RMSE versus the fourth-steel MAPE (left) and
  RMSE (right) against the 375 mN reference, one point per model; mean
  $\pm$ SD over five run seeds. The models with the lowest
  development-set error give the largest errors on the fourth steel.}
  \label{fig:S_reversal}
\end{figure}

\begin{figure}[htbp]
  \centering
  \includegraphics[width=\textwidth]{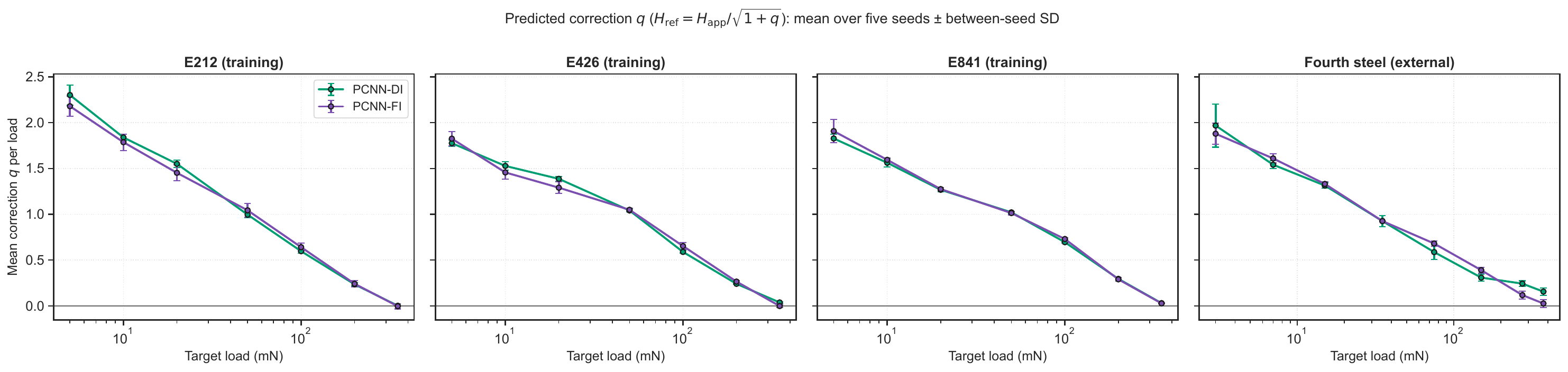}
  \caption{Per-load mean correction $q$
  ($\widehat{H}_{\mathrm{ref}} = H_{\mathrm{app}}/\sqrt{1+q}$) of the
  PCNN-DI and PCNN-FI on the three training steels and the fourth
  steel; mean $\pm$ between-seed SD over five run seeds.}
  \label{fig:S_q}
\end{figure}

\begin{figure}[htbp]
  \centering
  \begin{subfigure}[b]{0.32\textwidth}
    \centering
    \includegraphics[width=\linewidth]{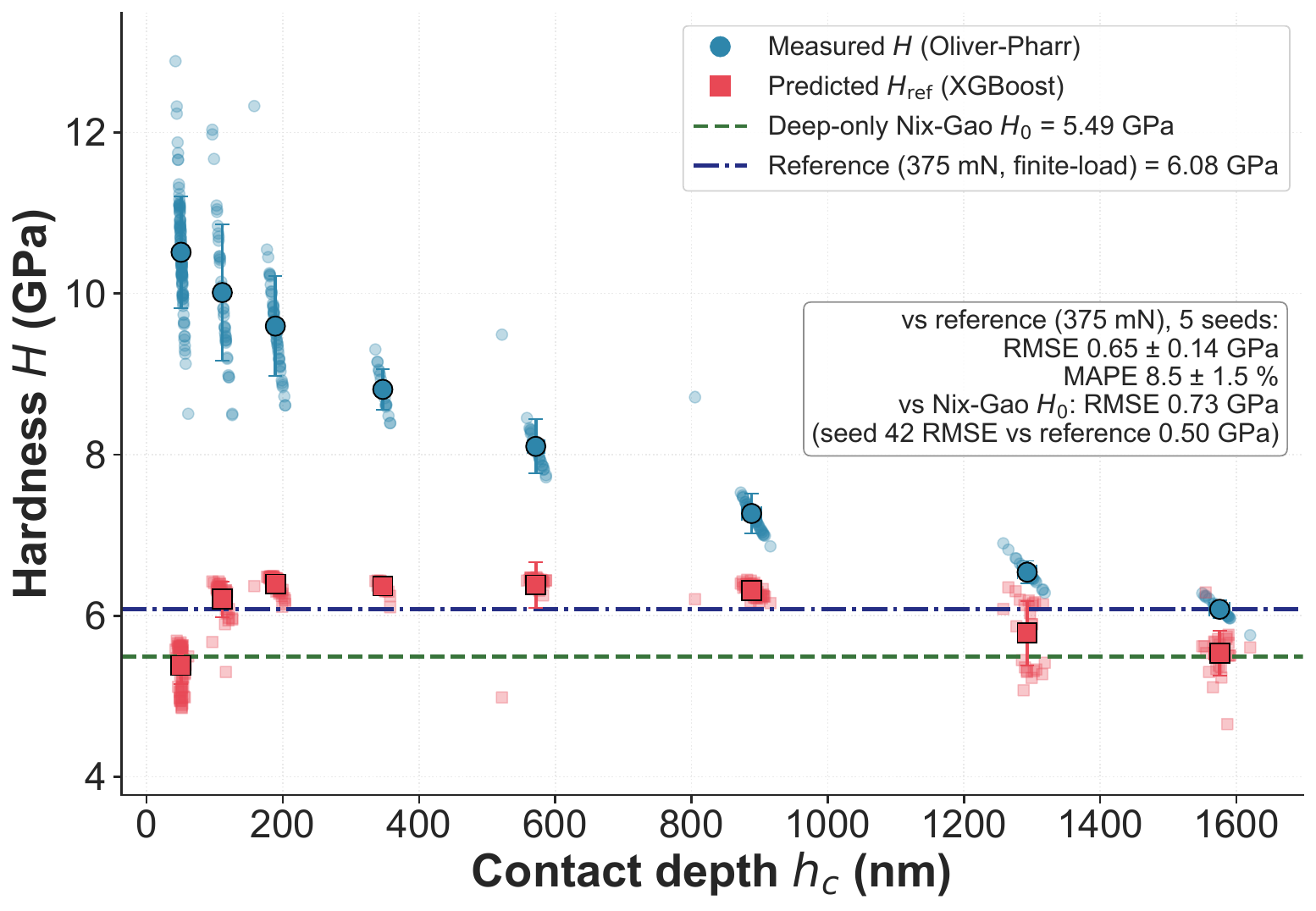}
    \caption{XGBoost}
  \end{subfigure}\hfill
  \begin{subfigure}[b]{0.32\textwidth}
    \centering
    \includegraphics[width=\linewidth]{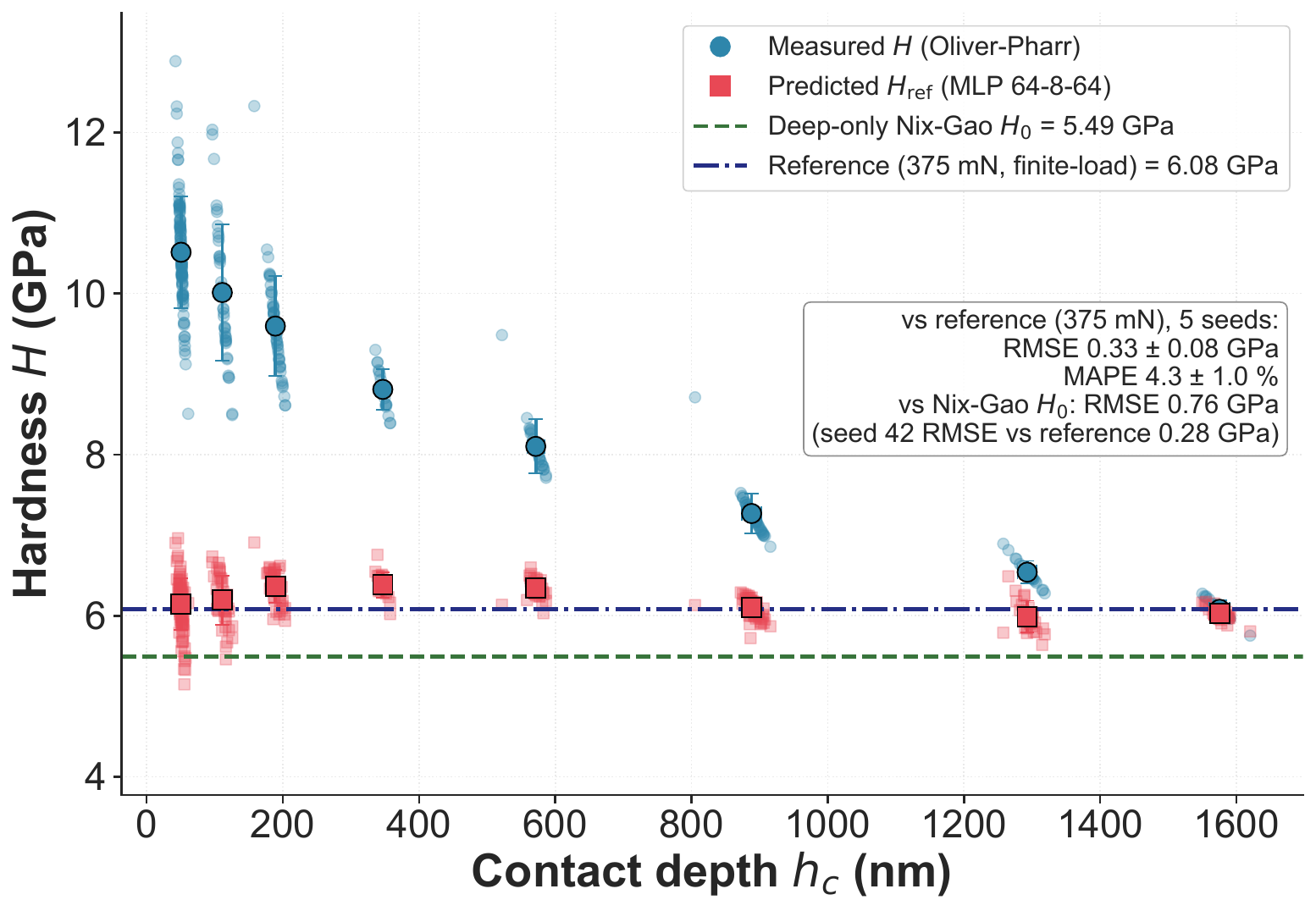}
    \caption{MLP 64-8-64}
  \end{subfigure}\hfill
  \begin{subfigure}[b]{0.32\textwidth}
    \centering
    \includegraphics[width=\linewidth]{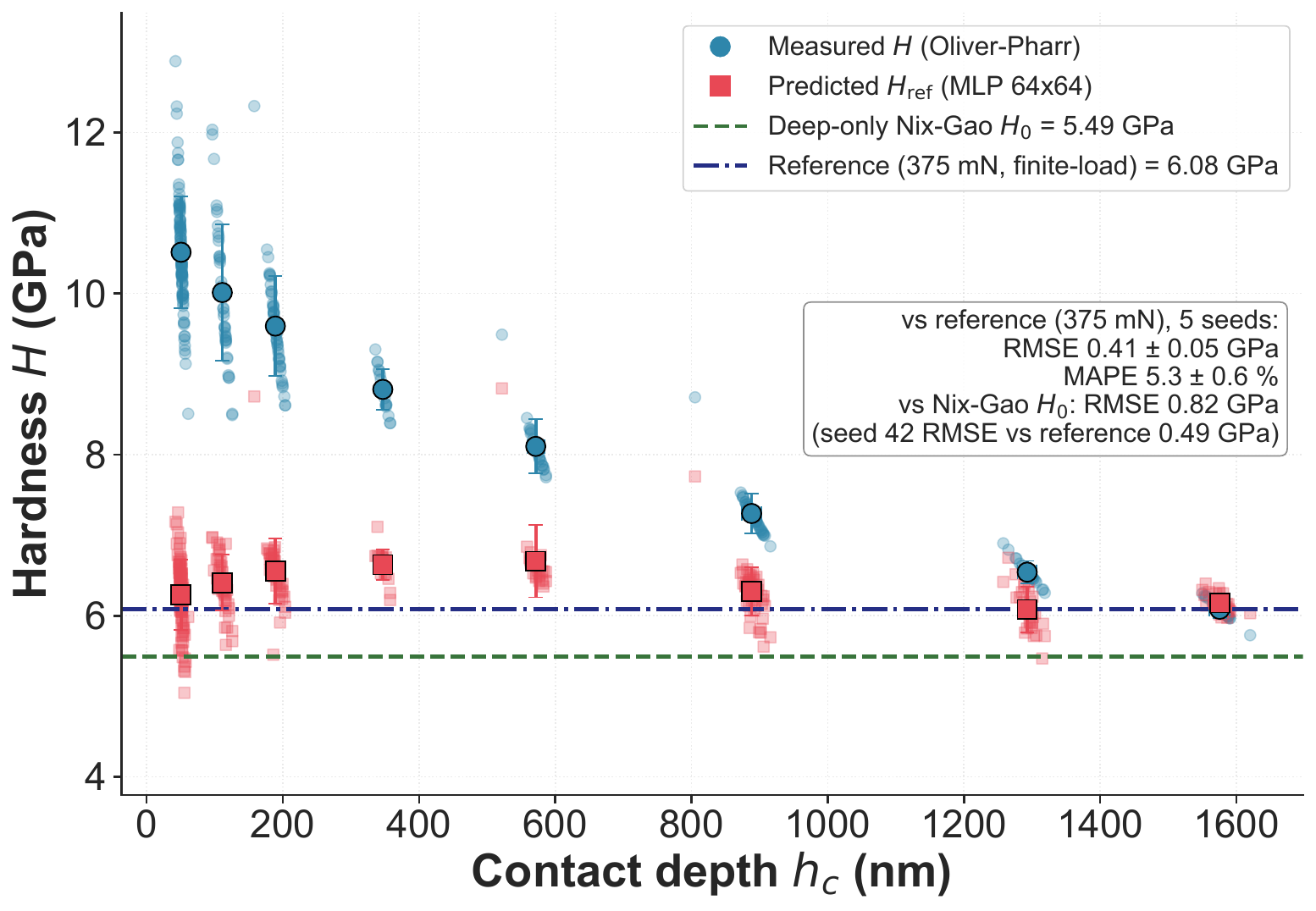}
    \caption{MLP 64$\times$64}
  \end{subfigure}
  \caption{Fourth steel: measured hardness and predicted high-load
  reference hardness versus contact depth for the three models not
  shown in the main text (points: seed~42; metrics in the boxes: five
  seeds). Dash-dotted line: 375 mN reference (6.08 GPa); dashed line:
  deep-only Nix--Gao $H_0$ (5.49 GPa).}
  \label{fig:S_fourth_other}
\end{figure}

\begin{figure}[htbp]
  \centering
  \includegraphics[width=0.7\textwidth]{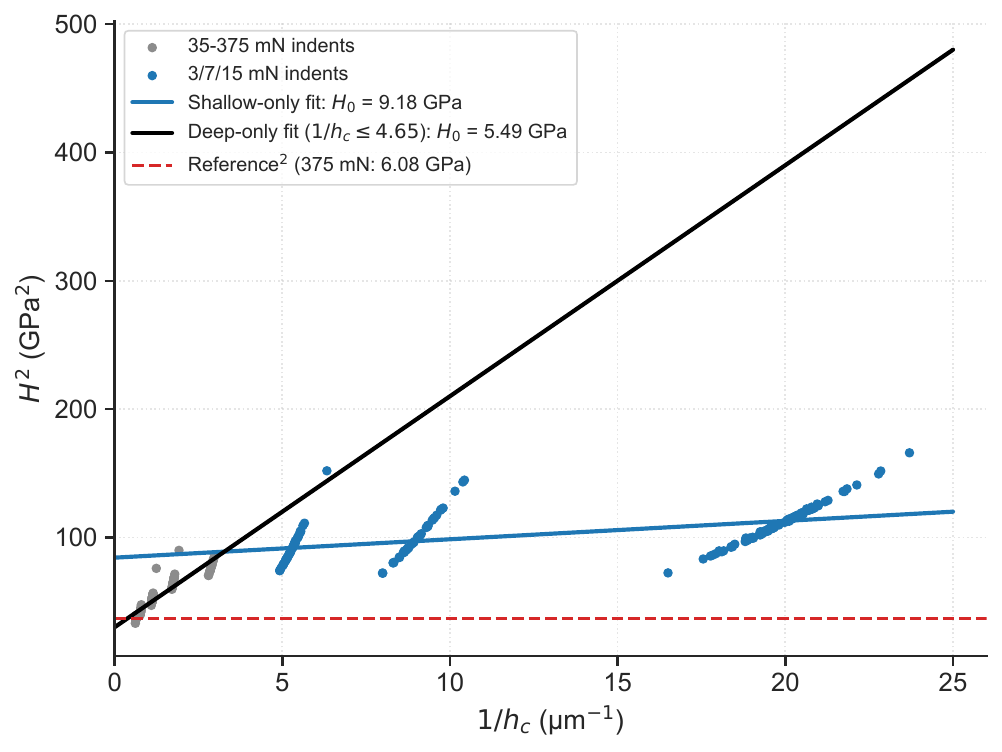}
  \caption{Fourth steel: Nix--Gao representation $H^2$ versus $1/h_c$.
  The fit to the 3, 7, and 15 mN indentations alone (shallow-only test)
  gives $H_0 = 9.18$ GPa; the deep-only fit ($1/h_c \le 4.65\
  \mu$m$^{-1}$, including the reference indentations) gives
  $H_0 = 5.49$ GPa. Dashed line: square of the 375 mN reference
  (6.08 GPa).}
  \label{fig:S_nixgao_shallow}
\end{figure}

\begin{figure}[htbp]
  \centering
  \begin{subfigure}[b]{0.8\textwidth}
    \centering
    \includegraphics[width=\linewidth]{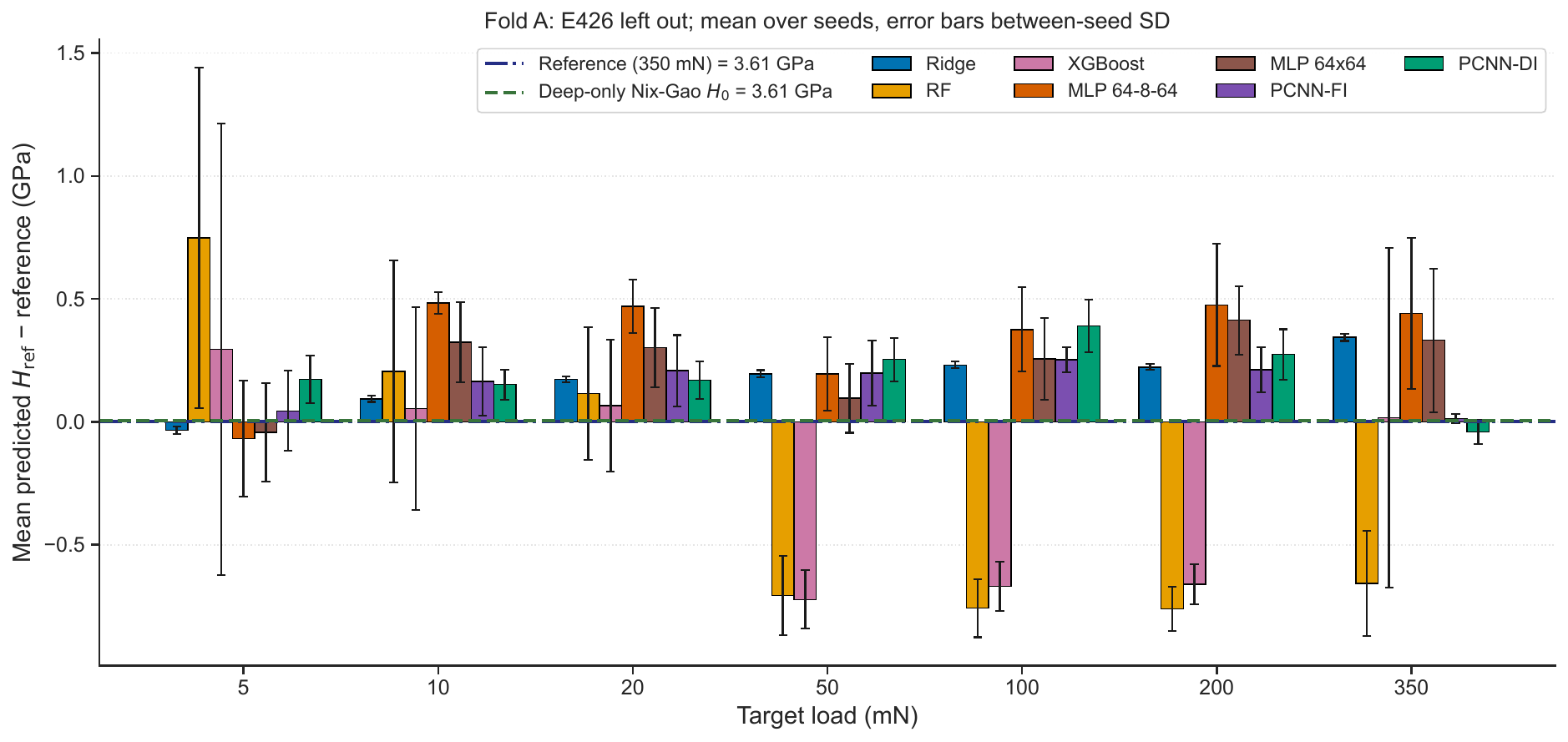}
    \caption{Fold A, E426 left out}
  \end{subfigure}

  \begin{subfigure}[b]{0.8\textwidth}
    \centering
    \includegraphics[width=\linewidth]{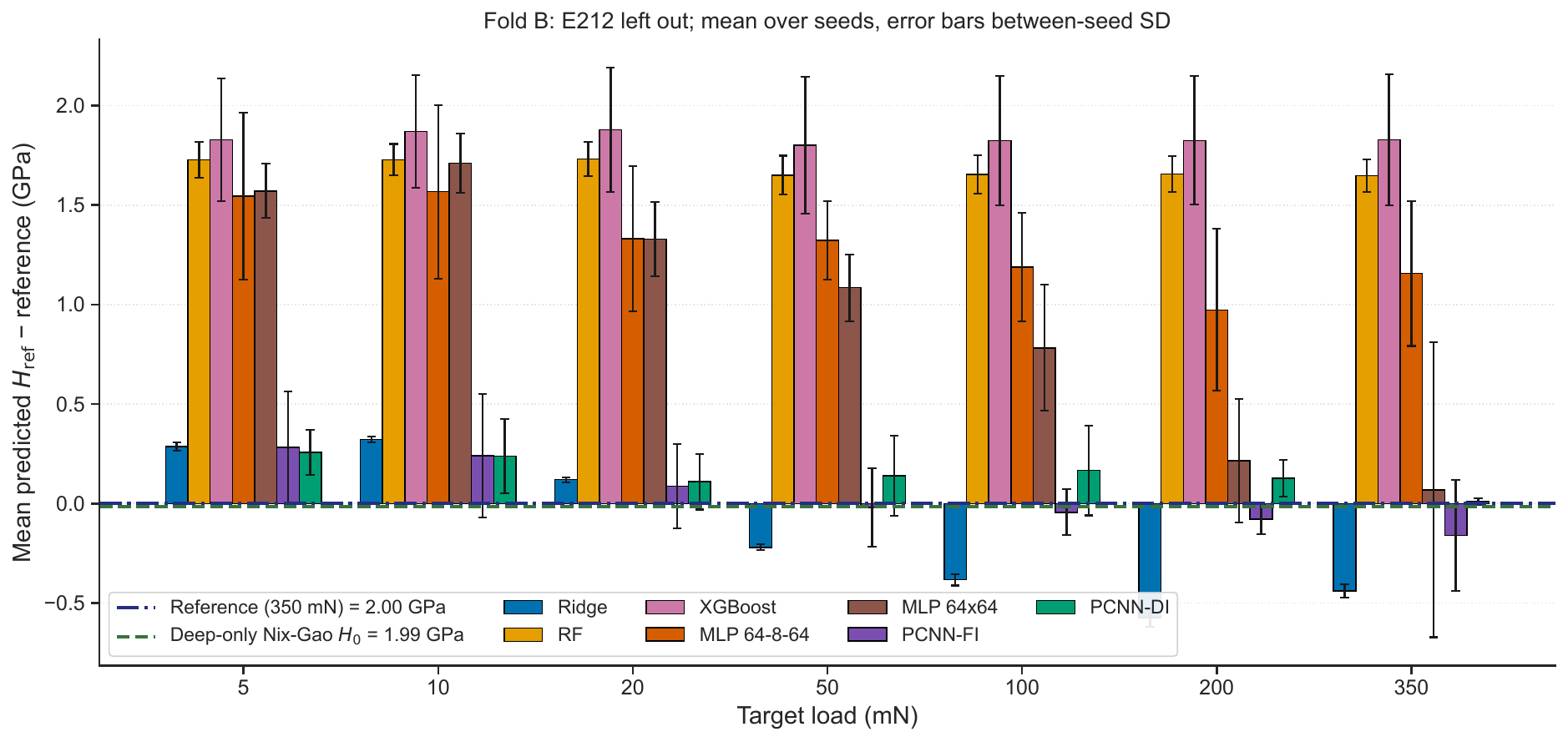}
    \caption{Fold B, E212 left out}
  \end{subfigure}

  \begin{subfigure}[b]{0.8\textwidth}
    \centering
    \includegraphics[width=\linewidth]{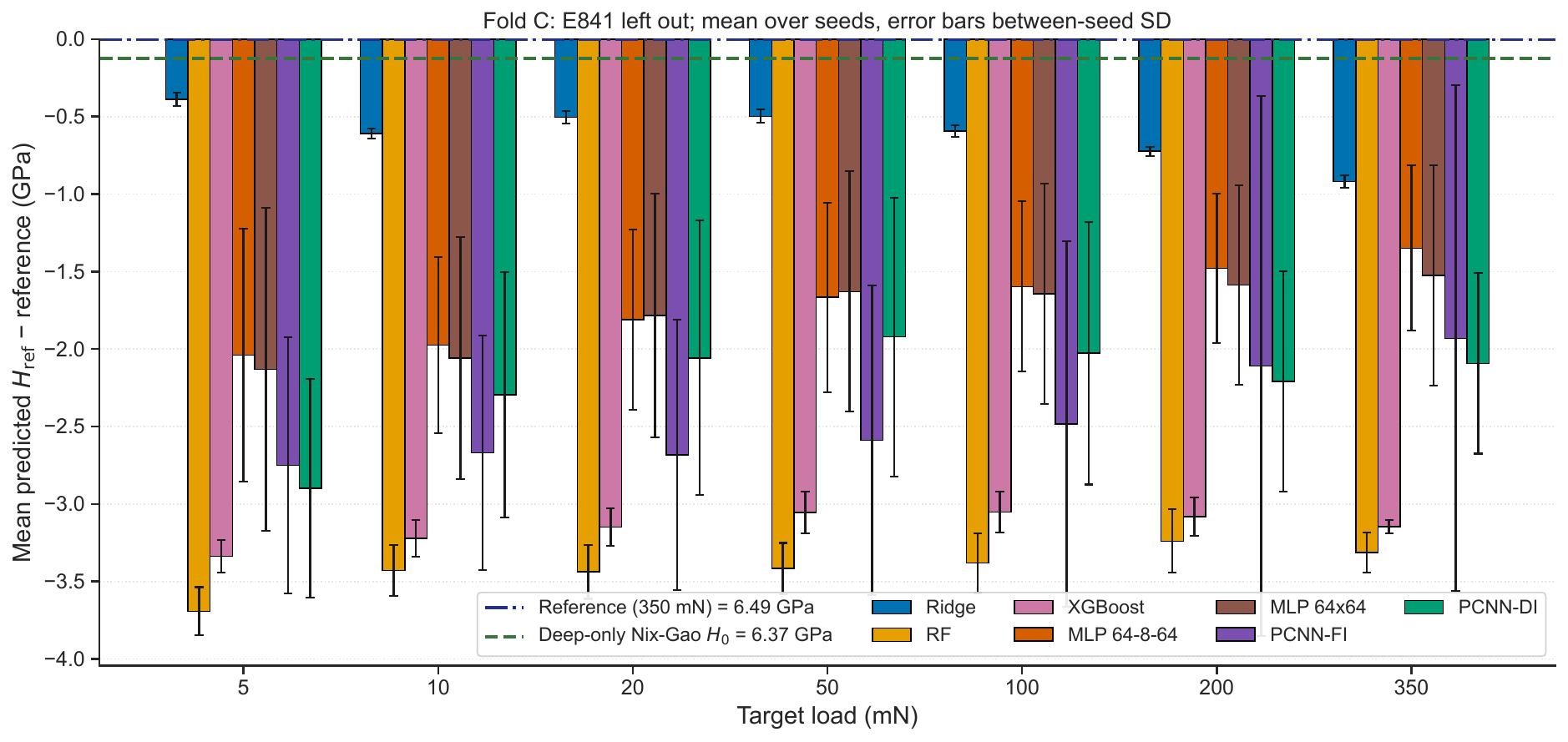}
    \caption{Fold C, E841 left out}
  \end{subfigure}
  \caption{Leave-one-steel-out check: per-load mean prediction minus
  the 350 mN reference of the left-out steel, all models; bars show the
  mean over five run seeds, error bars the between-seed SD.}
  \label{fig:S_loso_perload}
\end{figure}

\begin{figure}[htbp]
  \centering
  \includegraphics[width=\textwidth]{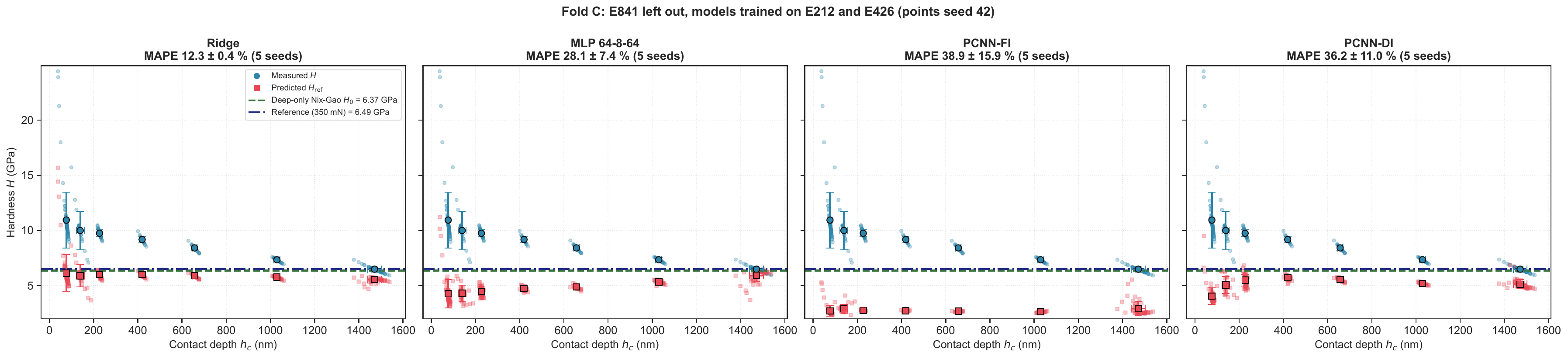}
  \caption{Leave-one-steel-out fold C (E841 left out, models trained on
  E212 and E426): measured hardness and predicted high-load reference
  hardness versus contact depth (points: seed~42; MAPE: five seeds).}
  \label{fig:S_foldC}
\end{figure}

\begin{figure}[htbp]
  \centering
  \begin{subfigure}[b]{0.32\textwidth}
    \includegraphics[width=\linewidth]{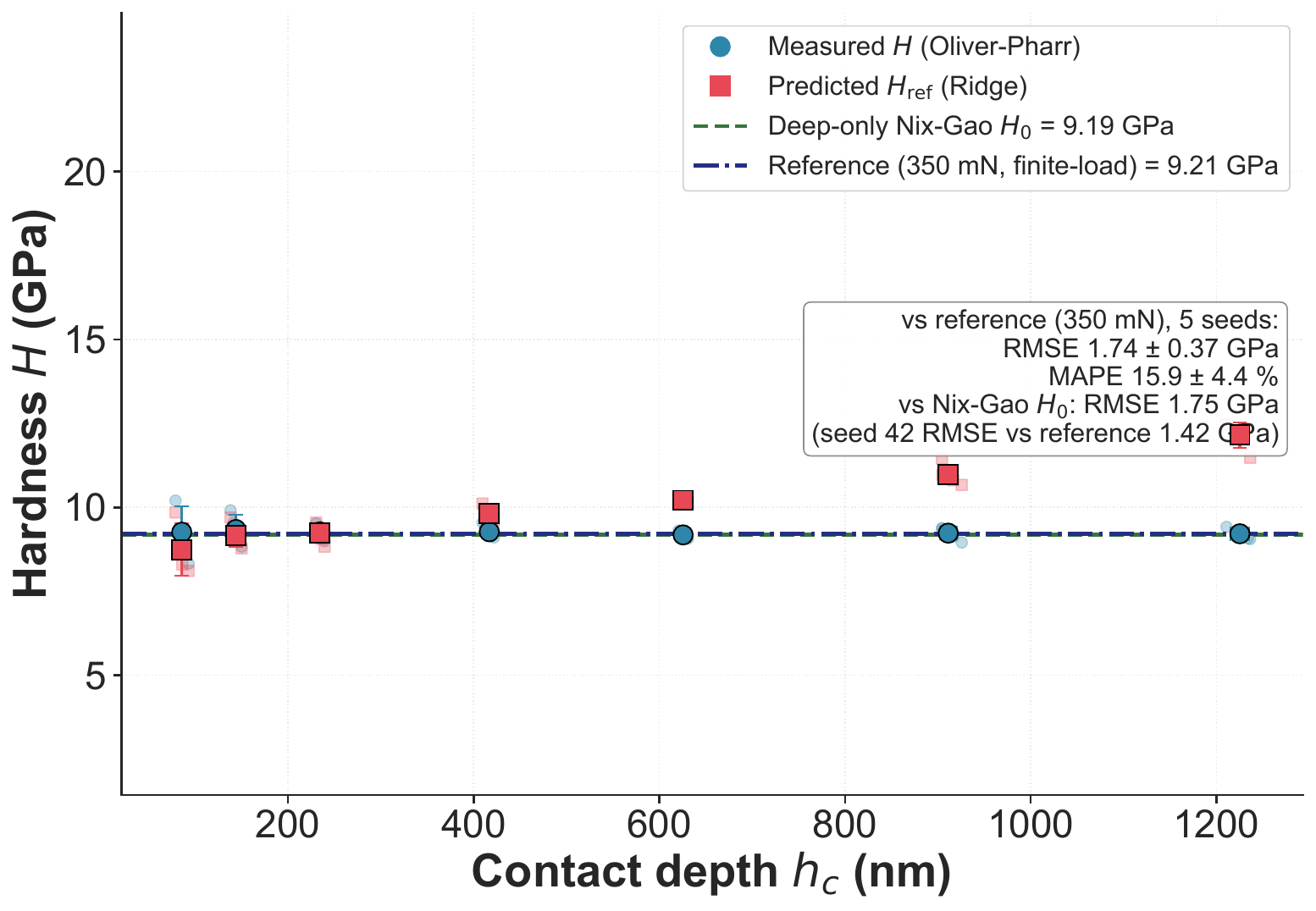}
    \caption{Ridge}
  \end{subfigure}\hfill
  \begin{subfigure}[b]{0.32\textwidth}
    \includegraphics[width=\linewidth]{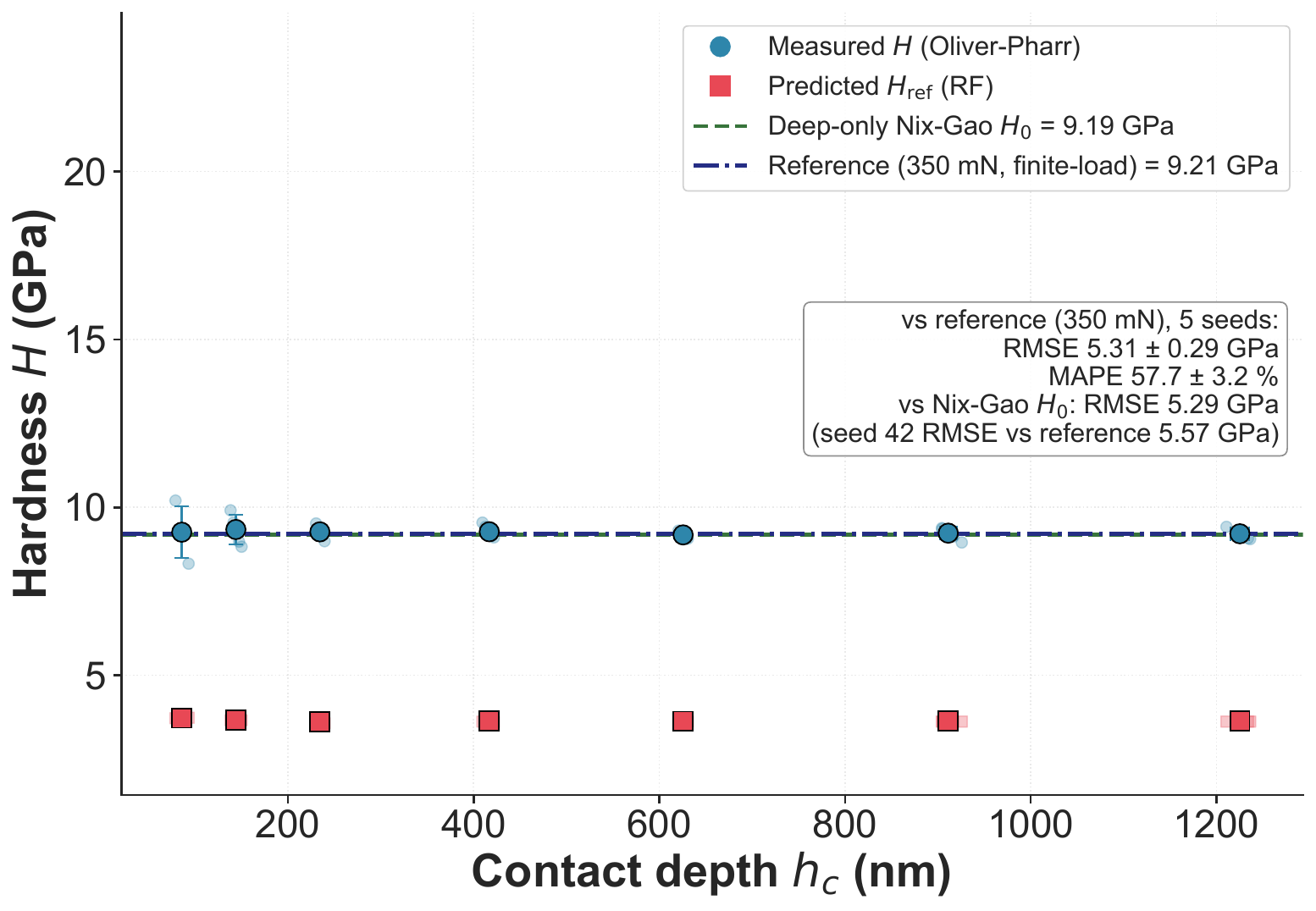}
    \caption{RF}
  \end{subfigure}\hfill
  \begin{subfigure}[b]{0.32\textwidth}
    \includegraphics[width=\linewidth]{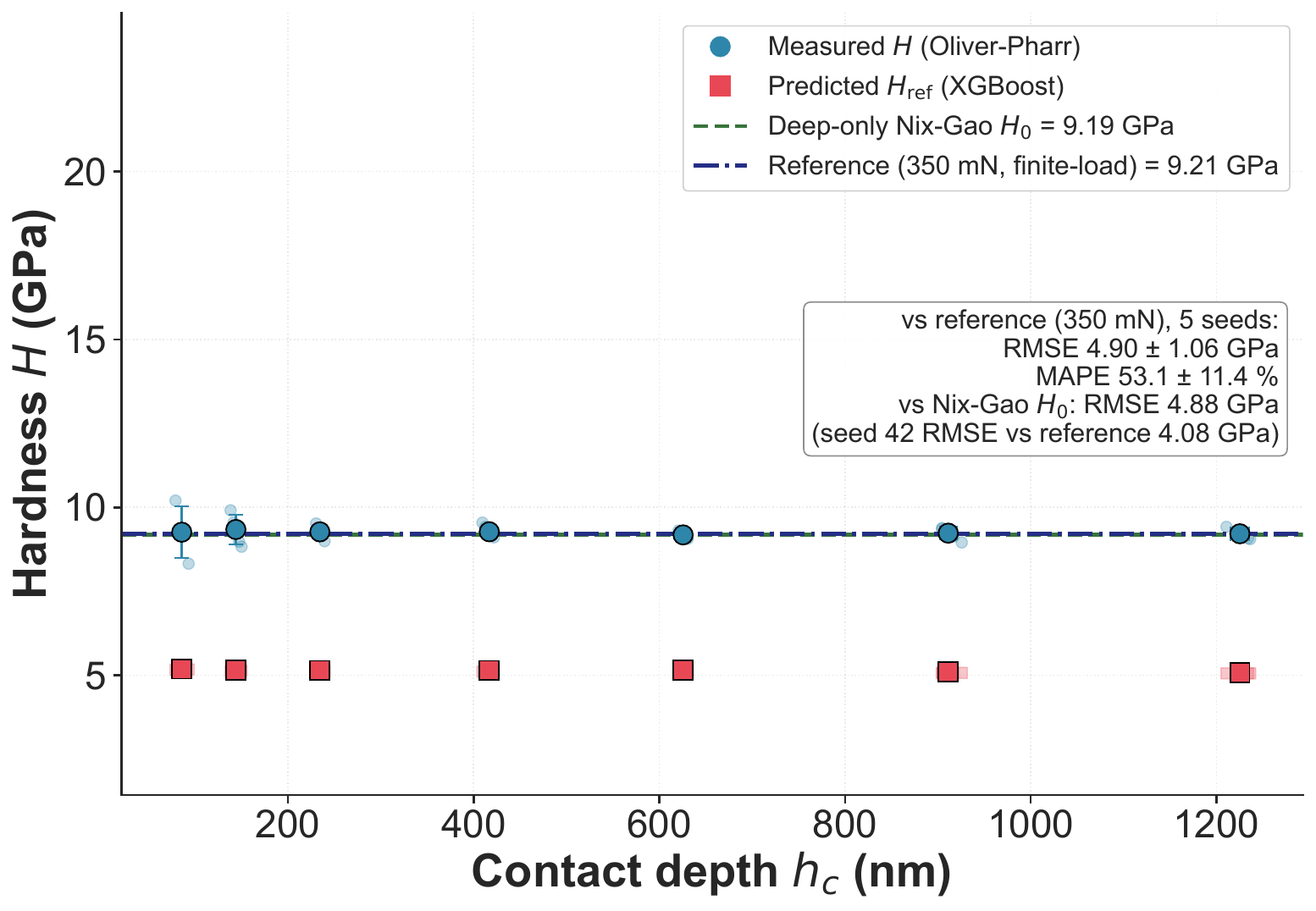}
    \caption{XGBoost}
  \end{subfigure}

  \begin{subfigure}[b]{0.32\textwidth}
    \includegraphics[width=\linewidth]{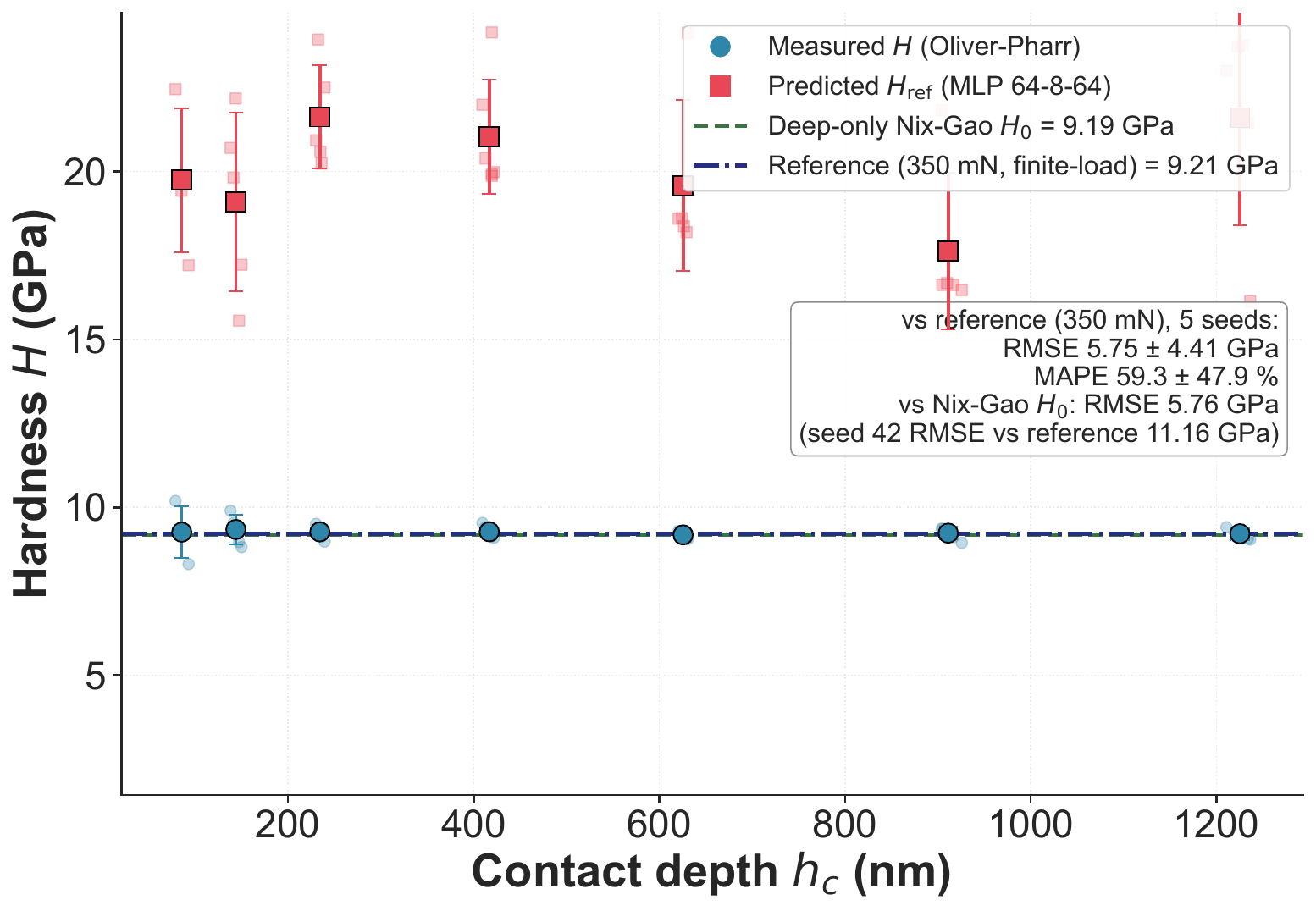}
    \caption{MLP 64-8-64}
  \end{subfigure}\hfill
  \begin{subfigure}[b]{0.32\textwidth}
    \includegraphics[width=\linewidth]{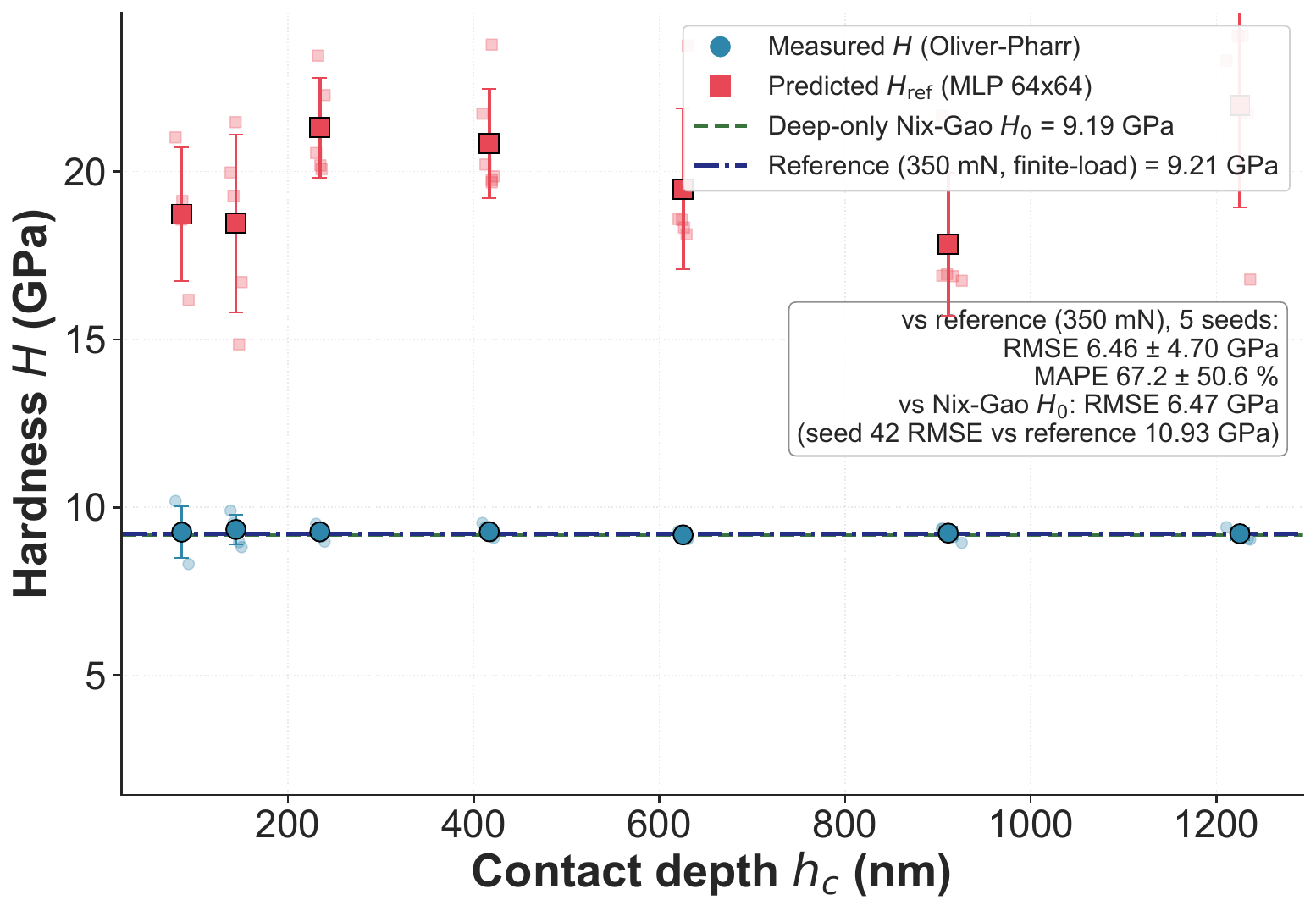}
    \caption{MLP 64$\times$64}
  \end{subfigure}\hfill
  \begin{subfigure}[b]{0.32\textwidth}
    \includegraphics[width=\linewidth]{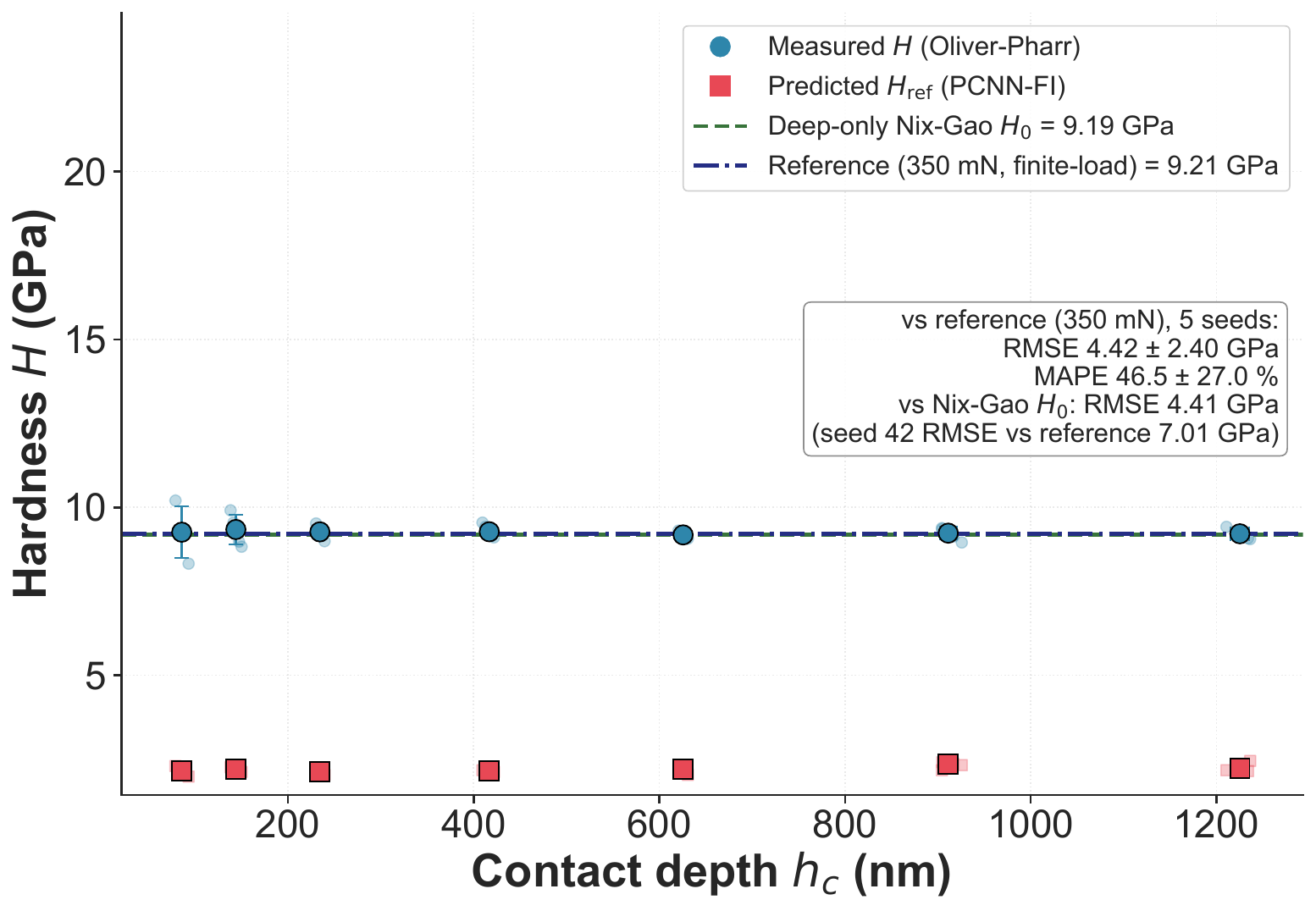}
    \caption{PCNN-FI}
  \end{subfigure}

  \begin{subfigure}[b]{0.32\textwidth}
    \includegraphics[width=\linewidth]{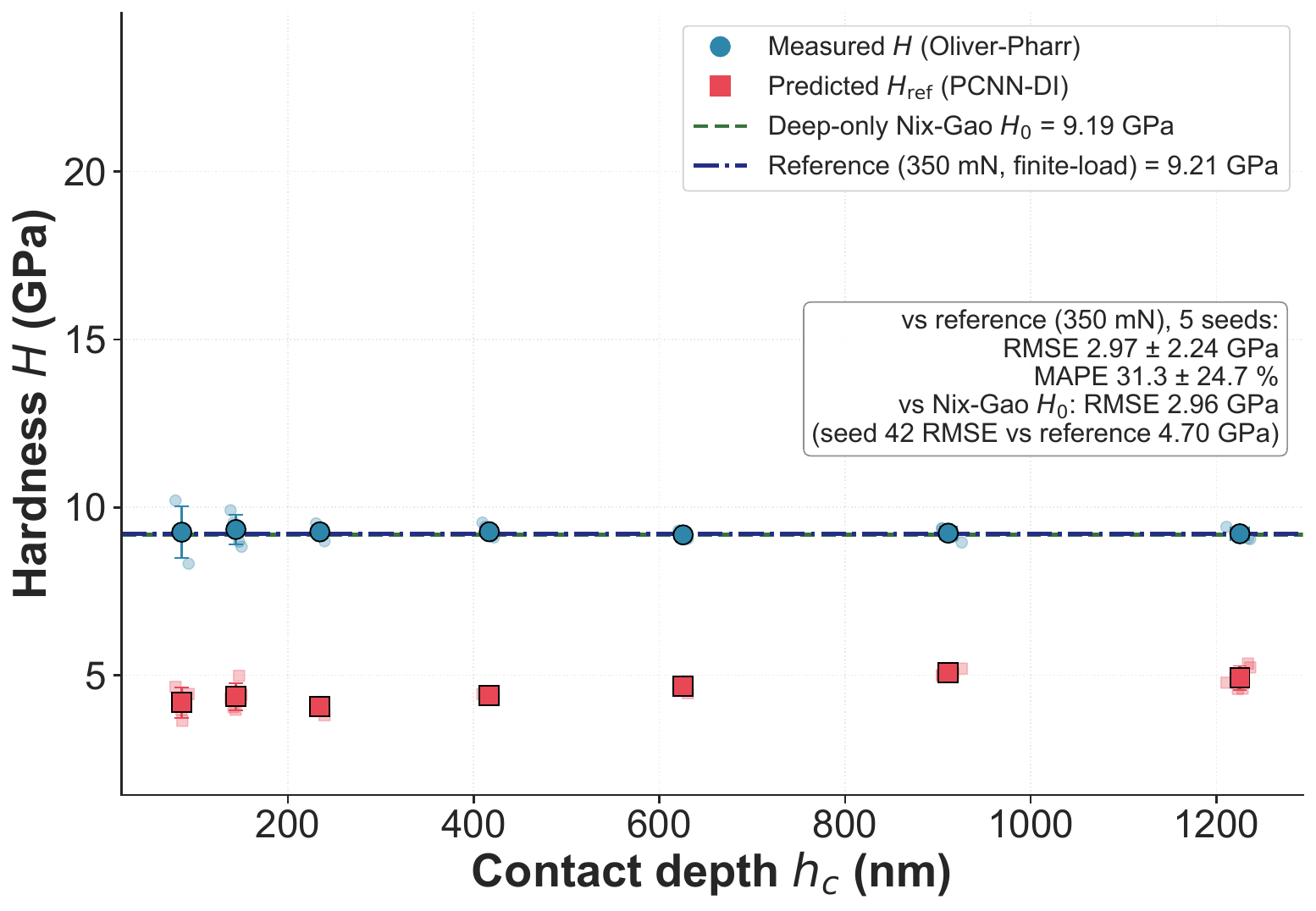}
    \caption{PCNN-DI}
  \end{subfigure}\hfill
  \begin{subfigure}[b]{0.66\textwidth}
    \includegraphics[width=\linewidth]{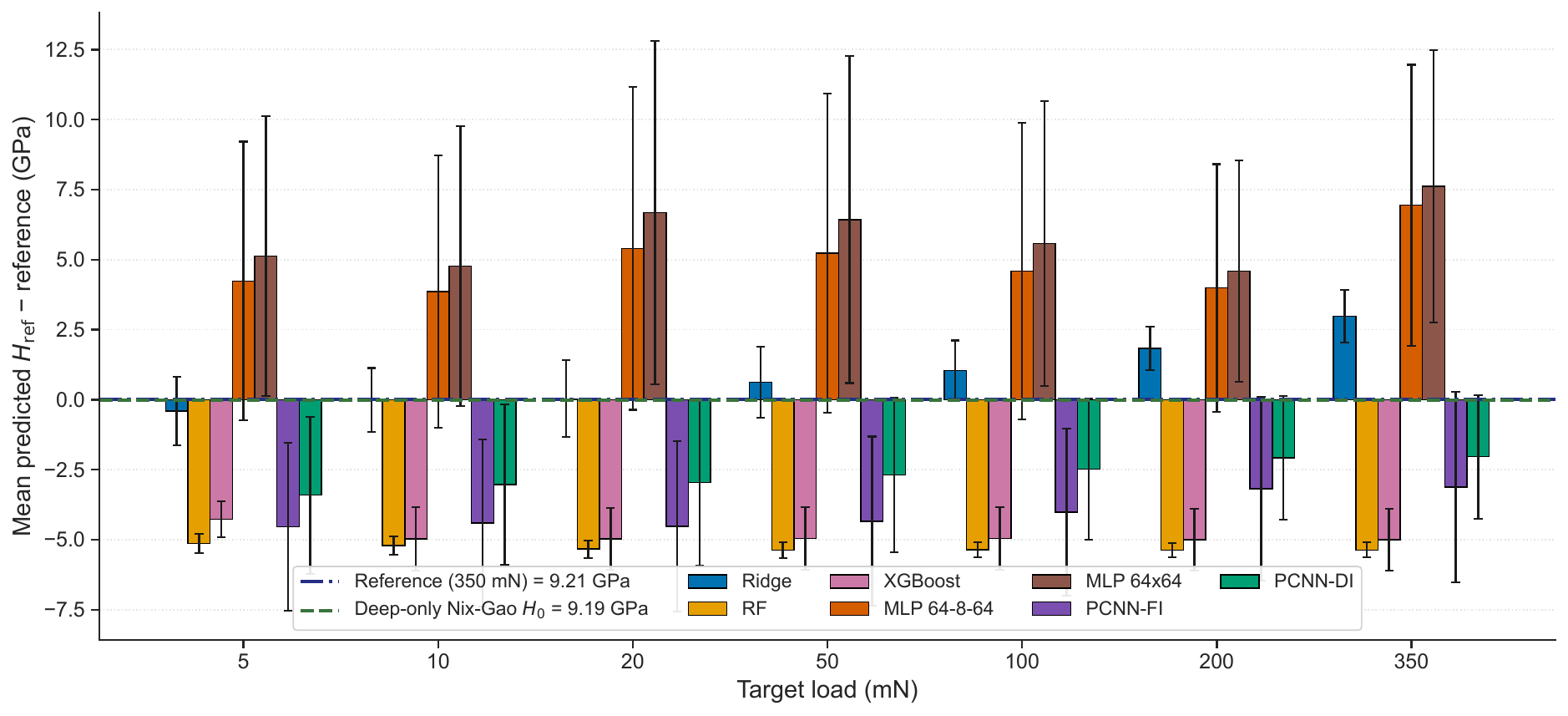}
    \caption{Per-load error, all models}
  \end{subfigure}
  \caption{Fused silica (out-of-domain test): (a--g) measured hardness
  and predicted high-load reference hardness versus contact depth
  (points: seed~42; metrics: five seeds); (h) per-load mean prediction
  minus the 350 mN reference (9.21 GPa), mean $\pm$ between-seed SD.}
  \label{fig:S_glass}
\end{figure}

\begin{figure}[htbp]
  \centering
  \includegraphics[width=\textwidth]{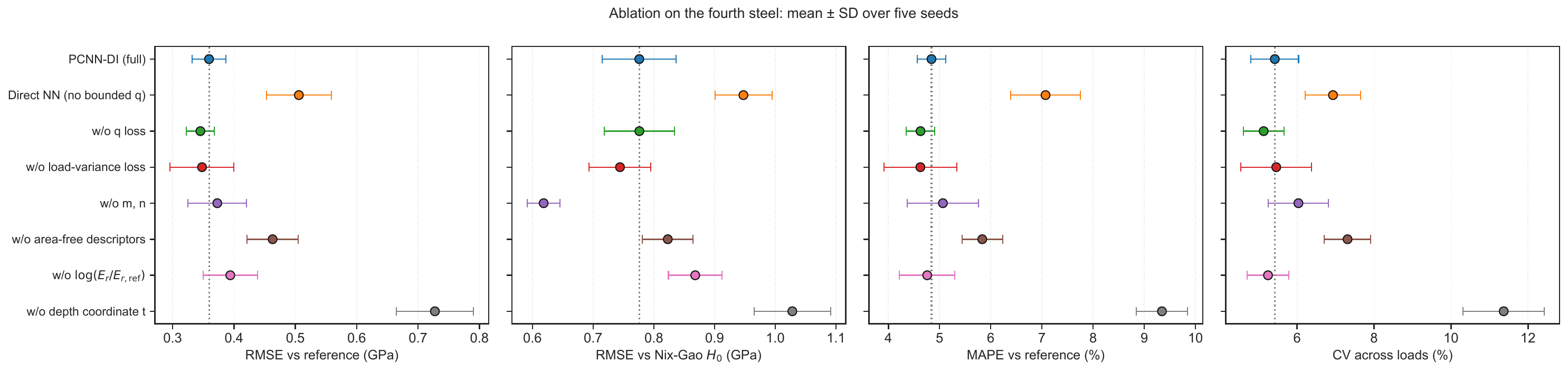}
  \caption{Ablation of the PCNN-DI on the fourth steel (ablation table of the
  main text): RMSE against the 375 mN reference, RMSE against the
  deep-only Nix--Gao $H_0$, MAPE, and coefficient of variation of the
  predictions over all indentations (CV), mean $\pm$ SD over five run
  seeds. Dotted lines mark the full PCNN-DI.}
  \label{fig:S_ablation}
\end{figure}

\begin{figure}[htbp]
  \centering
  \includegraphics[width=\textwidth]{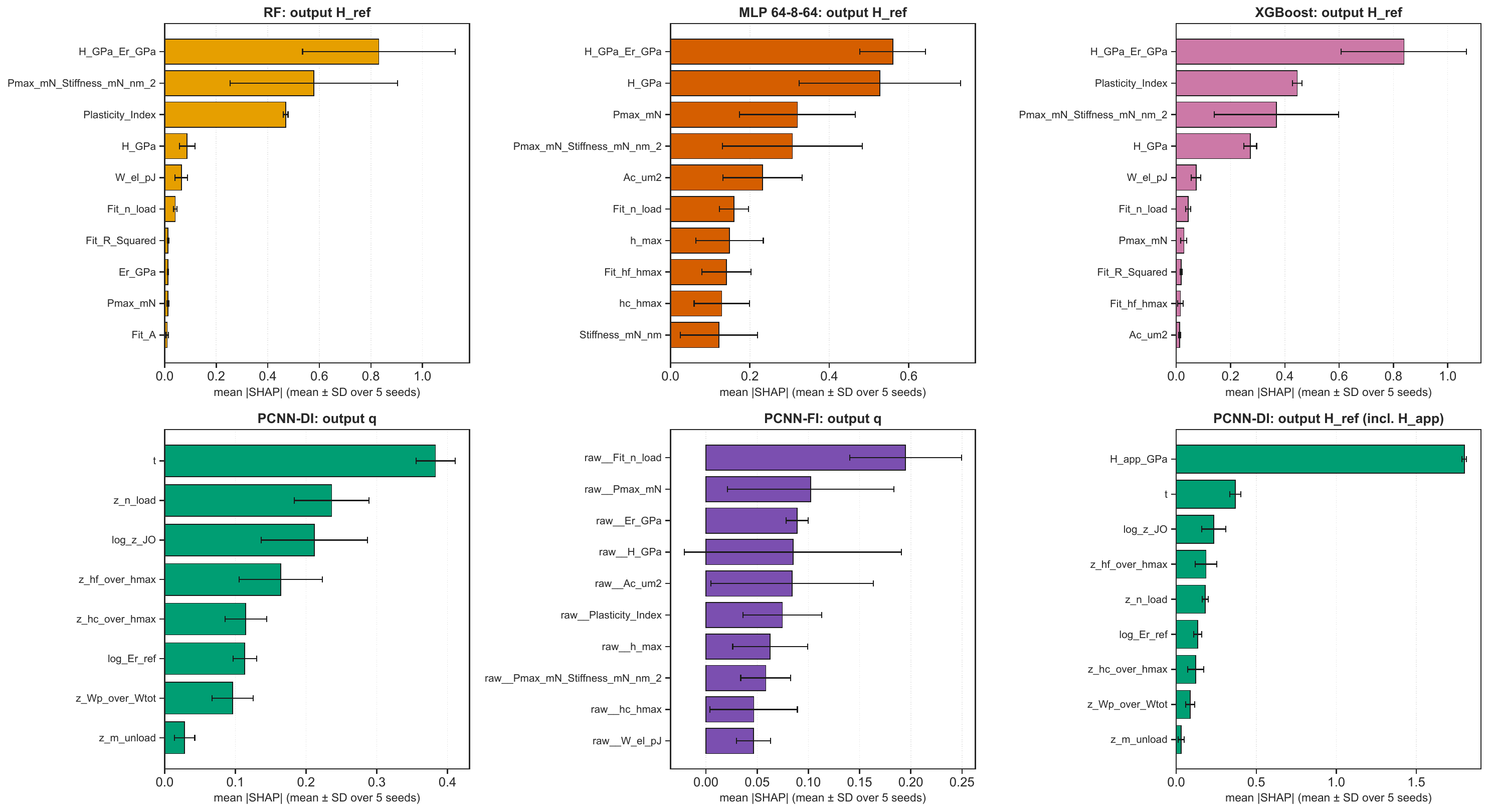}
  \caption{Mean absolute SHAP value per input (ten leading inputs),
  mean $\pm$ SD over five run seeds, for the Random Forest, MLP 64-8-64,
  and XGBoost (output $\widehat{H}_{\mathrm{ref}}$), the PCNN-DI and
  PCNN-FI (output $q$), and the PCNN-DI (output
  $\widehat{H}_{\mathrm{ref}}$, including $H_{\mathrm{app}}$).
  Explained rows: development set; background: k-means (50) of the
  original training rows.}
  \label{fig:S_shap_seeds}
\end{figure}







\end{document}